\documentclass[aps,prb,amsmath,superscriptaddress,amssymb,longbibliography,twocolumn, 10pt]{revtex4-2}

\usepackage{graphicx}
\usepackage{dcolumn}
\usepackage{amsmath}
\usepackage{bm}
\usepackage[breaklinks,colorlinks = true,linkcolor = red,urlcolor=cyan,citecolor=red]{hyperref}
\usepackage{multirow}
\usepackage{array}
\usepackage{booktabs}
\usepackage{ctable}
\usepackage{upgreek}
\usepackage{epsfig,psfrag,subfigure,amsopn}
\usepackage{mathrsfs}
\usepackage{amssymb}
\usepackage{amsbsy}
\usepackage{color}
\usepackage{xcolor}
\usepackage{cancel}
\usepackage{pifont}
\usepackage{marginnote}
\usepackage{float}
\usepackage{titlesec}

\begin{document}


\title{Variational wave-functions for correlated metals}
\author{Ankush Chaubey}
\affiliation{International Center Of Theoretical Science, Tata Institute of Fundamental Research, Bangalore 560089, India.}
\author{Harsh Nigam}
\affiliation{International Center Of Theoretical Science, Tata Institute of Fundamental Research, Bangalore 560089, India.}
\author{Subhro Bhattacharjee}
\affiliation{International Center Of Theoretical Science, Tata Institute of Fundamental Research, Bangalore 560089, India.}
\author{K. Sengupta}
\affiliation{ School of Physical Sciences, Indian Association for the Cultivation of Science,
2A and 2B Raja S.C.Mullick Road, Jadavpur, Kolkata-700032, West Bengal, India.}

\begin{abstract}
    We study a set of many-body wave-functions of Fermions that are naturally written using momentum space basis and allow for quantum superposition of Fermion occupancy, $\{n_{\bf k}\}$. This {enables} us to capture the fluctuations of the Fermi-surface {(FS)} -- the singularly most important signature of a metal. We bench-mark our results in one spatial dimensions (1D) to show that  these wave-functions allow for quantitative understanding of the Tomonaga-Luttinger liquid (TLL); computations of certain correlators using them can in fact be extended to larger systems sizes compared to conventional exact diagonalization (ED) allowing for a more systematic comparison with bosonization techniques. Finally we show that this basis  may be useful for obtaining fixed-point wave-function for strongly correlated metals {in dimensions greater that one}. In particular, we study the case of coherent (equal) superposition of elliptical FS {in continuum (2D) and on a} square lattice{. In case of the former, our variational wave-function systematically interpolates between the phenomenology of the Fermi liquid ground state, {\it i.e.}, finite single-Fermion residue at a sharp FS, to a non-Fermi liquid (NFL) with zero residue. In the NFL the jump in $\langle n_{\bf k}\rangle$ at the FS is replaced by a point of inflection (similar to a 1D TLL) whose contour is consistent with the Luttinger Theorem. In case of the square lattice, we} find highly anisotropic distribution of the quasi-particle residue, which, at finite resolution has an uncanny resemblance to the Fermi-arcs{, albeit at zero temperature,} seen in the pseudo-gap state of the cuprates. 
\end{abstract}

\maketitle

\section{Introduction} 

Our understanding of metallic phases squarely rest on the idea of Fermi-surface (FS). For non-interacting Fermions, it is sharply determined by the average single particle occupation in momentum space, $\langle n_{\bf k}\rangle$, which in turn depends on the single-particle energy $\varepsilon_{\bf k}-\mu$ where $\mu$ is the chemical potential, {and} $\varepsilon_{\bf k}$ is the single-particle dispersion. The shape and volume of the FS are determined by the underlying symmetry and filling respectively~\cite{ashcroft1976solid}. The corresponding ground state (GS) wave-function has a product structure in momentum space corresponding to non-trivial entanglement in real space~\cite{PhysRevLett.96.100503}. 

In the presence of a moderate short-ranged four Fermion interaction, the above description survives and forms the backbone of Landau-Fermi liquid (FL) theory~\cite{landau2013course,RevModPhys.66.129,polchinski1992effective} in spatial dimensions $d>1$. Indeed the single particle residue, $Z_{\bf k}$, obtained from the jump in $\langle n_{\bf k}\rangle $ at the FS, and the associated Luttinger theorem~\cite{PhysRev.118.1417,PhysRevLett.84.3370} provides a definition for the long-lived gapless electron-like Landau quasi-particles near the FS in such a FL  which then can be described via an effectively weakly interacting framework. In a FL, $0<Z_{{\bf k}_F}<1$, such that the FS remains sharply defined via the position of the jump in occupation inspite of fluctuations in $n_{\bf k}$ near the FS. One then expects that the corresponding GS is very well approximated via short-range entangled states in the momentum space across the non-interacting FS such that the real-space entanglement signatures are similar to that of the non-interacting case~\cite{PhysRevX.2.011012}. 

In this paper, we show that {\it quantum fluctuations} arising from the coherent superposition of wave-functions with different momenta occupation $\{n_{\bf k}\}$ can lead to a loss of FS along with the the loss of Landau quasi-particle. We explicitly construct variational wave-functions for such {\it quantum fluctuating} FS, which, we note, is  different from the dynamical fluctuation of a classical membrane. Instead, in this case, the location of the membrane (the FS) has a uncertainty in the momentum space. This  superposition is then closely related to the  dimer coverings in Rokhsar-Kivelson (RK) models~\cite{PhysRevLett.61.2376}, but applied to momentum space for the FS. Such wave-functions can form a starting point for understanding the loss of the Landau quasi-particle leading to a so-called non-Fermi liquid (NFL) phase. Such phases generically arise in one dimensional (1D) interacting Fermion systems and are collectively known as Tomonaga Luttinger liquids (TLL)~\cite{haldane1981luttinger,giamarchi_book}. The quasi-particle residue, for these TLLs, vanishes at the FS and is replaced by a point of inflection with $\langle n_{\bf k}\rangle\sim |{\bf k}-{\bf k}_{\rm F}|^p$ near the FS with $p>0$ being related to the Luttinger parameter~\cite{giamarchi_book} (also see below). Thus while the Landau quasi-particle is lost, the FS is still sharply defined in accordance with the Luttinger theorem~\cite{PhysRevLett.79.1110}. In higher dimensions ($d>1$) similar phenomena, {\it i.e.,} loss of quasi-particles can occur with a well defined {\it critical FS}~\cite{PhysRevB.78.035103,PhysRevB.2.4302,si2010quantum} which arises at Fermionic quantum critical point~\cite{PhysRevB.78.035103,PhysRevB.78.035104,zaanen2008quantum,RevModPhys.79.1015} between the large and small FSs in heavy Fermion systems~\cite{PhysRevLett.72.3262,RevModPhys.79.1015,gegenwart2008quantum,custers2003break,coleman2005quantum}  or Mott transition in correlated metals~\cite{PhysRevLett.95.177001,PhysRevLett.95.077001} where the entire FS disappears on approaching from the metallic side~\cite{PhysRevB.78.035103,PhysRevB.2.4302}.

For $d>1$, effective field theory calculations~\cite{lee2018recent} of Fermions (at finite density) interacting with gapless bosons near a metallic quantum critical point or in a gapless phase (such as in the U(1) quantum spin liquid with a spinon-FS~\cite{polchinski1994low,PhysRevB.80.165102}) indicate that due to the presence of the tangential scattering with the bosons, even at ultra-low energies, the Landau quasi-particle is killed possibly leading to a  NFL~\cite{hertz2018quantum,PhysRevB.81.035117,RevModPhys.79.1015,RevModPhys.73.797,PhysRevLett.70.3339} particularly in 2D~\cite{PhysRevB.40.11571,PhysRevB.50.14048,PhysRevLett.63.680,PhysRevB.80.165102,PhysRevB.82.075128,PhysRevB.82.045121,PhysRevB.88.245106,PhysRevB.90.045121,lee2018recent}. However, the fate of the infra-red (IR) fixed point of such theories is not quite settled due to lack of control over re-summation procedures; this is the case  even in large-$N_f$ limit with $N_f$ being the flavour of Fermions. This calls for new ideas in understanding such a NFL state which is parent to a large number of diverse strongly correlated phenomena; this is the central motivation for the variational wave-function approach that we study here.

The variational wave-functions that we propose here explicitly capture the partial or complete loss of the single-Fermion residue signalling the loss of the Landau quasi-particle at low energies. In 1D, we show that we can re-interpret the existing numerical and field-theoretic understanding for the 1D  TLL~\cite{giamarchi_book} for spinless Fermions within the above variational wave-functions (Figs.\ \ref{fig:amplitude_main} and \ref{fig_xxzcompare}). This can be achieved by re-writing such wave-functions in the basis of fluctuating $\{n_{\bf k}\}$ and comparing the average occupation, $\langle n_{\bf k}\rangle$ with predictions from bosonization. In addition, the entanglement signatures of such wave-functions can also be shown to match with those computed from TLL. Interestingly, following RK construction of dimer models, we can construct minimal projector Hamiltonians, albeit in momentum space, where the GS allows large fluctuations, both in 1D and 2D, in the occupation $\{n_{\bf k}\}$. These wave-functions, at partial filling, do not break any symmetry and are devoid of sharp FS like signatures in momentum space. Therefore, in accordance with Lieb-Schultz-Mattis (LSM)~\cite{lieb1961two,PhysRevLett.79.1110} theorem in 1D or their higher dimensional extensions~\cite{PhysRevLett.84.3370,PhysRevB.69.104431}, they possibly correspond to GSs of gapless phases of correlated metals (in higher dimensions, in principle, gapped topological order is also allowed). We study the generic properties of such wave-functions in 2D that correspond to smooth deformation of the FS as an example of fixed-point wave-functions for correlated metals without Landau quasi-particles and isotropically or anisotropically blurred FS. In particular, the anisotropic loss of the single-particle residue leads to Fermi-arc like phenomena, albeit at zero temperature, similar to underdoped cuprates and direction dependent Friedel oscillations.

In the rest of this paper, we provide the details of the  construction of the variational wave-functions starting with the single particle distribution in momentum space and use it to provide initial insights into their nature via selected examples in 1D and 2D. In section \ref{sec_basis} we provide the details of construction of the variational wave-function which is a straightforward extension of superposing single-particle Slater determinants. Such superposition generically leads to fluctuations in momentum space occupation $\{n_{\bf k}\}$ and this forms the right basis (Eq.\ \ref{eq_wftheta}) to capture the quantum fluctuation of the FS (Eq.\ \ref{eq_superposedfswf}) as is evident from the generic interacting Fermionic Hamiltonian, when written (Eq.\ \ref{eq_intHam_matrix}) in this basis. In Sec.\ \ref{sec_tll} we show that the new basis provides a useful momentum-space truncation scheme to understand the physics of TLL in a system of spinless Fermions in 1D. In particular, we characterize the relation between the fluctuations of the FS and the loss of the Fermionic quasi-particle residue by quantitatively comparing with bosonization for larger system sizes (Figs.\ \ref{fig_xxzcompare}, \ref{fig_qc} and \ref{fig_xxzcompare1}). Having obtained a description of the TLL in $d=1$, we turn to the general properties of the wave-functions of the type in Eq.\ \ref{eq_superposedfswf}. We mainly restrict ourselves to description of a class of effective projector models in Sec.\ \ref{tmod1} whose GSs allow {\it equal} superposition of various $\{n_{\bf k}\}$ distributions; we note that these wave-functions are qualitatively different from those describing a free Fermi gas GS. This naturally leads to possible fixed point Fermion wave-functions in Sec.\ \ref{fpt1} that are GSs similar to that of suitable RK-type projector Hamiltonians that describe strongly correlated NFLs; such NFL may be obtained near a metallic phase transition due to the Yukawa interactions with gapless bosons as discused above. Our study in Secs.\ \ref{tmod1} and \ref{fpt1} may be relevant for some of these systems. We discuss various properties of a selected class of such wave-functions emphasising their NFL features. Various technical details and supporting plots are given in the different appendices.

\section{The variational wave-function}
\label{sec_basis}

In order to define the wave-function, we consider a finite lattice of linear size $N$ with periodic boundary conditions.  In $d$-dimensions, the Brillouin zone (BZ) is $d$-tori with a grid of spacing $2\pi/N$ along each direction. For concreteness we restrict ourselves to $d=1, 2$ while extension to higher dimensions are straightforward. With such a discretized momentum grid, we can now re-write the $N_p$-particle occupation states (analogues to the Slater determinants) in the momentum-space occupation basis by defining an Ising variable $\theta_{\bf k}=\pm 1$ and using them to define
\begin{align}
    |\Psi [\theta]\rangle=(c^\dagger_{\bf k_B})^{(1+\theta_{{\bf k}_B})/2}\cdots (c_{\bf k_1}^\dagger)^{(1+\theta_{{\bf k}_1})/2}|0\rangle, 
    \label{eq_wftheta}
\end{align}
where the momenta are arranged from smallest to largest by assigning a particular ordering  in which ${\bf k}_1<\cdots<{\bf k}_B$ over the entire BZ. Here $c_{\bf k}^\dagger$ are spinless Fermion creation operators following usual Fermion algebra $\{c_{\bf k},c^\dagger_{{\bf k}'}\}=\delta_{{\bf k},{\bf k}'}$ and $\{c_{\bf k},c_{{\bf k}'}\}=0$.  In the above wave-function modes where $\theta_{\bf k}=+1$ are occupied while $\theta_{\bf k}=-1$ are empty. The above set of wave-functions form an orthonormal basis, {\it i.e.} $\langle\Psi[\theta]|\Psi[\theta']\rangle=\delta_{[\theta],[\theta']}$. The Slater-determinant form of the corresponding real-space representation and matrix elements of various many-body operator is easily represented in this basis and are shown in Appendix \ref{appen_basis} for completeness. In particular, the total particle number and the net momenta are given by
\begin{align}
        N_p=\sum_{\rm {\bf k}\in BZ}\frac{1+\theta_{\bf k}}{2},~~~~{\rm and}~~~~~{\bf k}_{\rm T}=\sum_{\rm {\bf k}\in BZ}~{\bf k}\frac{1+\theta_{\bf k}}{2}.
    \label{eq:particleno}
\end{align}
Note that the second relation in Eq.\ \ref{eq:particleno}  automatically implies that the momentum of a empty/filled band is zero. Generically, for a given filling ($\nu=N_p/N^d$), the number of above basis states increases as $^{N^d}C_{N_p}$ with $N$ and is therefore of little value for numerical calculations. However, it  provides an alternate way to construct $N_p$-particle Fock space basis states in terms of their momentum space occupation. In terms of these new variables $\theta_{\bf k}$,  a single FS stands for a closed volume in BZ separating $\theta_{\bf k}=+1$ (occupied) and $\theta_{\bf k}=-1$ (unoccupied) regions such that we recover the familiar FS wave  function.

Using the $N_p$-particle basis states in Eq.~\ref{eq_wftheta}, we can write down the generic state
\begin{align}
|\psi\rangle=\sum_{[\theta]|\sum_{\bf k}\theta_{\bf k}/N^d=\nu}\psi[\theta]|\Psi[\theta]\rangle, 
\label{eq_superposedfswf}
\end{align}
where $\psi[\theta]$ is the amplitude for the state with the occupancy distribution $[\theta]$ and each basis state obeys Eq.\ \ref{eq:particleno}. In this sense, Eq.~\ref{eq_superposedfswf} is a momentum space {\it wave-functional} which is equivalent to superposing Slater determinants. The above wave-functional is the central object in this work which captures the idea of quantum fluctuation of the FS. However, at this stage Eq.\ \ref{eq_superposedfswf} is too generic; in the remaining sections, we therefore focus on its application to particular cases.

To this end, we consider an interacting four (spinless) Fermion Hamiltonian $H = H_0 + H_I$, where 
\begin{eqnarray}
H_0 &=& \sum_{\textbf{k}}E_{\textbf{k}}c^{\dagger}_{\textbf{k}}c_{\textbf{k}} ,\nonumber\\
H_{I} &=& \sum_{\textbf{k}_1,\textbf{k}_2,\textbf{q}\neq 0}V_{\textbf{q}}~~c^{\dagger}_{\textbf{k}_1}c^{\dagger}_{\textbf{k}_2}c_{\textbf{k}_2-{\bf q}}c_{\textbf{k}_1+{\bf q}} , \label{eq_intham}  
\end{eqnarray} 
where $E_{\bf k}= (\varepsilon_{\bf k} -\mu)$ is the single-particle dispersion and $V_{\bf q}$ is the interaction kernel for the density-density interaction. In the basis of Eq.\ \ref{eq_wftheta}, the above Hamiltonian can be written in the $N_{p}$-particle sector as  
\begin{eqnarray}
H &=& \sum_{\{\theta\}}\sum_{\{\theta'\}}|\Psi[\theta]\rangle \Big[E[\theta] \delta_{[\theta],[\theta']}\nonumber\\
&& + E_I([\theta], [\theta'])(1-\delta_{[\theta],[\theta']})\Big]  \langle\Psi[\theta']|, \label{eq_intHam_matrix}\\
E[\theta] &=&\langle\Psi[\theta]|H_0|\Psi[\theta]\rangle ,
E_I([\theta], [\theta']) = \langle\Psi[\theta]|H_I|\Psi[\theta']\rangle,\nonumber 
\end{eqnarray}
where the first term in the expression of $H$ denotes the non-interacting part which, as expected, is diagonal in the momentum space occupation basis. The details of the form of $E[\theta]$ is given by Eq.\ \ref{free_ham_matrix_element1} in Appendix \ref{appen_Hamiltonian}. From Eq.\ \ref{eq_intHam_matrix}, one can easily see that the $[\theta]$ configuration which yields minimum energy is given by (within the constraint of Eq.\ \ref{eq:particleno})  choosing ${\rm Min}\left(E[\theta]\right)$. The corresponding distribution $[\theta]$ gives rise to the FS of the free electron gas which typically is a smooth but sharply defined surface in the thermodynamic limit ($N\rightarrow\infty$) dictated by the ultra-violet(UV) symmetry and energetics via the form of $E_{\bf k}$. 
 
The second term in Eq.~\ref{eq_intHam_matrix} corresponds to the four-Fermion interaction which has off-diagonal contributions.  It's explicit form is given by Eq.~\ref{eq_off_diagonal_ham_AC}. Thus interactions lead to generic superposition of distributions of $[\theta]$ having the same filling and total momentum (Eq.~\ref{eq:particleno}). It is easy to see that in the case of the four Fermion interactions, the matrix elements of $H_I$ are non-zero only between two distributions of $[\theta]$ that differ by occupation at four momenta (up to overall momentum conservation); this has been shown explicitly in  Appendix \ref{appen_Hamiltonian}. The matrix corresponding to $H_I$ therefore generically leads to a sparse $^{N^d}C_{N_p}\times\, 
 ^{N^d}C_{N_p}$ matrix in this basis.

Other possible alternatives forms of the interaction stem from the electrons interacting with a gapless bosonic mode, $\phi_{\bf q}$, (phonon/gauge field/collective modes) via typically a Yukawa type interaction $\sim \sum_{\bf k,q}V({\bf q})\phi_{\bf q}c^\dagger_{\bf k+q}c_{\bf k}$; these interacting Hamiltonians also have a sparse off-diagonal structure once the bosons are integrated out. Such Yukawa interactions are renormalization group (RG) relevant and can lead to NFLs in d=2~\cite{PhysRevB.40.11571,PhysRevB.50.14048,PhysRevLett.63.680,PhysRevB.80.165102,PhysRevB.82.075128,PhysRevB.82.045121,PhysRevB.88.245106,PhysRevB.90.045121,lee2018recent}. The central distinguishing feature of such Yukawa interactions in comparison to the short range four-Fermion interactions is the presence of tangential scatterings; their RG relevance indicates the collective effect of such scattering which would necessarily change the Fermion distribution about the non-interacting FS. This leads to the quantum fluctuation of the FS as suggested by the form of Eq.~\ref{eq_intHam_matrix}.

Below, we assume a generic form of $\langle\Psi[\theta]|H_I|\Psi[\theta']\rangle$ and understand the nature of the resultant GSs. We note that the simplest case where the four-Fermion interactions can already lead to a TLL~\cite{haldane1981luttinger,giamarchi_book} in $d=1$. So, before we proceed to study the generic properties of the wave-function in Eq.\ \ref{eq_superposedfswf}, we benchmark it to show that the TLL can indeed be understood in terms of fluctuating FS. Note however, in $d=1$, fluctuations in the occupancy necessarily leads to disconnected segments of $\{n_{\bf k}\}$ which is not the generic case in higher dimensions.

\section{Interacting Fermions in 1D: The TLL} 
\label{sec_tll}

\begin{figure*}
    \centering
    \includegraphics[width=2\columnwidth]{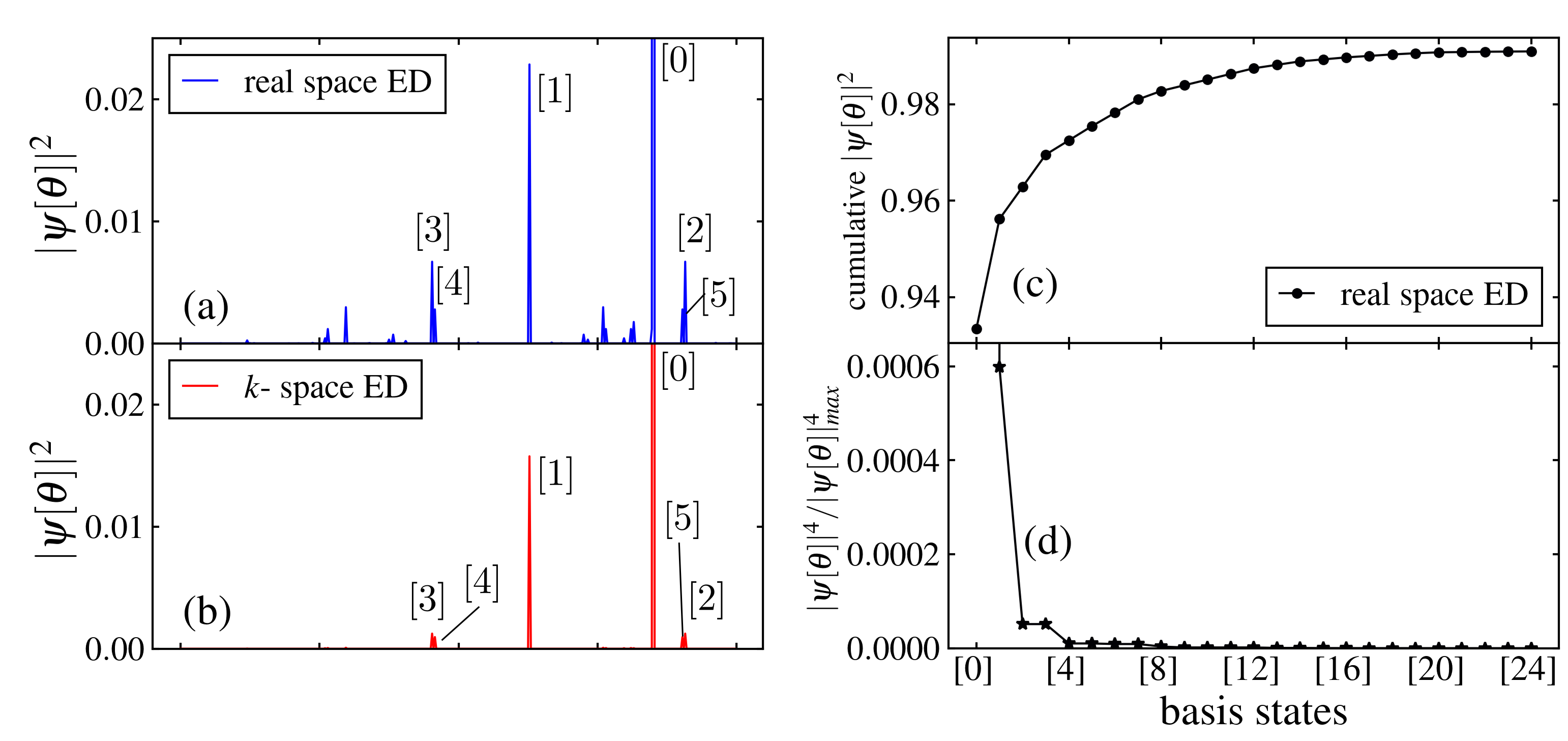}
    \caption{ (a) The probabilities of the GS wave-function obtained from ED of the half filled Hubbard model ($N=14, t=1, V_0=0.3$) for spinless Fermions (see main text and Appendix \ref{appen_ed1d} for details) in the basis given by Eq.\ \ref{eq_wftheta} (horizontal axis). The relevant momenta in the first BZ are denoted by $k=-\pi+\alpha\pi/7$ with $\alpha=0,1,\cdots,13$. In this labelling {scheme}, the six states with leading contributions are shown as $[0]\equiv [00001111111000]$ being the non-interacting GS and the rest corresponding to the particle-hole excitations above it. These are $[1]= [00010111110100]$, $[2]= [00011011110010]$, $[3]=[00100111101100]$, $[4]= [00011101110001]$ and $[5] = [01000111011100]$. The states are arranged in terms of the increasing number of particle-hole excitations which can also be quantified via their root mean square momenta deviation from the FS as defined in Table \ref{tab_normalisedwieights} in Appendix \ref{appen_ed1d}. Note that the vertical scale has been truncated for improved visibility of the contributing basis states. The normalised weights are given in Table \ref{tab_normalisedwieights}. (b) The exact diagonalization done using the momentum space cut-off ${\bf q}_c$ ($n_c = 3\left(q_c=\frac{2\pi}{N}n_c\right),~n_0 = 1.5\left(q_0=\frac{2\pi}{N}n_0\right)$ (see main text for details). (c) The cumulative weights of the different basis states. (d) The IPR for the different basis states. The IPR has been normalised with respect to the maximum contribution (vertical axis truncated for better visibility) from $[0]$.}
    \label{fig:amplitude_main}
\end{figure*}

At finite but moderate $V_{\bf q}$ the  Hamiltonian in Eq.~\ref{eq_intham} supports a TLL. This is realized, in particular, in a microscopic model of spinless Fermions at half filling ($N_p/N=\nu=1/2$) consisting of nearest neighbour hopping such that the single particle dispersion (in Eq.\ \ref{eq_intham}) is given by  $E_{\bf k}= - 2 t \cos k$. In what follows, we shall use an interaction potential $V_{\bf q}=V_0 e^{-q^2/(2q_0^2)}$, where $q_0$ denotes an upper cutoff for momentum around the FS. This, in real space corresponds to a density-density interactions, {\it i.e.}, $\sum_{i,j, i\ne j} n_i n_{j} V_{|i-j|}$, where $V(x)= V_0 \exp[-x^2/(2 a^2)]$, where $a \sim 1/q_0$ is the lattice spacing. We note that the nearest neighbor density-density interaction is obtained from this model in the limit $a\to 0$. 

We use exact diagonalization (ED) technique to study the GS properties of these interacting Fermions on a finite size chain (for details, see Appendix\ \ref{appen_ed1d}) for nearest neighbour density-density interaction $V_0\sum_{i}n_in_{i+1}$. The GS for a finite chain can indeed be written as a superposition of the Fermionic wave-functions as envisaged in Eq.\ \ref{eq_superposedfswf}. This is shown in Fig.\ \ref{fig:amplitude_main} for $N=14$ and $\nu=1/2$ where we plot the overlap probabilities, $|\psi(\theta)|^2$, of the exact GS wave-function, obtained using ED, with $|\psi(\theta)\rangle$.  Fig.\ \ref{fig:amplitude_main}(a) shows the the six leading order contributions to $|\psi(\theta)|^2$ obtained through real space ED. These states, as expected, consist of the non-interacting GS and  particle-hole excitations above it. Further details of these overlaps are shown in Table\ \ref{tab_normalisedwieights} in Appendix\ \ref{appen_ed1d} which include up to $25$ basis states ordered in terms of their increasing RMS momenta and indicating number of particle-hole excitations about the FS. The cumulative weights of these states and their inverse participation ratios are shown in Fig.\ \ref{fig:amplitude_main}(c) and (d) respectively.

\begin{figure*}
    \centering
\includegraphics[width=2\columnwidth]{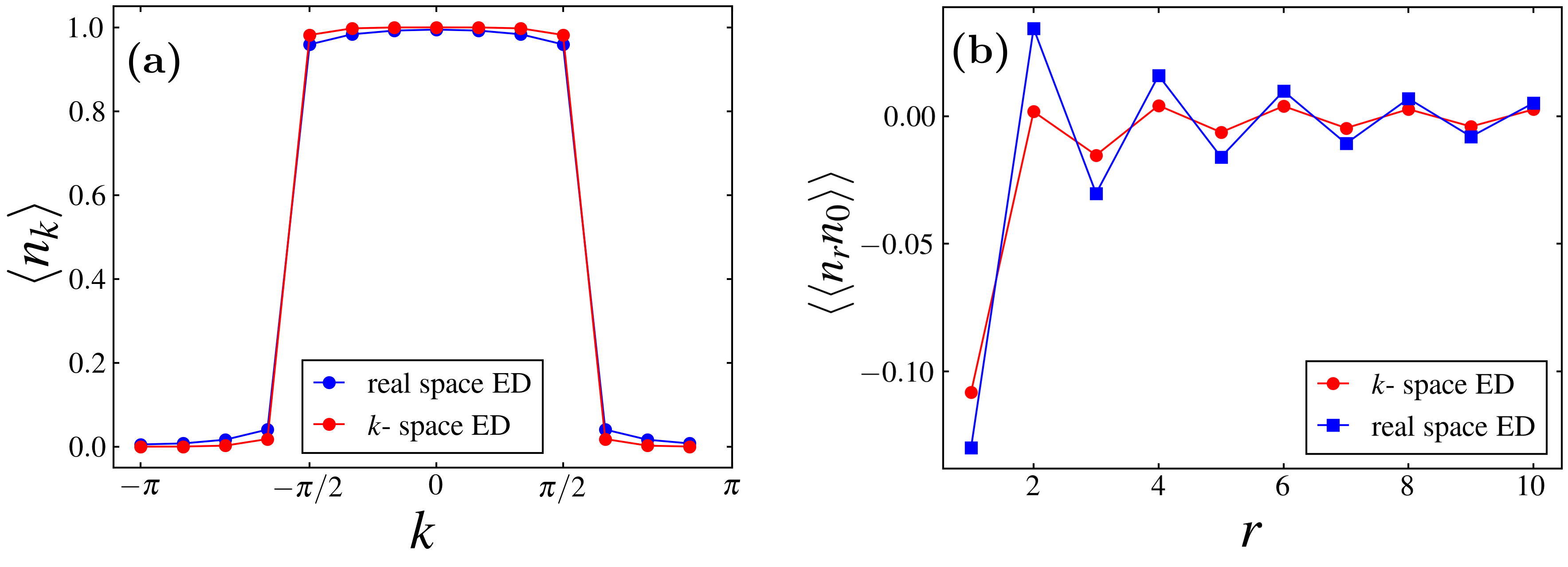}
\caption{Comparison of (a)  $\langle n_{\bf k}\rangle$ calculated from real space ED and $k$-space ED ($N=14$) and (b) the two-point {connected correlator $\langle \langle n_{\bf r} n_0\rangle \rangle$} in real space ($N=22$). The other parameters are $t=1.0,~V_0=0.3,~n_c=3\left(q_c=\frac{2\pi}{N}n_c\right),~n_0=1.5\left(q_0=\frac{2\pi}{N}n_0\right)$.}   
\label{fig_xxzcompare}
\end{figure*}

It is clear from Fig.~\ref{fig:amplitude_main} that only the basis states $|\Psi[\theta]\rangle$ whose distribution in $[\theta]$ differs from that of the non-interacting FS (denoted by $[0]$ in Fig.\ \ref{fig:amplitude_main}) only near its FS has appreciable overlap with the interacting GS wave-function. In fact Figs.~\ref{fig:amplitude_main}(c) and (d) show that the weight of the GS wave-function is saturated appreciably solely by basis states with particle-hole excitations near the non-interacting FS. Taking cue from this, we performed the ED directly in the occupation basis (Eq.\ \ref{eq_wftheta}) in momentum space (for details, see Appendix\ \ref{appen_ed1d}).  This is  analogous to momentum-space ED adapted to the FS by taking $V_{\bf q}=V_0 e^{-q^2/(2q_0^2)}$ with a soft (${\bf q_0}$) as well as a hard cutoff on ${\bf q}={q}_{c}\hat{\bf x}$ for $H$. The resultant contributions of the basis states is shown in Fig.\ \ref{fig:amplitude_main}(b); these serve as a benchmark of the momentum space ED in terms of the weights of different configurations since they can be compared with their real-space counterparts. We can further benchmark the momentum space ED by comparing different correlators with those obtained from the real space ED. In this work we calculate the average momentum occupation, $\langle n_{\bf k}\rangle$, the two-point connected correlator, 
\begin{eqnarray} 
\langle\langle n_{\bf r}n_{0}\rangle\rangle &=& \langle n({\bf r}) n(0)\rangle - \langle n({\bf r}) \rangle \langle n(0)\rangle, \label{tpcc} 
\end{eqnarray} 
 and the bipartite von-Neumann entanglement entropy, $S_{vN}$.

\begin{figure}
\includegraphics[width=1\columnwidth]{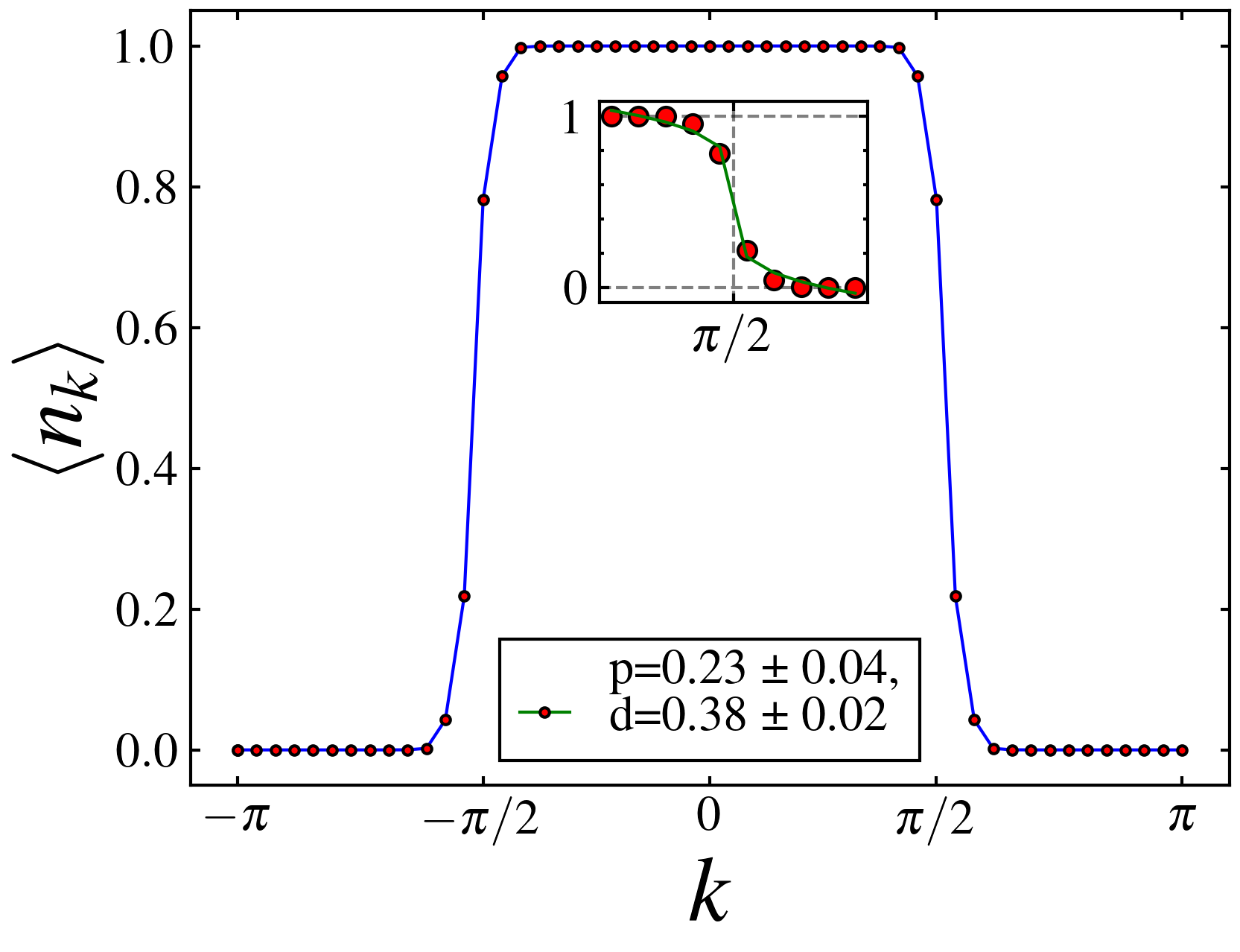}%
    \caption{ 
    A plot of the Fermi distribution function $\langle n_{\bf k}\rangle$ as a function of $k$; the inset shows the curve fitting of $\langle n_{\bf k}\rangle$ (near the Fermi point) with known bosonization result (near the Fermi point: $\langle n_{\bf k}\rangle = d~\text{sign}(k_f-k) |k-k_f|^p + 1/2$ with $p=(K+K^{-1})/2 -1$ \cite{giamarchi_book} where $K$ is the Luttinger parameter and $d$ is a constant). For this plot, $t=1.0,~V_0=0.6,~n_0=1.5,~n_c=5\left( q_c=\frac{2\pi}{N}n_c\right)$ and $N=50$.}
     \label{fig_qc}
\end{figure}

\begin{figure*}
    \centering

\includegraphics[width=2\columnwidth]{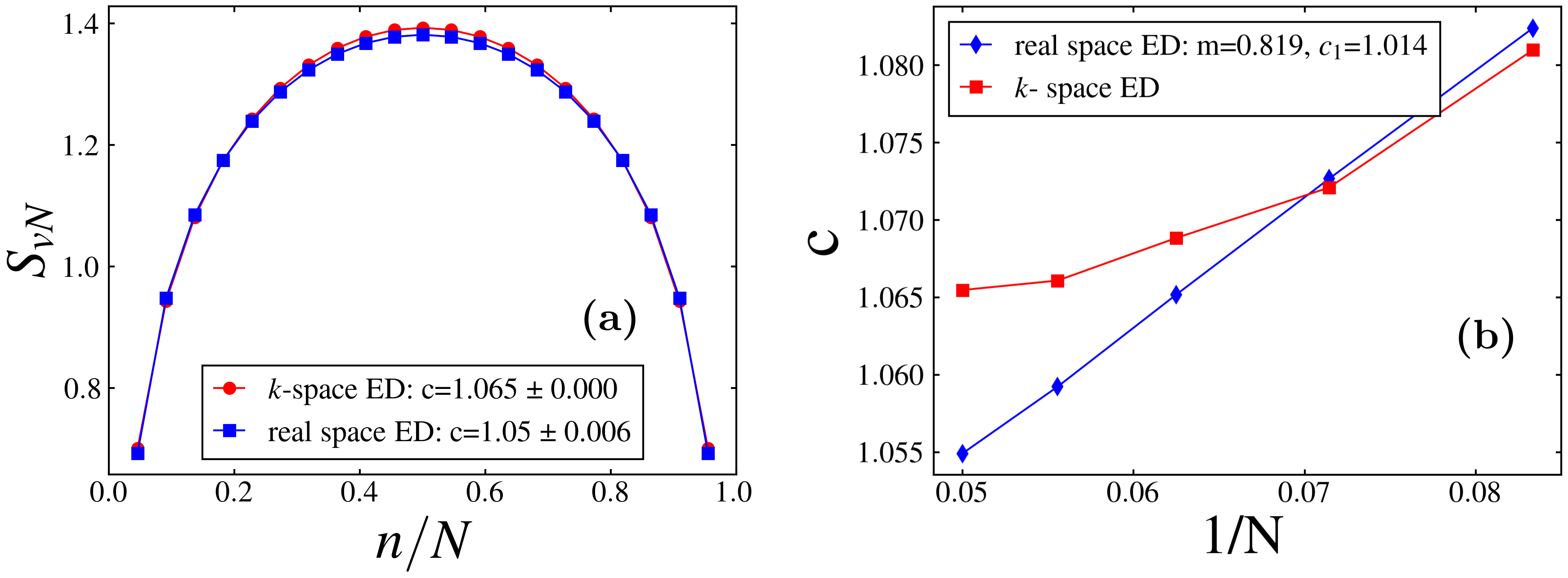}
   
\caption{Comparsion of (a) Entanglement Entropy, (b) Central charge variation with $1/N$ (we have done linear curve fit ($c= m (1/N) + c_1$) for real space data) for the two ED methods. The system parameters are  $t=1.0,~V_0=0.3,~n_c=3\left(q_c=\frac{2\pi}{N}n_c\right),~n_0=1.5\left(q_0=\frac{2\pi}{N}n_0\right)$ and $N=22$.}   
\label{fig_xxzcompare1}
\end{figure*}

The plot of $\langle n_{\bf k}\rangle$, shown in Fig.\ \ref {fig_xxzcompare} (a), provides a comparison between the two methods. The results agree well and in fact the momentum space calculations with a cut-off about the FS can be used for much larger systems sizes as shown in Fig.\ \ref{fig_qc} for $N=50$ and $q_c=2\pi/10$. This, therefore, can be compared with the standard bosonization results for $\langle n_{\bf k}\rangle$ which is devoid of a jump and has a form $\langle n_{\bf k}\rangle\sim |{\bf k}-{\bf k}_{\rm F}|^p$ with $p=(K+1/K)/2-1$ with $K$ being the Luttinger parameter~\cite{giamarchi_book}. An estimation of the point of inflection, obtained from such a comparison, is in accordance with the Luttinger theorem. Thus the basis in Eq.\ \ref{eq_wftheta} effectively captures the TLL GS which has a sharp FS denoted by the point of inflection, but no Landau quasi-particle. This further confirms some advantage in using the  basis in Eq.\ \ref{eq_wftheta} and validates the fact that the low energy physics is dominated by FS fluctuations.

For the two-point correlator in real space (Fig.\ \ref{fig_xxzcompare}(b)) and the entanglement (Fig.\ \ref{fig_xxzcompare1}), the advantage of our momentum space ED however is much less clear at this point. For example, while the two-point correlator can be calculated for larger system sizes, it suffers from spurious oscillations arising from the cut-off, ${\bf q}_c$, for parameters similar to Fig.\ \ref{fig_qc}. For the entanglement entropy, on the other hand, we are still restricted to smaller sizes due to calculational costs. In addition, also due to the cut-off, the estimation of the central charge (Fig.\ \ref{fig_xxzcompare1}(b)) is somewhat crude. Therefore, while promising, our momentum-space ED scheme with cut-off around the FS requires more systematic investigation in future.

\section{Exactly solvable toy model for GS with FS superposition} 
\label{tmod1} 

Having shown that the basis in Eq.\ \ref{eq_wftheta} may be useful to capture the physics of TLL in 1D, in this section, we study another particularly useful question-- are there parent Hamiltonians which allows massive FS fluctuation in the GS ? The positive answer is provided in a simple limit of the Hamiltonian in Eq.\ \ref{eq_intHam_matrix}. This is  obtained when, for a particular filling $\nu(=N_{\rm P}/N^d)$,
\begin{align}
    \frac{E[\theta]}{E_I([\theta],[\theta'])}=-2\left(^{N^d}C_{N_p}-1\right),
    \label{eq_rkcondition}
\end{align}
where $E[\theta]$ and $E_I([\theta'],[\theta])$ are respectively the diagonal (non-interacting) and off-diagonal (interacting) matrix elements (Eq.\ \ref{eq_intHam_matrix}). In this limit, the Hamiltonian then can be re-written as a sum of projectors $h([{\theta}], [{\theta'}])$ as 
\begin{align}
    H_{\rm RK}
    &=\mathcal{C}\sum_{[\theta],[\theta'],[\theta]\neq
[\theta']} h([{\theta}],[{\theta'}]), 
\label{eq_rkham}
\end{align}
where $\mathcal{C}>0$ is a filling dependent positive constant and 
\begin{align}
    h([{\theta}],[{\theta'}]) = \left(\frac{\left(|\Psi[\theta]\rangle-|\Psi[\theta']\rangle\right)}{\sqrt{2}}\right)\left( \frac{\left(\langle\Psi[\theta]|-\langle\Psi[\theta']|\right)}{\sqrt{2}} \right), 
\end{align}
with $h^2([\theta],[\theta']) = h([\theta],[\theta'])$. We note that the condition in Eq. \ref{eq_rkcondition} may be obtained in systems with all-to-all-interactions.

The Hamiltonian in Eq.\ \ref{eq_rkham} can be solved in the same spirit as the projector dimer model Hamiltonians by Rokhsar and Kivelson (RK)~\cite{PhysRevLett.61.2376}. The difference between the two models comes from the fact that in the present case the $[\theta]$ basis is orthonormal unlike the hardcore dimer-covering basis of the quantum dimer models. The GS wave-function is then easily read off; it is given by the state which are annihilated by $h([\theta],[\theta'])$ for all pairs of $([\theta],[\theta'])$ and constitutes an equal and completely symmetric superposition

\begin{align}
    |\psi_{\rm RK}\rangle = \frac{1}{\sqrt{N_{\rm RK}}}\sum_{[\theta]}|\Psi[\theta]\rangle, 
    \label{eq_rkstate}
\end{align}
with $N_{\rm RK}$ being the number of states in the superposition. Notably $\langle\Psi_{\rm RK}|\hat n_{\bf k}|\Psi_{\rm RK}\rangle=\nu$ for all ${\bf k}\in { \rm BZ}$. Thus, an equal superposition of all $|\psi(\theta)\rangle$ in momentum space leads to loss of any sharp structure and hence the FS.  In fact, this is not very surprising because the condition in Eq.\ \ref{eq_rkcondition} mixes distributions of $[\theta]$ belonging to different momentum sectors and hence the resultant Hamiltonian does not have translation symmetry. 

\begin{figure}
    \centering
\includegraphics[width=0.9\columnwidth]{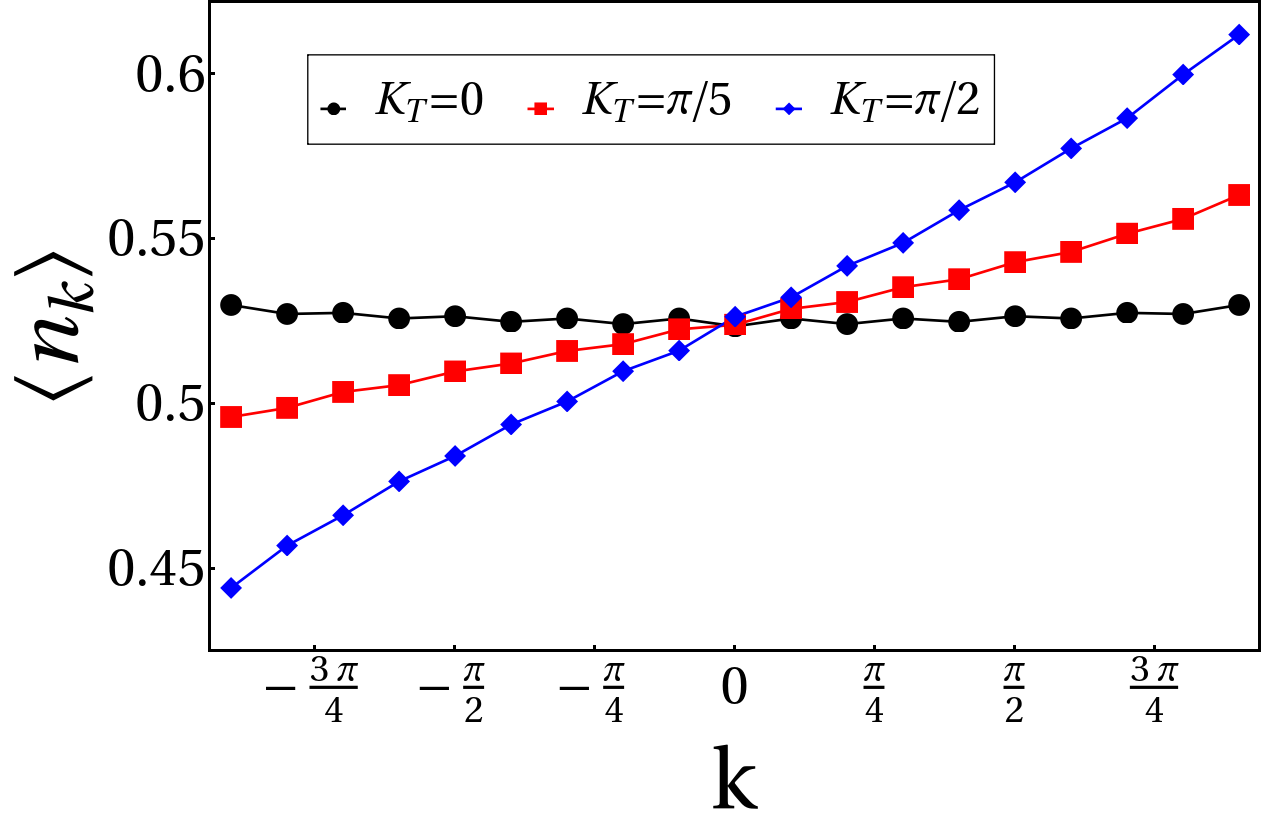}
    \caption{Plot of $\langle n_{\bf k} \rangle $ as a function of $k$ for the constrained {RK type projector} GSs ($|\psi_{\rm RK}^{{\bf k}_{\rm T}}\rangle$), here we have taken 20 sites for different  ${\bf k}_{\rm T}$ for periodic boundary conditions (PBC). Note that $n_{\pi}$(=0) is not shown as in the plot.}
\label{fig:rkmomentumresolved}
\end{figure}

Hence, while it is interesting to study the properties of the generic projector Hamiltonian in Eq.\ \ref{eq_rkham} ({\it e.g.}, we plot $\langle n_{\bf r}\rangle=\langle\Psi_{\rm RK}|\hat n_{\bf r}|\Psi_{\rm RK}\rangle$ in Fig.\ \ref{fig_rk_nr} in Appendix \ref{appen_rk} along with few other details), we shall consider a sub-class of such superposed states where different total momentum sectors do not mix. This is obtained by considering distribution of $[\theta]_{{\bf k}_{\rm T}}$ with the same total momentum ${\bf k}_{\rm T}$. We also put in an extra constraint that $\theta_{{\bf k}=\pi}=-1$, {\it i.e.}, the ${\bf k}=\pi$ mode is always empty for convenience (see discussion in Appendix \ref{appen_rk}). The GS is therefore an equal superposition of such basis states $|\Psi[\theta]_{{\bf k}_{\rm T}}\rangle$ similar to Eq.~\ref{eq_rkstate} which we denote as $|\psi_{\rm RK}^{{\bf k}_{\rm T}}\rangle$. It is fairly easy to show that $\langle \psi_{\rm RK}^{{\bf k}_{\rm T}}|n_{\bf r}|\psi_{\rm RK}^{{\bf k}_{\rm T}}\rangle=\nu$, {\it i.e.}, constant (Eq.~\ref{eq_real_space_occupancy_constraint_RK_2}); moreover, such a constrained RK GS is translation invariant unlike for the full RK case.

The average momentum-space occupation, $\langle n_{\bf k} \rangle$, for such constrained RK GSs is plotted as a function of $k$ in Fig.\ \ref{fig:rkmomentumresolved} for a 1D Fermion system with $N=20$ and $N_p=10$, {\it i.e.}, $\nu=1/2$ for different $K_T$. The slope of $\langle n_{\bf k}\rangle$ is proportional to $K_T$; in particular it is flat for the zero momentum sector with a value tending to $\nu$ for large $N$. For finite systems $\langle n_{\bf k}\rangle$ shows oscillations (see Appendix \ref{appen_rk}), but the amplitude of these oscillations vanishes with increasing system size. This indicates absence of sharp FS-like signature for $|\psi_{\rm RK}^{{\bf k}_{\rm T}}\rangle$ for large $N$ even in presence of translation symmetry. Notably such symmetric wave-functions are not allowed to be unique at partial filling due to Lieb-Schultz-Mattis (LSM) theorem~\cite{lieb1961two} and hence the above wave-function most likely corresponds to the  GS of a gapless phase constituting a realization of a NFL in a toy model.

\section{Fixed point wave-functions: Fermi surface superposition in two dimensions} 
\label{fpt1} 

The RK construction of the superposition of FS provides important clues to generating variational wave-functions in higher dimensions. It shows that a generic superposition completely destroys all features of the metallic GS. This raises an interesting question -- can we superpose FS but still partially retain properties of metals-- both weakly and strongly correlated ones? In this section we investigate this question by constructing a series of variational wave-functions in 2D which interpolates between non-fluctuating FS in the free Fermi-gas limit and the systematically fluctuating FS (subset of Eq.~\ref{eq_superposedfswf}); we expect it to capture the phenomenology of correlated metals. We note here that it is possible to write appropriate RK-like parent projector Hamiltonians for which some of the wave-functions (those with equal superposition) that we discuss below are exact GSs. Such wave-functions can be thought as a IR fixed point wave-functions describing the GS of correlated metals.

One way to obtain such wave-functions is by choosing $\psi[\theta]$ such that various microscopic symmetries beyond translation are implemented manifestly via $[\theta_{\bf k}]$. For example, under time reversal $\theta_{\bf k}\rightarrow \theta_{-\bf k}$ while for a generic spatial symmetry $\mathcal{R}$, we get $\theta_{\bf k}\rightarrow \theta_{\mathcal{R}[\bf k]}$. To incorporate the symmetries we then need to choose $\psi[\theta]$ in Eq.\ \ref{eq_superposedfswf} such that the distribution of $[\theta]$ and $[\theta']$ that are connected by symmetry transformation are weighted equally, i.e.  $\psi[\theta]=\psi^*[\theta']$ for TR and $\psi[\theta]=\psi[\theta']$ with spatial symmetries. We shall focus on the case of TR symmetric systems which is achieved by considering $[\theta]$ to be symmetric about ${\bf k}=0$ and $\psi[\theta]$ to be real. Other symmetries are implemented in a similar manner. We focus on two situations-- (1) isotropic, and, (2) square lattice which have different spatial symmetries.

\begin{figure}
    \centering
\includegraphics[width=.8\columnwidth]{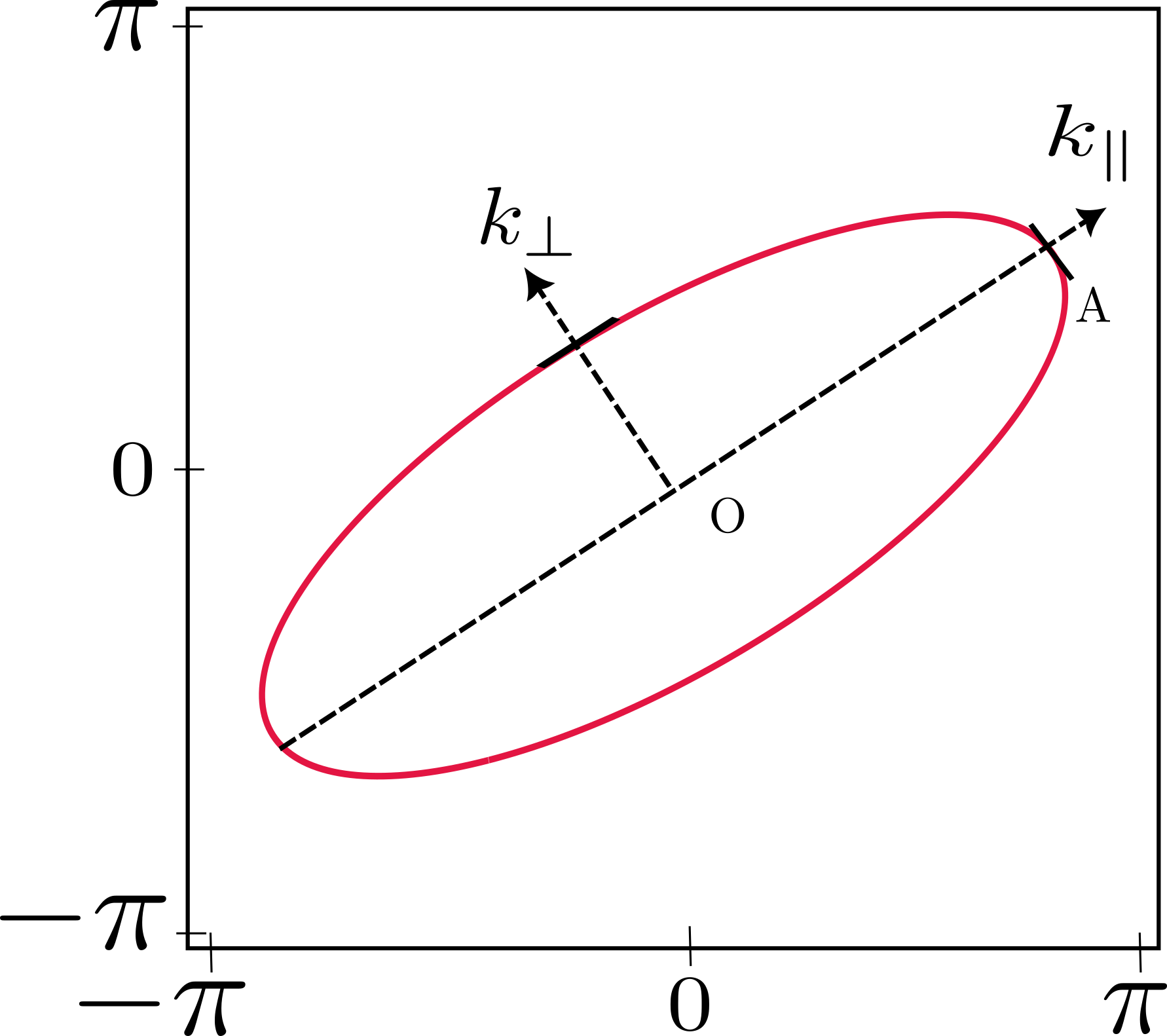}
    \caption{Schematic Figure of an elliptical FS (Eq.~\ref{eq_ellipse_eqn}) with major axis along $k_\parallel$ and eccentricity $\epsilon$.} 
    \label{fig_ellipse}
\end{figure}

\begin{figure*}
\begin{subfigure}
    {
\includegraphics[width=0.9\columnwidth]{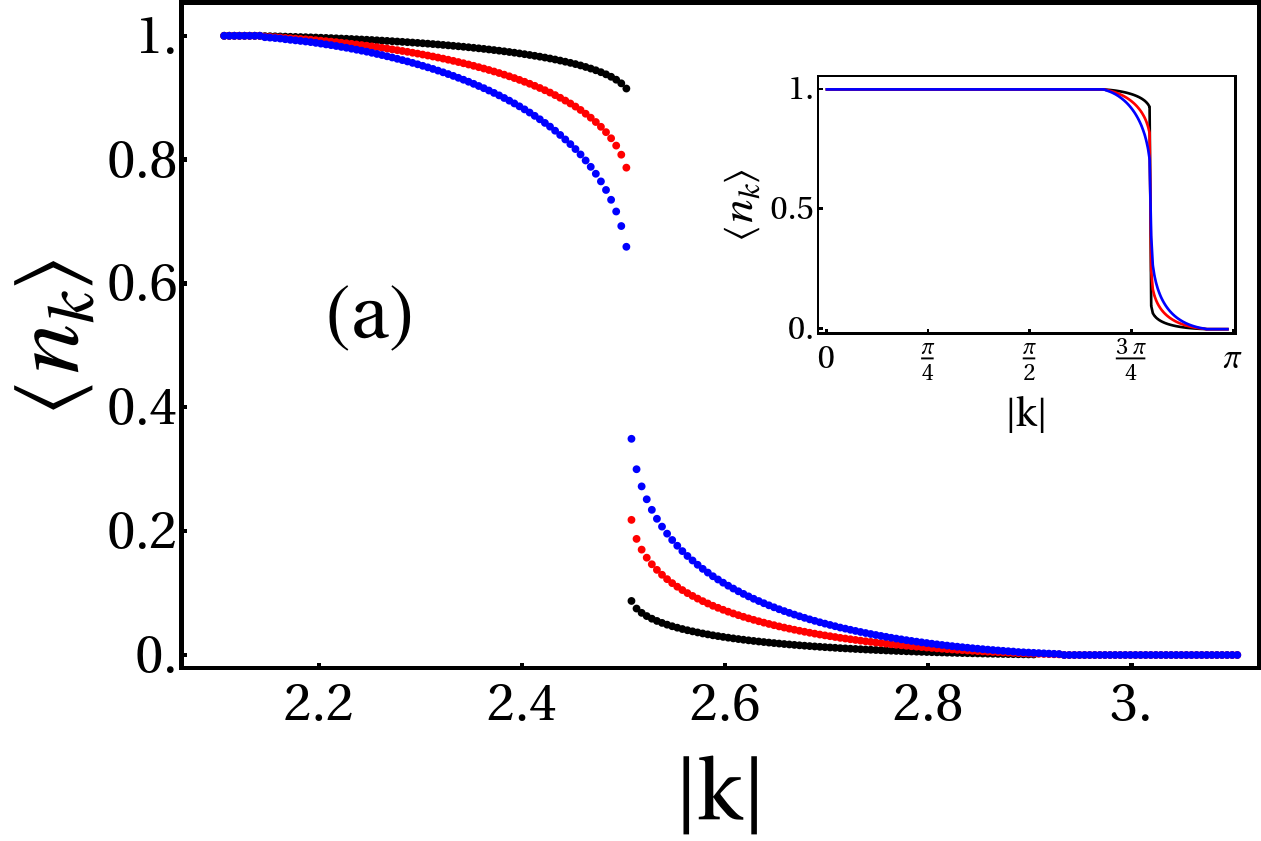}
     \label{fig:fs_nk_isotropic_FL}}
\end{subfigure}
\hfill
\begin{subfigure}{
\includegraphics[width=0.9\columnwidth]{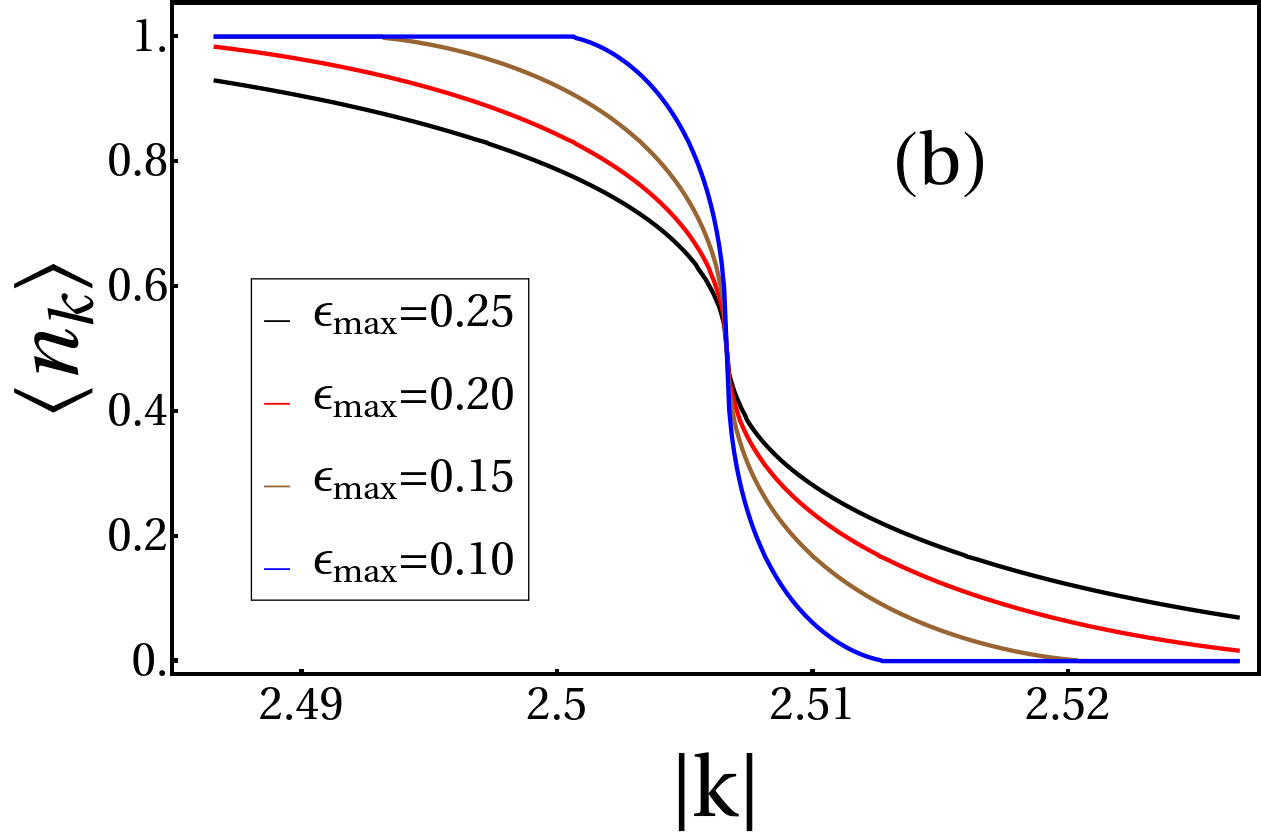}
\label{fig_isotropic_NFL_nk}}
\end{subfigure}
\caption{ (a) Plot of $\langle n_{\bf k} \rangle$(in the vicinity of FS) as a function of $k$ for Eq.~\ref{eq_FL_distribution} shown for $\epsilon_0=0, \epsilon_{\rm max}=0.7$ and different values of ${Z}_{\epsilon_0, {\hat{\bf k}}_0}=0.8$ (black), $0.5$(red), and $0.2$ (blue). The Inset shows $\langle n_{\bf k} \rangle$ as a function of $k\in(0,\pi)$. The jump in the occupancy of $\langle n_k\rangle$ equals ${Z}_{\epsilon_0, {\hat{\bf k}}_0}$. (b) Plot of $n_{\bf k}$ as a function of $k$ for the isotropic NFL (Eq.\ \ref{eq_NFL_distribution_1}) in the vicinity of FS $(k_F=\sqrt{2\pi})$.}
\label{fig_isotropic_nk}
\end{figure*}

Finally, Eq.\ \ref{eq_wftheta} contains arbitrary distributions of $\{n_{\bf k}\}$ over the entire BZ subject to the constraints given by Eq.\ \ref{eq:particleno}. We note that at low energies, various interaction terms can only cause scattering near the FS~\cite{RevModPhys.66.129} similar to the case of TLL in the previous section. In contrast, in higher dimensional systems with Yukawa interactions that hosts NFLs, the RG relevant tangential scatterings~\cite{PhysRevB.80.165102} around the FS lead to a large number of Fermion-hole excitations. These excitations ultimately lead to the breakdown of the Landau quasi-particle. However, the proliferation of the such Fermion-hole excitation, particularly on integrating out the boson, would lead to change in the distribution of $\{n_{\bf k}\}$ via low energy processes. For the rest of this paper, we assume that these low energy processes lead to smooth deformation of the FS and basis states of the form in Eq.\ \ref{eq_wftheta} which represents such smooth deformation have non-zero amplitude in the GS wave-function that we focus on. 

Noting this, we consider the basic fluctuations of an elliptic distribution of $[\theta]$ which describes an elliptic FS at filling, $\nu$, with eccentricity $\epsilon$ and major axis along $\hat{\bf k}$ in the 2D momentum space. An example is shown in Fig.\ \ref{fig_ellipse}. We denote such a sub-set of these basis state wave-function (Eq.\ \ref{eq_wftheta}) as $|\epsilon,\nu,\hat{\bf k}\rangle$ (see Appendix\ \ref{appen_ellipses} for details). The generic wave-function for superposition of such {\it elliptical} FS is given by
\begin{align}\label{eq_superposed_ellipses}
    |\Psi_{\nu} \rangle= \int_{0}^{\epsilon_{\rm max}} d\epsilon\,\,\int_0^{\pi} d\hat{\bf k}\,\, \psi(\epsilon,\hat{\bf k})| \epsilon,\nu,\hat{\bf k}\rangle, 
\end{align}
where the first integral is over the eccentricity between $[0,\epsilon_{\rm max}(\leq 1)]$ for the elliptical distribution with major axis along $\hat{\bf k}$ and the second integral denotes the sum over all possible directions of the major axis. To impose the filling constraint, all the participating ellipses for which the amplitude $\psi(\epsilon,\hat{\bf k})$ is non-zero, have same area and for the rest of this work we set $\nu=1/2$. 

It is fairly easy to show that such wave-functions have uniform Fermion density in real space, {\it i.e.}, $\langle \Psi_{e} \vert n_{\bf r} \vert \Psi_{e} \rangle  = \nu$, unlike the generic RK wave-functions (Eq.\ \ref{eq_rkstate}) (see Appendix\ \ref{appen_ellipses}). However, not surprisingly, depending on the form of $\psi(\epsilon, \hat{\bf k})$, its single particle signatures can be rather different from that of a FL. This is seen by calculating the single particle residue, $Z_{\epsilon,\hat{\bf k}}$, defined via the single particle correlator~\cite{landau1980course} for an elliptical FS, {\it i.e.}, a basis state $|\epsilon,\nu=1/2, \hat{\bf k}\rangle\equiv|\epsilon, \hat{\bf k}\rangle$ (details in Appendix\ \ref{sec_quas_residue}),
\begin{align}
Z_{\epsilon,\hat{\bf k}}=\lim_{\delta \to +0}\frac{\delta^2}{\Big(\frac{k_{\parallel}^2}{a^2}+\frac{k_{\perp}^2}{a^2(1-\epsilon^2)}-\mu\Big)^2+\delta^2} , 
\label{eq_quas_residue}
\end{align}
where $\mu$ fixes the filling, $\nu$, for the  elliptic FS of eccentricity $\epsilon$ and area $A=\pi \mu a^2\sqrt{1-\epsilon^2}$ and $k_\parallel (k_\perp)$ are momentum components resolved along (perpendicular to) the major axis along $\hat{\bf k}$ as shown in Fig.\ \ref{fig_ellipse}. For a single elliptic FS, the residue shows a jump at the perimeter of the ellipse as expected, signalling the presence of the FS which can then be used to define long-lived low energy quasi-particles. For the superposed elliptical FS (Eq.\ \ref{eq_superposed_ellipses}) then, the {\it averaged} residue is given by
\begin{align}
{Z}=\int_0^{\epsilon_{\rm max}} d\epsilon \int_0^{\pi} d\hat{\bf k} |\psi(\epsilon,\hat{\bf k})|^2 Z_{\epsilon,\hat{\bf k}}. 
\label{eq_netz}
\end{align}
Importantly, Eq.\ \ref{eq_netz} recovers the FL form for 
\begin{align}
    \psi(\epsilon, \hat{\bf k}) =&\sqrt{{Z}_{\epsilon_0, {\hat{\bf k}}_0}}~\delta(\epsilon-\epsilon_0)~\delta(\hat{\bf k}-\hat{\bf k}_0) + \sqrt{\frac{1-{Z}_{\epsilon_0, {\hat{\bf k}}_0}}{\pi \epsilon_{\rm max}}}, 
\label{eq_FL_distribution}
\end{align}
for ${Z}_{\epsilon_0, {\hat{\bf k}}_0}\in [0,1]$. The resultant momentum space occupancy of the Fermions $\langle n_{\bf k}\rangle$ is shown in Fig.~\ref{fig:fs_nk_isotropic_FL} where we chose $\epsilon_0=0$ ({\it i.e.}, circular FS) for convenience. The quasi-particle residue, by construction, is ${Z}_{\epsilon_0, {\hat{\bf k}}_0}$ and represents the non-interacting limit for ${Z}_{\epsilon_0, {\hat{\bf k}}_0}=1$. Further details are given in Appendix\ \ref{sec_quas_residue}.

\begin{figure}
    \centering
\includegraphics[width=1\columnwidth]{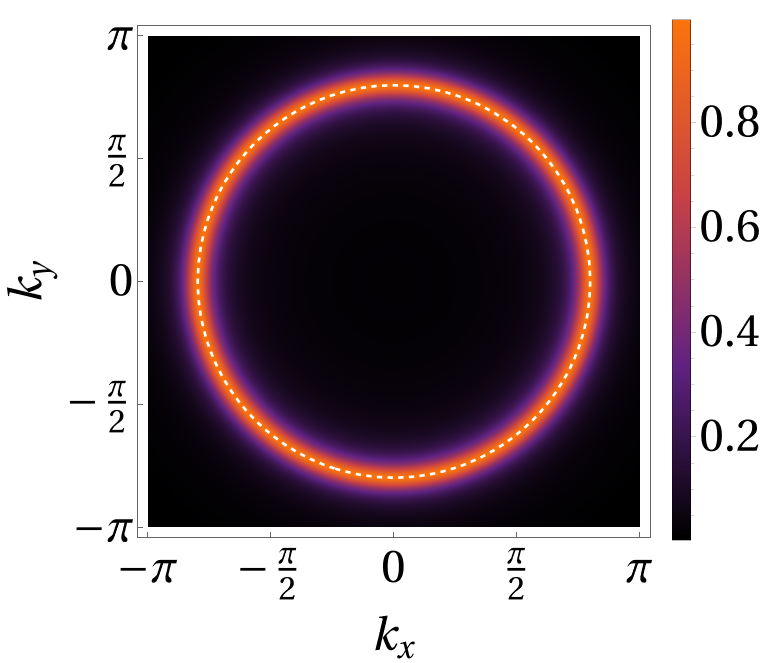}
    \caption{Plot of the single particle residue $Z$ {as a function of $k_x$ and $k_y$ for the superposed FS corresponding to the wave-function in Eq.\ \ref{eq_NFL_distribution_1} for $\epsilon_{\rm max}=0.2$ and $\delta=0.12$. Note that the residue is uniform and is $\leq 1$; it $\to 0$  as $\delta\to0$  for $\epsilon_{\rm max}\neq 0$. The maximum value of $Z$ (for finite $\delta$) is at the perimeter of circle of radius=$\sqrt{2\pi}$ (dashed contour) that also marks the position of the point of inflection of $n_{\bf k}$. See text for details.}
    \label{fig:fs_Zk_isotropic}} 
\end{figure}

The above form of the wave-function (Eq.\ \ref{eq_FL_distribution}) naturally leads us to investigate the case where the weight of the $\delta$-function goes to zero such that 
\begin{eqnarray}\label{eq_NFL_distribution_1}
    & \psi(\epsilon, \hat{\bf k})  = \frac{1}{\sqrt{\epsilon_{\rm max}}}\frac{1}{\sqrt{\pi}} .
 \end{eqnarray}
The resultant plot of $\langle n_{\bf k}\rangle$ is shown in Fig.\ \ref{fig_isotropic_NFL_nk} for different extent of superposition of eccentricities obtained by changing $\epsilon_{\rm max}$ (further details in Appendix\ \ref{sec_isotropic_NFL}). Notably for finite $\epsilon_{\rm max}$, the jump in the occupancy is systematically smeared out as is evident by comparing the two panels of Fig.\ \ref{fig_isotropic_nk}. This is further confirmed by calculating the residue using Eqs.\ \ref{eq_quas_residue} and \ref{eq_netz} as shown in  \ref{fig_Zk_delta}(b) for various $\epsilon_{\rm max}\neq 0$ which shows that $Z=0$. However, instead the residue is replaced by a point of inflection (see Fig.\ \ref{fig_isotropic_NFL_cusp} in Appendix\ \ref{fitness_Qp}) somewhat similar to the TLL in 1D. The position of the point of inflection as well as the finite resolution jump of $\langle n_{\bf k}\rangle$ is shown in Fig.~\ref{fig:fs_Zk_isotropic}. Notably the contour of the inflection point encloses an area in accordance with the Luttinger theorem in spite of the single particle residue being zero. Thus the presence of the inflection point sharply defines the position of the FS, while $Z=0$ indicates the absence of Landau quasi-particles similar to the critical FS~\cite{PhysRevB.78.035103} or marginal FLs~\cite{PhysRevLett.65.2306,PhysRevLett.63.1996}. Similar to the TLL, the average momentum mode occupancy is a power law, {\it i.e.,} $\langle n_{\bf k}\rangle \sim |{\bf k}_{\rm F}-{\bf k}|^p$ with $p\geq0$ (see Fig.\ \ref{fig_fitzalpha}) and  $p$ increases monotonically with $\epsilon_{\rm max}$. This power-law form is consistent with the scaling form suggested in Ref. \cite{PhysRevB.78.035103}.

\begin{figure}
\includegraphics[width=1\columnwidth]{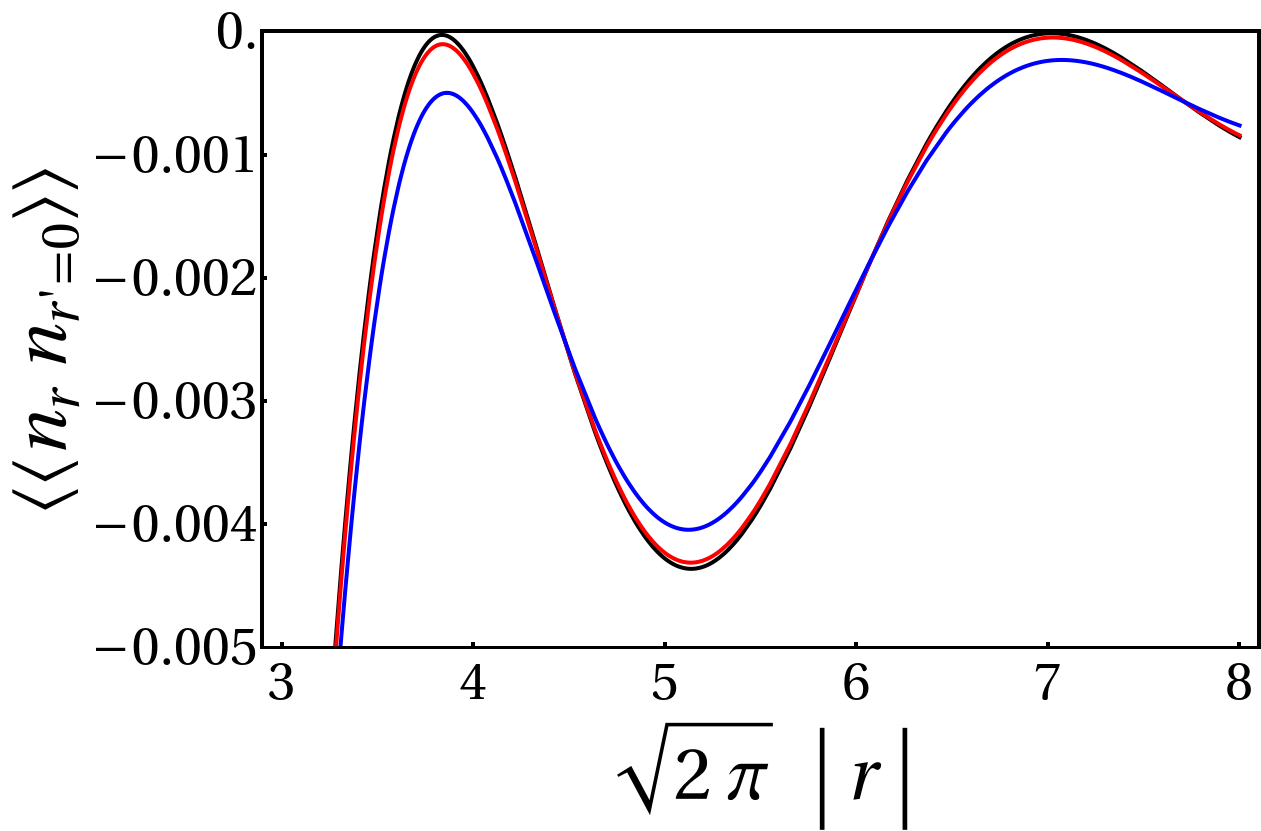}\caption{Plot of the density-density correlation {$W({\bf r},{\bf 0}) =\langle \langle n_{\bf r} n_0\rangle \rangle$ as a function of $r=|{\bf r}|$} corresponding to state Eq.~\ref{eq_FL_distribution} for $\epsilon_{\rm max}=0.75$ and different values of ${Z}_{\epsilon_0, {\hat{\bf k}}_0}=0.95$ (black), $0.80$(red) and $0.00$ (blue).}
\label{fig:fs_n_r_n_0_Isotropic}
\end{figure}

\begin{figure}
    \centering
\includegraphics[width=1\linewidth]{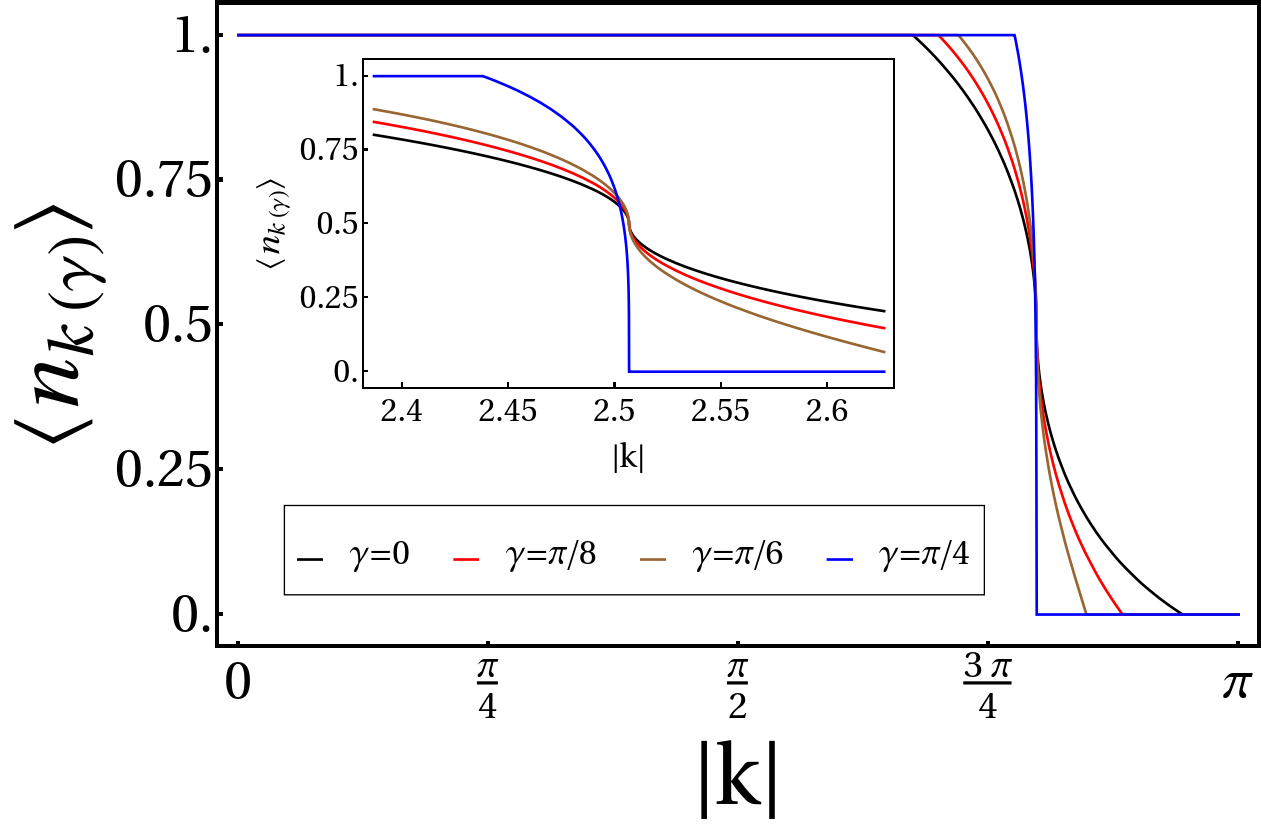}
\caption{Plot of $\langle n_{\bf k} \rangle$ as a function of $k$, where {\bf k} = $ (k\cos{\gamma}~,~k\sin{\gamma})$ for the superposed state mentioned in Eq.~\ref{eq_superposed_ellipses_2} for different $\gamma$ and 
$\epsilon_{\rm max}=0.7$. The inset shows $\langle n_{\bf k} \rangle $ as a function of $k$ 
in the vicinity of the FS.}
\label{nk_theta}
\end{figure}

\begin{figure*}
\begin{subfigure}{
\includegraphics[width=0.65\columnwidth]{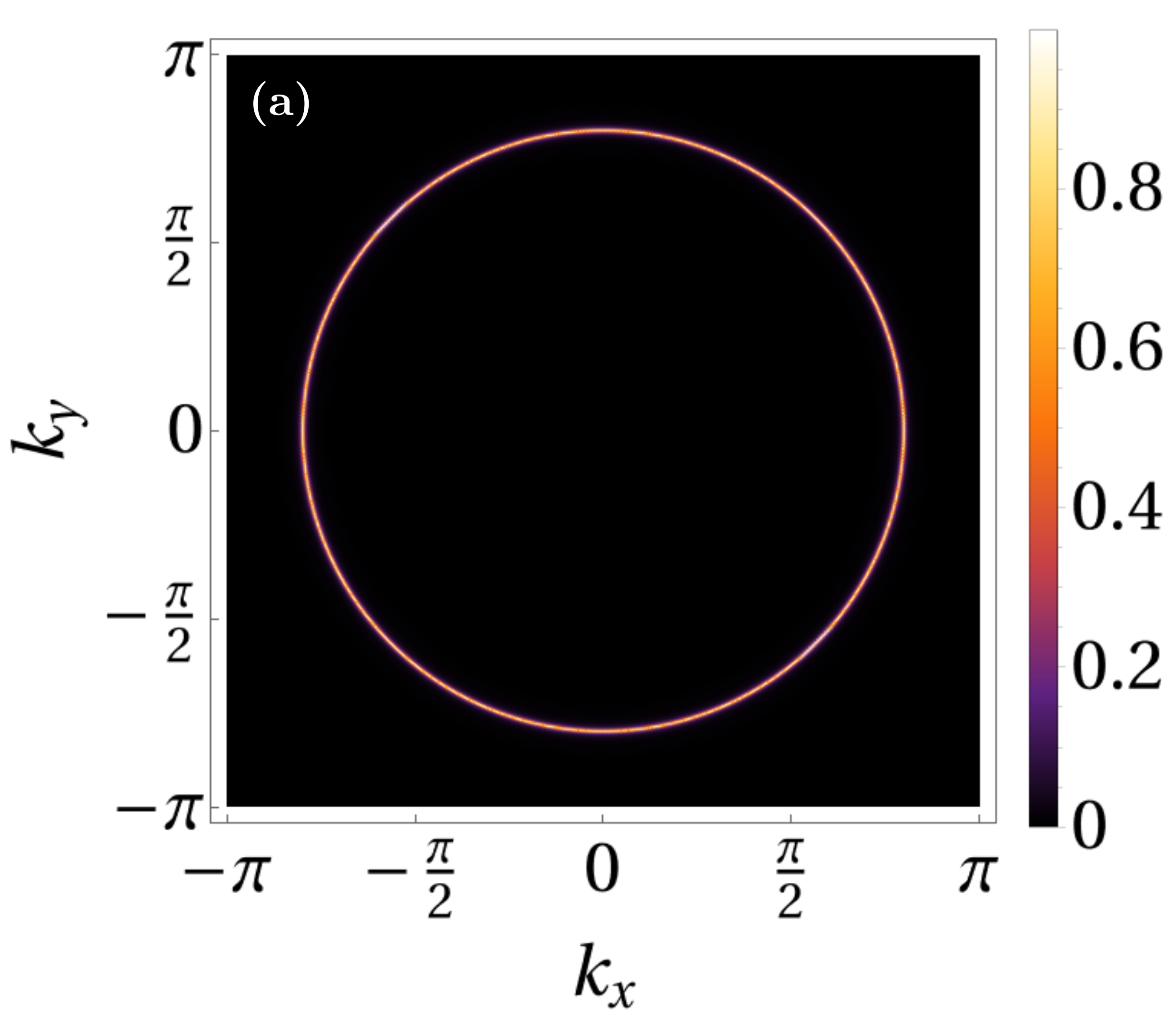}
}
\end{subfigure}
\begin{subfigure}{
\includegraphics[width=0.65\columnwidth]{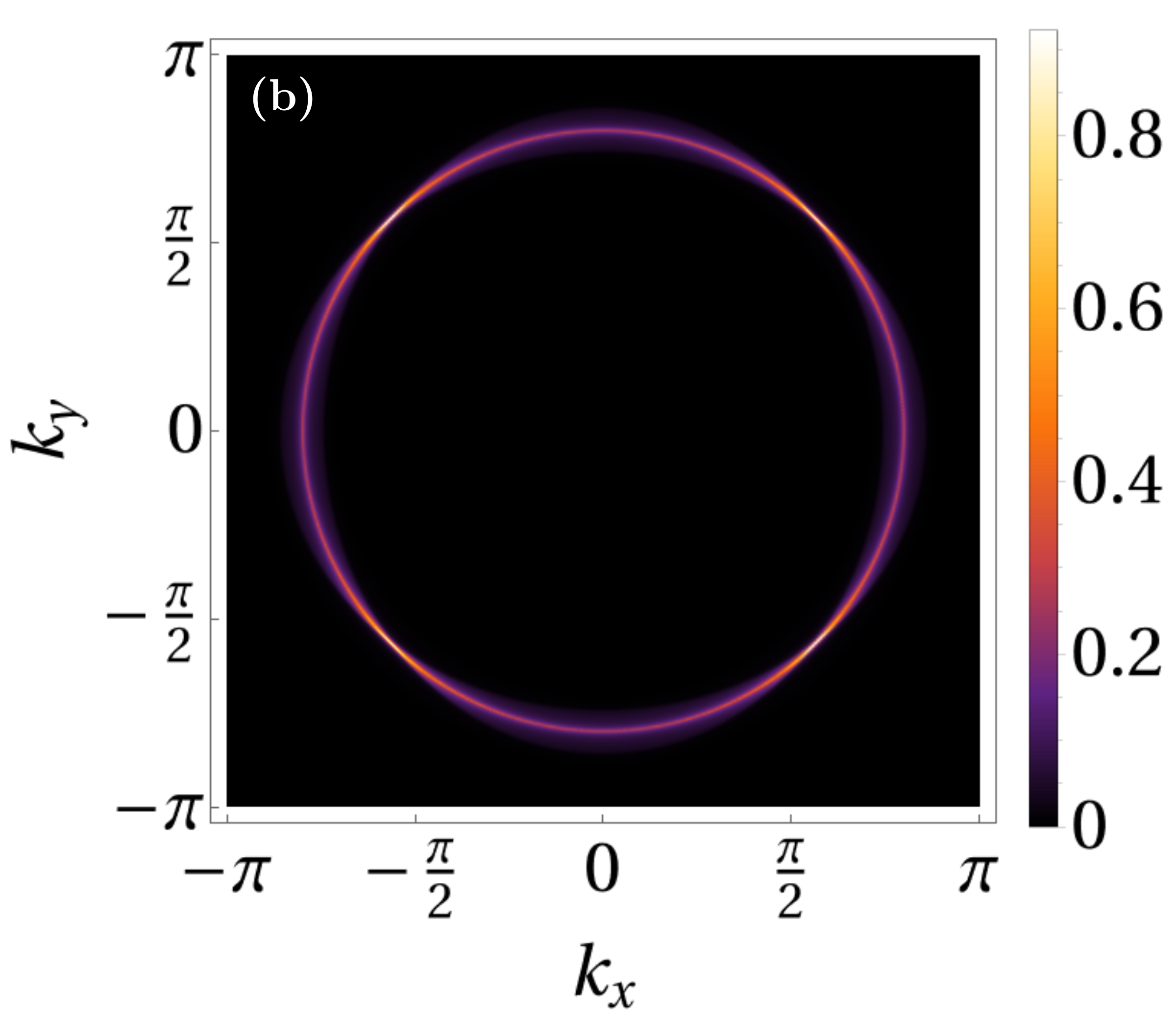}
}
\end{subfigure}
\begin{subfigure}{
\includegraphics[width=0.65\columnwidth]{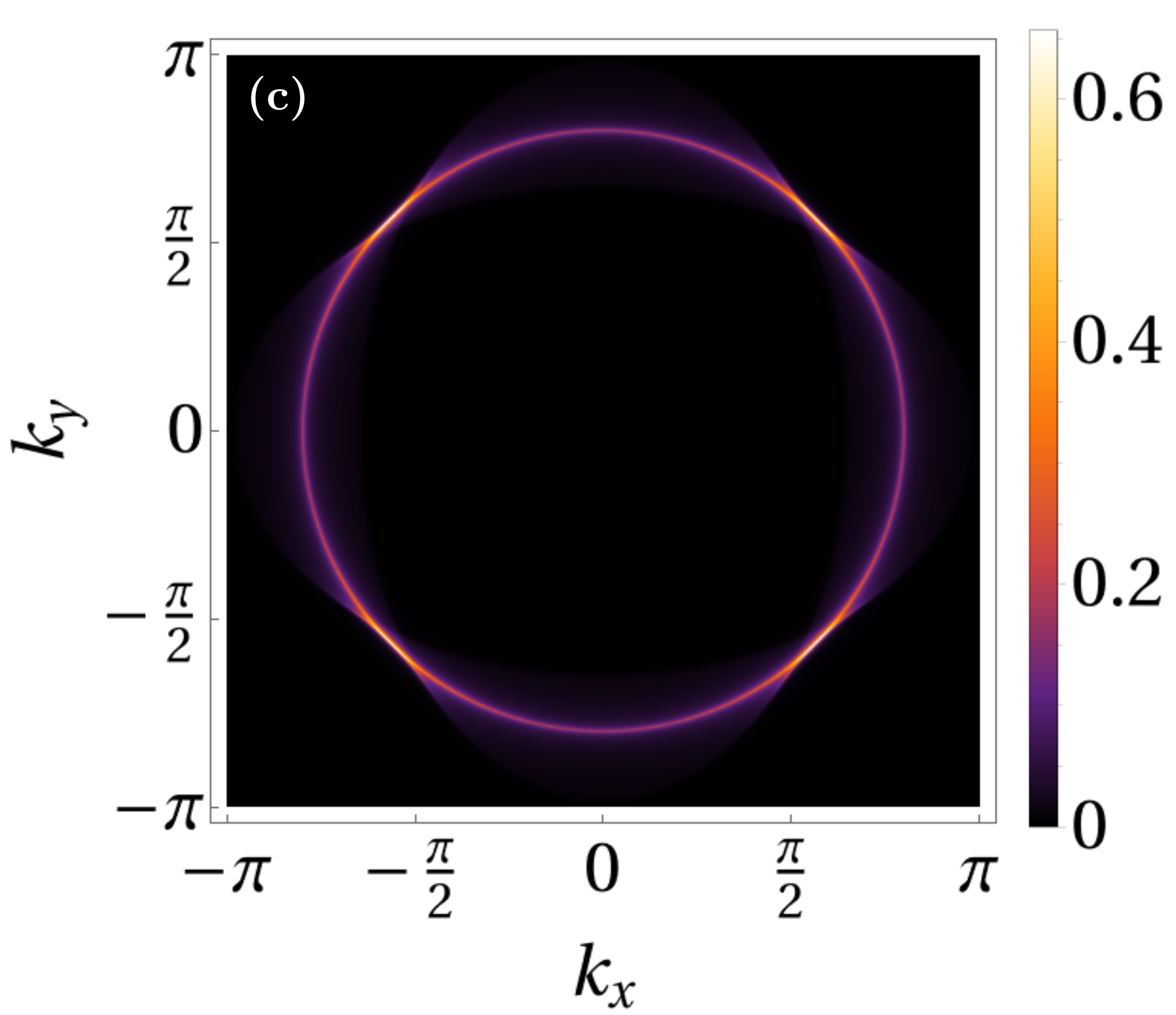} 
}
\end{subfigure}
\caption{Plot of $Z$ as a function of $k_x$ and $k_y$ for the superposed states (Eq.~\ref{eq_superposed_ellipses_2}). The parameters used are $\delta=0.01 $ and (a) $\epsilon_{\rm max}=0.10$ , (b) $\epsilon_{\rm max}=0.50$ , and  (c) $\epsilon_{\rm max}=0.75$. The variations of $Z$ with the resolution $\delta$ is shown in Fig.\ \ref{fig_quasi-particle_superposed_correlator_diff_delta}.}
\label{fig_quasi-particle_superposed_correlator}
\end{figure*}

Fig.\ \ref{fig:fs_n_r_n_0_Isotropic} shows the two-point connected correlator $W({\bf r},0)= \langle\langle n_{\bf r}n_0\rangle\rangle$ as a function of $r=|{\bf r}|$(see Appendix\ \ref{appen_density2}  for detailed expressions) for the wave-functions given by Eq.\ \ref{eq_FL_distribution} for $\epsilon_{\rm max}=0.75$ and different $Z_{\epsilon_0,\hat{\bf k}_0}$ (including $Z_{\epsilon_0,\hat{\bf k}_0}=0$ whence we get Eq.\ \ref{eq_NFL_distribution_1}). The only difference between the non-interacting limit ($Z_{\epsilon_0,\hat{\bf k}_0}\rightarrow 1$) and the completely superposed limit ($Z_{\epsilon_0,\hat{\bf k}_0}\rightarrow 0$) seems to be in the amplitude of oscillations while other features survive the loss of single-Fermion quasi-particle as signalled by the vanishing residue.

Finally, we move to the case of the square lattice where we only have a four-fold rotational symmetry. This should be reflected in the choice of the amplitudes in Eq.\ \ref{eq_superposed_ellipses} via constraining them for different ellipses related by a four-fold rotation as discussed above. For illustrative purposes, in this case, we choose the major axis of the ellipses to lie along the Cartesian directions $\hat{\bf x}$ and $\hat{\bf y}$ such that we now have
\begin{align}
    \psi[\epsilon,\hat{\bf k}]=\sum_{\bf{\boldsymbol\alpha}=\hat{\bf x},\hat{\bf y}}\psi(\epsilon)\delta(\hat{\bf k}-\hat{\boldsymbol \alpha}), 
\end{align}
where $\psi(\epsilon)$ is a smooth real (to ensure TR) function of $\epsilon$ such that Eq.~\ref{eq_superposed_ellipses} reduces to
\begin{align}\label{eq_superposed_ellipses_2}
    |\Psi_{s} \rangle= \int_{0}^{\epsilon_{\rm max}} d\epsilon\,\, \psi(\epsilon)\frac{\left(| \epsilon,\hat{\bf x}\rangle+ |\epsilon,\hat{\bf y}\rangle \right)}{\sqrt{2}}.
\end{align}
This is nothing but equally superposing ellipses of eccentricity up to $\epsilon_{\rm max}$, but with major axis only restricted to horizontal and vertical directions with equal amplitude for any given $\epsilon$. We note that dynamic lattice nematic fluctuations of the type discussed in context of cuprate superconductors~\cite{kivelson1998electronic} can possibly lead to such superpositions of the FS. Choosing different functions $\psi(\epsilon)$ result in superposing the above set of ellipses with varying amplitudes and leads to different wave-functions with manifest TR and lattice symmetries. Consider, for example, $\psi(\epsilon)=\frac{1}{\sqrt{\epsilon_{\rm max}}}$ that correspond to an equal superposition of all such ellipses (see Appendix\ \ref{fitness_Qp} for further details).

The momentum space occupation is shown in Fig.\ \ref{nk_theta} for different directions $\gamma$ (the angle with respect to $k_x$). As can be seen from the figure (see inset), in contrast to Fig.\ \ref{fig_isotropic_nk},  both the position of the rapid change of $\langle n_{\bf k}\rangle$ and its form depends on $\gamma$ (being sharpest for $\gamma=\pm \pi/4$). This is reflected in the finite resolution quasi-particle residue (Eq.\ \ref{eq_netz}) plotted in Fig.\ \ref{fig_quasi-particle_superposed_correlator} for different extent of superposition of eccentricities, $\epsilon_{\rm max}$ and finite resolution, $\delta$ (Eq.\ \ref{eq_quas_residue}). Notably, on taking $\delta\rightarrow 0$ (see Fig.\ \ref{fig_Zk_delta}(c) in Appendix \ref{sec_quas_residue}) the residue goes to zero for all momentum points. However, unlike the isotropic case, it goes to zero anisotropically with the contrast of the anisotropy increasing with $\epsilon_{\rm max}$; this is revealed by comparing the finite resolution leftmost ($\epsilon_{\rm max}=0.1$) and the rightmost ($\epsilon_{\rm max}=0.75$) plots of Fig.\ \ref{fig_quasi-particle_superposed_correlator}. In other words, now we get $\langle n_{\bf k}\rangle\sim |{\bf k}_{\bf F}-{\bf k}|^{p(\gamma)}$ such that for a given $\epsilon_{\rm max}$, the exponent $p(\gamma)$ depends on the angle; this is in accordance with the scaling form expected in Ref. \cite{PhysRevB.78.035103}. However, we note that the above simple power-law breaks down very close to the $\gamma=\pi/4$ (and equivalent points).

\begin{figure*}
\begin{subfigure}{
\includegraphics[width=0.95\columnwidth]{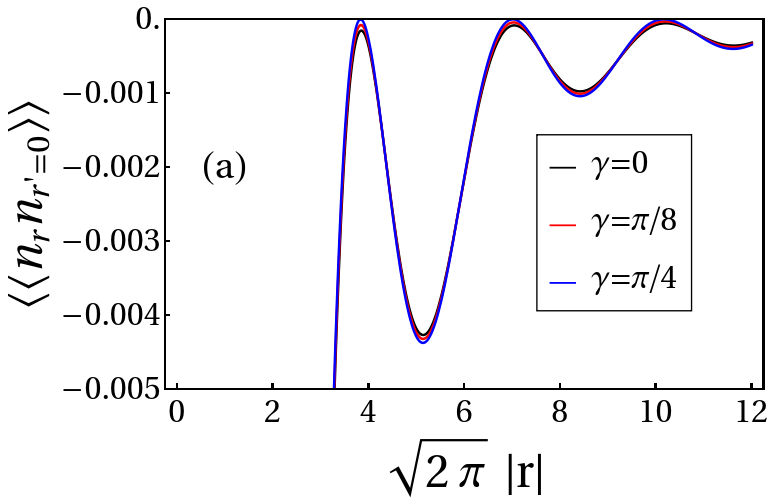}}
\end{subfigure}
\begin{subfigure}{
\includegraphics[width=0.95\columnwidth]{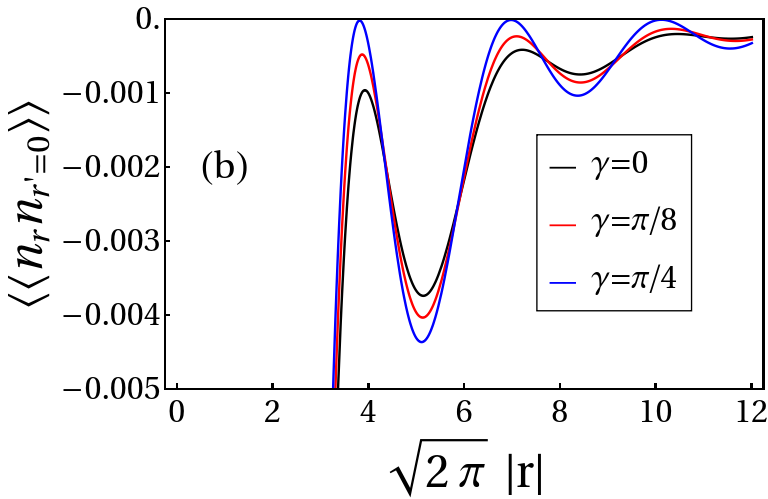}}
\end{subfigure}
\caption{Plot of $\langle\langle n_{\bf r} n_{{\bf r'=0}}\rangle\rangle$ (W({\bf r},{\bf 0})) as a function of $r=|{\bf r}|$ for the superposition of the elliptical FSs (Eq.~\ref{eq_superposed_ellipses_2}) in different directions($\tan\gamma = \frac{y}{x} $ ) for  (a) $\epsilon_{\rm max}$=0.5 and (b) $\epsilon_{\rm max}$=0.75. }
\label{fig_cuts_superposed_correlator}
\end{figure*}

Concentrating on Fig.\ \ref{fig_quasi-particle_superposed_correlator}(c), we find that the largest residue at finite resolution, $Z_{\rm max}(\delta)$, always occur along the $k_x=\pm k_y$ lines with the maximum at $(\sqrt{\pi},\sqrt{\pi})$ (Fig.\ \ref{Z_sqrt_pi_sqrt_pi}) for all $\epsilon_{\rm max}$ (Figs.\ \ref{fig_quasi-particle_superposed_correlator} and \ref{Z_delta_.001}) and goes to zero parametrically more slowly compared to the other directions such that at finite resolution we get a smeared residue around these directions as shown in Fig.~\ref{fig_quasi-particle_superposed_correlator}. One can ascribe a {\it length} to such segments of finite resolution residue as a function of both $\delta$ and $\epsilon_{\rm max}$ as mentioned in Fig.~\ref{arc_length_same_delta} of Appendix\ \ref{fitness_Qp}. Finally turning to the density-density-correlator, $\langle\langle n_{\bf r} n_{{\bf r}'}\rangle\rangle$, the Friedel oscillations become highly direction dependent as shown in Fig.\ \ref{fig_cuts_superposed_correlator}. This completes the phenomenology of the one and two particle static correlations of the many-Fermion wave-function obtained by superposition of ellipses. 

\section{Summary and outlook} 

\begin{figure}
\includegraphics[width=0.9\columnwidth]{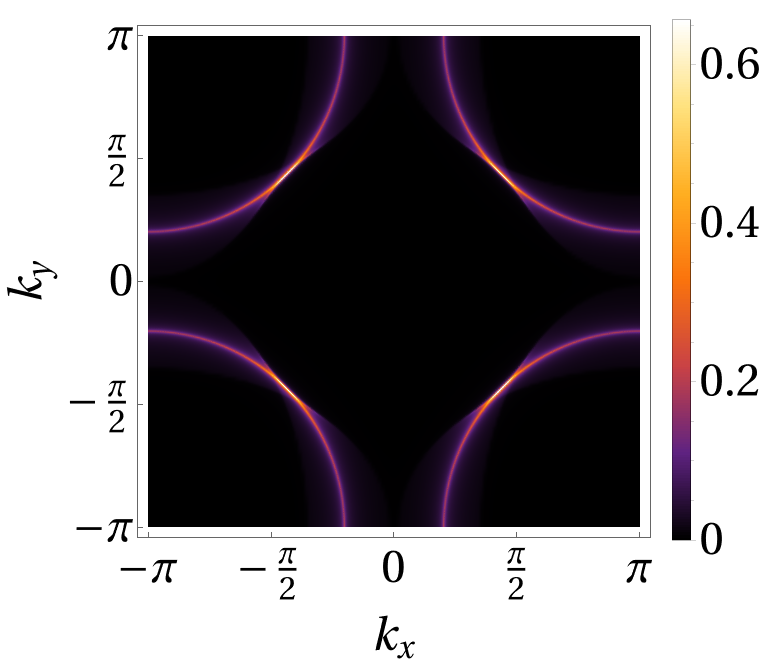}
\caption{Plot of $Z$ as a function of $k_x$ and $k_y$ for superposed states (Eq.~\ref{eq_superposed_ellipses_2}) for a hole FS centered around ${\bf k}=(\pi,\pi)$.}
\label{fig_isotropic_NFL_ZK_pi_pi}
\end{figure}

Having described the construction of many-Fermion variational wave-functions via superposition of FS that can possibly describe correlated metals including NFLs, we now summarize {our main results}. 

The variational wave-function of the form in Eq.\ \ref{eq_superposedfswf} provide interesting insights into the two basic phenomenology of NFL metals-- (1) the lack of Landau quasi-particles, and (2) the well defined FS-- that are obtained from quantum superposition of momentum space occupation. In view of the anisotropic loss of Landau quasi-particles on square lattice, it is interesting to re-plot Fig.\ \ref{fig_quasi-particle_superposed_correlator} for the hole FS centered around ${\bf k}=(\pi,\pi)$. This is shown in Fig.\ \ref{fig_isotropic_NFL_ZK_pi_pi} where the similarity with Fermi-arcs in underdoped cuprates~\cite{norman1998destruction} is apparent. {However in cuprates such phenomena is observed at finite temperature while the present construction is strictly for the GS, {\it i.e.}, $T=0$; in underdoped cuprates, a superconductor is obtained at lower temperatures.} It has been suggested  ~\cite{PhysRevB.78.035103} that Fermi arcs can arise as a finite temperature effect of critical FS with angle-dependent scaling exponents. It would indeed be promising to investigate if similar low energy phenomenology can be obtained starting from the wave-function presented here.

Turning back to structure of the basis (Eq.\ \ref{eq_wftheta}), consider a distribution $[\theta]$ representing a fully filled FS, $|\Psi[\theta]\rangle$, at filling $\nu$. Now consider a related state, $|\Psi[\theta']\rangle$, at the same filling obtained from $[\theta]$ by taking a particle at ${\bf k}_1$ from inside the FS and putting it at an empty mode at ${\bf k}_2$. Then, superposing the above two results in an EPR state of a particle and a hole at the two momenta in the background of the rest of the Fermions. At low energies, both ${\bf k}_1$ and ${\bf k}_2$ lie within an annulus around the FS-- hence local in momentum space. Thus, the GSs obtained by superposing the FS may be roughly thought as made up of {\it local} (in momentum space) particle-hole EPR pairs across the FS. It would be interesting to explore if this point provides advantage in understanding metallic phases using density-matrix-renormalization-group (DMRG) methods in momentum space~\cite{PhysRevB.53.R10445}.

Extending the present formulation to spinful Fermions can be obtained by defining a augmented momentum-space grid $\{{\bf k},\sigma\}$ with $\sigma=\uparrow\downarrow$. Extension of the present formulation to the spinful case then provides for careful comparison with Gutzwiller wave-functions for correlated metals~\cite{PhysRev.137.A1726}. Finally, the present variational ground-states immediately lead to questions pertaining to the low energy spectrum both within and beyond the RK type Hamiltonians. At this point we note that the variational wave-functions (Eq.~\ref{eq_superposedfswf} or \ref{eq_superposed_ellipses}), after implementing the symmetries, represent symmetric wave-functions at fractional filling, $\nu$. Application of LSM theorem~\cite{lieb1961two,PhysRevLett.79.1110} would suggest that this corresponds to the GS of a gapless phase as is indeed the case of the half filled Hubbard model as our calculations explicitly show. In higher dimensions, application of flux threading arguments of Ref. \cite{PhysRevLett.84.3370} indicate that it is possible to generate orthogonal states with momentum difference $2\pi\nu N^{d-1}$. For Hamiltonians which are short ranged (in real space) two dimensional generalization of LSM theorem~\cite{PhysRevLett.84.3370,PhysRevB.69.104431} would lead to a similar conclusion as in 1D about the fate of the present wave-functions in regards to them being valid GSs of gapless phases. Addressing these issues would decide the feasibility of the present approach to understand the physics of correlated metals.

\begin{acknowledgments}
The authors acknowledge useful discussions with F. Alet, K. Damle, R. Moessner, P. Majumdar, S. Mukerjee, M. Randeria, S. Roy, S. L. Sondhi, R. Sensarma and N. Trivedi. SB acknowledges funding  by Swarna Jayanti fellowship of SERB-DST (India) Grant No. SB/SJF/2021-22/12; DST, Govt. of India (Nano mission), under Project No. DST/NM/TUE/QM-10/2019 (C)/7 and Max Planck Partner group Grant at ICTS. AC, HN and SB acknowledge the support of the Department of Atomic Energy, Government of India, under project no. RTI4001. KS thanks DST for support through SERB project JCB/2021/000030.
\end{acknowledgments}

\appendix

\section{The details of the \texorpdfstring{$[\theta]$}{} basis}
\label{appen_basis}

{In this section, we chart out several properties of the $|\Psi[\theta]\rangle$ (Eq.~\ref{eq_wftheta}) basis. The real-space representation of the many-body state in Eq.\ \ref{eq_wftheta}, $\Psi[\theta]\left(\{{\bf r}\}\right)=\langle\{{\bf r}\}|\Psi[\theta]\rangle$, can be written as} 
\begin{align}
\left|\begin{array}{ccccc}
s_{{\bf k}_1}\phi_{{\bf k}_1}({\bf r}_1) & s_{{\bf k}_2}\phi_{{\bf k}_2}({\bf r}_1) & \cdots & \cdots & s_{{\bf k}_B}\phi_{{\bf k}_B}({\bf r}_1)\\
s_{{\bf k}_1}\phi_{{\bf k}_1}({\bf r}_2) & s_{{\bf k}_2}\phi_{{\bf k}_2}({\bf r}_2) & \cdots & \cdots & s_{{\bf k}_B}\phi_{{\bf k}_B}({\bf r}_2)\\
\cdots & \cdots & \cdots & \cdots & \cdots \\
\cdots & \cdots & \cdots & \cdots & \cdots \\
s_{{\bf k}_1}\phi_{{\bf k}_1}({\bf r}_{N_p}) & s_{{\bf k}_2}\phi_{{\bf k}_2}({\bf r}_{N_p}) & \cdots & \cdots & s_{{\bf k}_B}\phi_{{\bf k}_B}({\bf r}_{N_p})\\
    \end{array}\right|, \label{sladet1}
\end{align}
where $s_{\bf k}=(1+\theta_{\bf k})/2$, $\{{\bf r}\}\equiv ({\bf r}_1,{\bf r}_2,\cdots, {\bf r}_{N_p})$ is the position basis for particles corresponding to filling $\nu$, and $\phi_{\bf k}({\bf r})$ are the corresponding single-particle wave-functions. Note that by changing of $\nu$, the  the number of rows and columns are appropriately changed by the constraint in Eq.\ \ref{eq:particleno}.

Turning to the various many-body operators, consider the action of $c^{\dagger}_{\bf q}$ on this state,
\begin{eqnarray}
&& c^{\dagger}_{\bf q}  \vert \Psi[\theta]\rangle  = c^{\dagger}_{\bf q}  \prod_{\bf{k}}(c^{\dagger}_{\bf{k}})^{\frac{1+\theta_{\bf k}}{2}} \vert 0 \rangle  \label{opeq1} \\
&& =\delta_{\theta_{\bf q},-1}\left( \prod _{\bf k'\neq q}\delta_{\theta_{\bf k'},\theta'_{\bf k'}}\right)\delta_{\theta_{\bf q},-\theta'_{\bf q}}(-1)^{\sum_{\bf k > q}\frac{1+\theta_{\bf k}}{2} }\vert \Psi[\theta']\rangle .\nonumber
\end{eqnarray}
Similarly
\begin{eqnarray} 
    c_{\bf q}|\Psi[\theta]\rangle &=& \delta_{\theta_{\bf q},1}\left( \prod _{\bf k'\neq q}\delta_{\theta_{\bf k'},\theta'_{\bf k'}}\right)\delta_{\theta_{\bf q},-\theta'_{\bf q}}\nonumber\\
    && \times (-1)^{\sum_{\bf k > q}\frac{1+\theta_{\bf k}}{2} }\vert \Psi[\theta']\rangle. \label{opeq2} 
\end{eqnarray}
Therefore the matrix elements are given by
\begin{align}
\left(c^{\dagger}_{\bf q}\right)_{[\theta],[\theta']} =&\left( \prod _{\bf k'\neq q}\delta_{\theta_{\bf k'},\theta'_{\bf k'}}\right)\delta_{\theta_{\bf q},-\theta'_{\bf q}}\nonumber\\
&~~~~\times\left(\frac{1-\theta_{\bf q}}{2}\right)\exp\left[i\pi{\sum_{\bf k > q}\frac{1+\theta_{\bf k}}{2} }\right] ,\label{matcr1}
\end{align}
and
\begin{align}
    \left(c_{\bf q}\right)_{[\theta],[\theta']} =&\left( \prod _{\bf k'\neq q}\delta_{\theta_{\bf k'},\theta'_{\bf k'}}\right)\delta_{\theta_{\bf q},-\theta'_{\bf q}}\nonumber\\
&~~~~\times\left(\frac{1+\theta_{\bf q}}{2}\right)\exp\left[i\pi{\sum_{\bf k > q}\frac{1+\theta_{\bf k}}{2} }\right].\label{matan1} 
\end{align}

These results are easily extended to matrix elements of multiple creation and annihilation operators. For two particles, represented by corresponding Fermion creation operators,
\begin{align}
\left(c^{\dagger}_{\bf q_1}  
c^{\dagger}_{\bf q_2}\right)_{[\theta],[\theta']} =&\left( \prod _{{\bf k}'\neq {\bf q}_1,{\bf q}_2}\delta_{\theta_{\bf k'},\theta'_{\bf k'}}\right)\delta_{\theta_{\bf q_1},-\theta'_{\bf q_1}}~\delta_{\theta_{\bf q_2},-\theta'_{\bf q_2}}\nonumber\\
&\times{\rm sign}({\bf q}_1,{\bf q}_2)\left(\prod_{i=1}^2\frac{1-\theta_{{\bf q}_i}}{2}\right)\nonumber \\
   & \times\prod_{i=1}^2 \exp\left[{i\pi \sum_{{\bf k}> {\bf q}_i}\frac{1+\theta_{\bf k}}{2}}\right], \label{twoparteq} 
\end{align}
where in the same ordering as used in Eq.\ \ref{eq_wftheta}, {${\rm sign}({\bf q_1}, {\bf q_2})$ reads}
\begin{align}
    {\rm sign}({\bf q}_1,{\bf q}_2)=\left\{\begin{array}{cl}
    +1 &\forall~{\bf q}_1>{\bf q}_2\\
    0  & \forall~{\bf q}_1={\bf q}_2\\
    -1 &\forall~{\bf q}_1<{\bf q}_2.\\
    \end{array}\right . \label{signeq} 
\end{align}
This shows that the matrix elements encode the sign structure associated with the anticommutation of Fermion operators. Continuing for $N$-particle, we have 

\begin{align}
    \left(\prod_{i=1}^N c^{\dagger}_{\bf q_i}\right) =c^{\dagger}_{\bf q_1}c^{\dagger}_{\bf q_2}.....c^{\dagger}_{\bf q_{N-1}}c^{\dagger}_{\bf q_N},
\end{align}
\begin{widetext}
\begin{align}
    \left(\prod_{i=1}^N c^{\dagger}_{\bf q_i}\right)_{[\theta],[\theta']} =& \left(\prod_{\bf{k'} \neq \bf{q_i} \forall i}\delta_{\theta_{\bf{k'}},\theta'_{\bf{k'}}}\right)\left(\prod_{i=1}^{N}\delta_{\theta_{{\bf q_i}},-\theta'_{{\bf q_i}}}\right) \times\prod_{i,j, i>j}^{N}{\rm sign} ({\bf q}_j,{\bf q}_i)\left(\prod_{i=1}^{N}\frac{1-\theta_{\bf{ q_i}}}{2}\right) \times\prod_{i=1}^{N} \exp\left[i\pi\sum_{{\bf{k}}> {\bf q}_i} \frac{1+\theta_{\bf{{k}}}}{2}\right]. \label{nparteq}
\end{align}

We note that the above matrix elements are valid provided both $[\theta]$ and $[\theta']$ obey the constraint relation of Eq.\ \ref{eq:particleno} with $0\leq\nu\leq1$. For particle-hole bilinears in momentum space the matrix elements are given by 
\begin{align}
    \left(c^{\dagger}_{\bf k+q}c_{\bf k}\right)_{[\theta],[\theta']}=&\left(\prod_{\bf k'\neq k+q,k}\delta_{\theta_{\bf k'},\theta'_{\bf k'}}\right)\delta_{\theta_{\bf k+q},-\theta'_{\bf k+q}}\delta_{\theta_{\bf k},-\theta'_{\bf k}}\times{\rm sign}({\bf k+q},{\bf k})\left(\frac{1-\theta_{\bf k+q}}{2}\right)\left(\frac{1+\theta_{\bf k}}{2}\right)\nonumber\\
    &\times\exp\left[i\pi\sum_{{\bf k+q}>{\bf k}'>{\bf k}}\frac{1+\theta_{\bf k'}}{2}\right]+\delta_{{\bf q},0}\left(\prod_{\bf k'}\delta_{\theta_{\bf k'},\theta'_{\bf k'}}\right)\frac{1+\theta_{\bf k}}{2} , \label{matbil}
\end{align}
where the last term gives the diagonal contribution for the momentum space occupation $n_{\bf k}=c^\dagger_{\bf k}c_{\bf k}$.
\section{Hamiltonian in the \texorpdfstring{$[\theta]$}{} basis}
\label{appen_Hamiltonian}

Having constructed the wave-function in the $[\theta]$ basis, we now consider the action of interacting Fermionic Hamiltonian $H$ (Eq.\ \ref{eq_intham}) on these wave-functions. The matrix elements of the non-interacting part, $H_0$, in this basis is given by {
 \begin{eqnarray}
\langle\Psi[\theta']|H_0|\Psi[\theta]\rangle 
 &=&\left(\sum_{\textbf{k}}E_{\textbf{k}} \left(\frac{1+\theta_{\textbf{k}}}{2}\right)\right)\langle\Psi[\theta']\vert\Psi[\theta]\rangle = E[\theta]~\delta_{[\theta'],[\theta]} .\label{free_ham_matrix_element1}
\end{eqnarray}}
Thus $H_0$ is diagonal in the $[\theta]$ basis, as discussed in the main text. 

Next, we consider the matrix elements for interacting part $H_I$, given by the second term of Eq.\ \ref{eq_intham}. These matrix elements can be written as
\begin{align} \label{eq_off_diagonal_ham_AC}
\langle\Psi[\theta']|H_I|\Psi[\theta]\rangle&=E_I([\theta'], [\theta])\nonumber\\ 
 &=\sum_{{\bf k}_1}\sum_{{\bf k}_2}\sum_{{\bf q}\neq 0}~ V({\bf q})~ \left(\prod_{{\bf k} \neq {\bf k}_1, {\bf k}_2, {\bf k}_1+{\bf q}, {\bf k}_2-{\bf q}} \delta_{\theta_{\bf k} \theta'_{\bf k}}\right)\left(\delta_{\theta_{{\bf k}_1} ,-{\theta'}_{{\bf k}_1}}\delta_{\theta_{{\bf k}_2},-{\theta'}_{{\bf k}_2}}\delta_{\theta_{{\bf k}_1+{\bf q}},- {\theta'}_{{\bf k}_1+{\bf q}}}\delta_{\theta_{{\bf k}_2-{\bf q}},-\theta'_{{\bf k}_2-{\bf q}}}\right)\nonumber \\ 
&~~~~~~~~~~~~~~~~~~~~\times\left(\frac{1-\theta_{\bf k_1}}{2}\right) \left(\frac{1-\theta_{\bf k_2}}{2}\right)\left(\frac{1+\theta_{\bf k_1+q}}{2}\right) \left(\frac{1+\theta_{\bf k_2-q}}{2}\right)
 \nonumber\\
 &~~~~~~~~~~~~~~~~~~~\times {\rm sign}({\bf k}_1,{\bf k}_2)~{\rm sign}({\bf k}_2-{\bf q}, {\bf k}_1
+{\bf q})\times
 \exp\left[i\pi\left(\chi_{\bf{k_1}} + \chi_{\bf{k_2}} +\chi_{\bf{k_2-q}} +\chi_{\bf{k_1+q}}\right)\right]  \\
 &=\sum_{{\bf k}_1}\sum_{{\bf k}_2}\sum_{{\bf q}\neq 0}~ V({\bf q})~ \left(\prod_{{\bf k} \neq {\bf k}_1, {\bf k}_2, {\bf k}_1+{\bf q}, {\bf k}_2-{\bf q}} \delta_{\theta_{\bf k} \theta'_{\bf k}}\right)\left(\delta_{\theta_{{\bf k}_1} ,-{\theta'}_{{\bf k}_1}}\delta_{\theta_{{\bf k}_2},-{\theta'}_{{\bf k}_2}}\delta_{\theta_{{\bf k}_1+{\bf q}},- {\theta'}_{{\bf k}_1+{\bf q}}}\delta_{\theta_{{\bf k}_2-{\bf q}},-\theta'_{{\bf k}_2-{\bf q}}}\right)\nonumber \\ 
&~~~~~~~~~~~~~~~~~~~~\times\left(\frac{1-\theta_{\bf k_1}}{2}\right) \left(\frac{1-\theta_{\bf k_2}}{2}\right)\left(\frac{1+\theta_{\bf k_1+q}}{2}\right) \left(\frac{1+\theta_{\bf k_2-q}}{2}\right)\nonumber\\
&~~~~~~~~~~~~~~~~~~~~~\times\exp\left[i\pi\left(\tilde\chi_{\bf{k_1}} + \tilde\chi_{\bf{k_2}} +\tilde\chi_{\bf{k_2-q}} +\tilde\chi_{\bf{k_1+q}}\right)\right] , 
\end{align}
where, $\chi_{{\bf k_i}} = \sum_{\bf k>k_{ i}} \frac{1+\theta_{\bf k}}{2}$ for ${\bf k_i}\in\{ {\bf k_1, k_2, k_2-q, k_1+q} \}$ and
\begin{align}\label{eq_phase_factor_KS}
 &   \tilde{\chi}_{\bf k_1+q} = \sum_{\bf k_i> k_1+q}~\frac{1+\theta_{\bf k_i}}{2} ,~~ \tilde{\chi}_{\bf k_2-q}  = \sum_{\bf k_i> k_2-q}~\frac{1+\theta_{\bf k_i}}{2}  -\frac{\left(1-{\rm sign}\left({\bf (k_2-q)-(k_1+q)}\right) \right)}{2} ,~~ \nonumber \\ 
 &  \tilde{\chi}_{\bf k_2} = \sum_{\bf k_i> k_2}~\frac{1+\theta_{\bf k_i}}{2}  
 -\frac{\left(1-{\rm sign}\left({\bf k_2-(k_2-q)}\right) \right)}{2} -\frac{\left(1-{\rm sign}\left({\bf k_2-(k_1+q)}\right) \right)}{2} ,~~ \nonumber \\   
 &  \tilde{\chi}_{\bf k_1} = \sum_{\bf k_i> k_1}~\frac{1+\theta_{\bf k_i}}{2}-\frac{\left(1-{\rm sign}\left({\bf k_1-k_2)}\right) \right)}{2} -\frac{\left(1-{\rm sign}\left({\bf k_1-(k_2-q)}\right) \right)}{2}  -\frac{\left(1-{\rm sign}\left({\bf k_1-(k_1+q)}\right) \right)}{2}. \nonumber \\   
\end{align}
In Eq.~\ref{eq_phase_factor_KS}, all the sign factors neutralize the effects of each other in the contribution. So we have, 
{
\begin{eqnarray}
  \sum_{i=1}^{4}\tilde{\chi}_{\bf k_i} &=& \sum_{i=1}^{4}\chi_{{\bf k_i}} -\frac{\left(1-{\rm sign}\left({\bf (k_2-q)-(k_1+q)}\right) \right)}{2} -\frac{\left(1-{\rm sign}\left({\bf k_1-k_2)}\right) \right)}{2} ,\label{signeq2} 
\end{eqnarray}}
\end{widetext}
which implies
\begin{align}
    (-1)^{\sum_{i=1}^{4}\tilde{\chi}_{{\bf k_i}}} =& {\rm sign}\left({\bf (k_2-q)-(k_1+q)}\right) \nonumber\\
& \times {\rm sign}\left({\bf k_1-k_2)}\right) (-1)^{\sum_{i=1}^4\chi_{\bf k_i}}.
\end{align}
 Eq.\ \ref{eq_off_diagonal_ham_AC} thus provides matrix elements for $H_I$. The full Hamiltonian can now be written as in Eq.\ \ref{eq_intHam_matrix} which can be put into a more concise form as {
\begin{align}
    H & = \sum_{[\theta],[\theta']} M\left([\theta],[ \theta']\right)|\Psi[\theta]\rangle\langle\Psi[\theta']| ,  \label{eq_sfs} \\ 
 M([\theta],[\theta']) &=E[\theta]~\delta_{[\theta],[\theta']} +E_I([\theta],[\theta'])\left(1-\delta_{[\theta],[\theta']}\right). \nonumber
\end{align}}

\section{Details of the Exact diagonalization in one spatial dimension.}
\label{appen_ed1d}
{To perform exact diagonalization (ED) on the XXZ spin-1/2 Hamiltonian, we begin by defining the Hamiltonian itself. The XXZ Hamiltonian for a chain of $N$ spins can be written as in Eq.~\ref{eq_XXZ_Ham}. Now to construct the Hamiltonian matrix for the entire system, we use the tensor product (Kronecker product) to combine the Pauli matrices ($S^\alpha = \sigma^\alpha /2$) for different sites. Once the Hamiltonian matrix is constructed, we need to diagonalize it to find its eigenvalues and eigenvectors. We performed this process using numerical linear algebra libraries like \textit{NumPy} in Python (we use sparse matrix technique to do it).}

\begin{table*}
        \centering
\begin{tabular}
{|p{0.25\textwidth}|p{0.2\textwidth}|p{0.25\textwidth}|p{0.25\textwidth}|}
\hline 

  State & $ {\rm k}_{\rm RMS}$ & $\displaystyle |\Psi _{\theta } |^{2}_{XXZ}$ \ \ (real space ED) & $\displaystyle |\Psi _{\theta } |^{2}_{q_{c}}$ \ ( k-space ED) \\
\hline 

$\displaystyle |0\rangle =[0 0 0 0 \textbf{\color{red}{1}} 1 1 1 1 1 \textbf{\color{red}{1}} 0 0 0]$ & 0.00 & 0.93336 & 0.97935   \\
\hline 
 $\displaystyle |1\rangle=[0 0 0 \textbf{\color{red}{10}} 1 1 1 1 1 \textbf{\color{red}{01}} 0 0] $ & 0.54 & 0.02284 & 0.01575    \\
\hline 
 $\displaystyle |2\rangle = [0 0 0 \textbf{\color{red}{110}} 1 1 1 1 \textbf{\color{red}{001}} 0] $ &  0.99 & 0.00669 & 0.00124  \\ 
  \hline 
 $\displaystyle |3\rangle = [0 0 \textbf{\color{red}{100}} 1 1 1 1 \textbf{\color{red}{011}} 0 0]$ & 0.99 & 0.00669 & 0.00124   \\
\hline 
$\displaystyle |4\rangle = [0 0 0 \textbf{\color{red}{1110}} 1 1 1 \textbf{\color{red}{0001}} ]$ & 1.38 & 0.00296& 0.00008  \\
\hline 
$\displaystyle |5\rangle = [0 \textbf{\color{red}{1000}} 1 1 1 \textbf{\color{red}{0111}} 0 0 ]$ &  1.38 & 0.00296 & 0.00008 \\
\hline 
\end{tabular}
\caption{{The probability density corresponding to different basis states representing particle-hole excitations over the FS as obtained through real space (XXZ) (Column $3$) and the momentum-space with a fixed $q_c$ (Column $4$) ED methods for the system discussed in Fig.\ \ref{fig:amplitude_main} of the main text. The red numbers in column 1 indicate position of the FS in $|0\rangle$ and chart out the position of the particle-hole excitations near the FS for other states. Column $2$ indicates the RMS momenta with respect to FS defined as  ${\rm k}_{\rm RMS}=\sqrt{\langle k^2 \rangle} - \sqrt{\langle k^2 \rangle}_{FS}$,  where ($\sqrt{\langle k^2 \rangle} = \sqrt{\sum_k k^2 (1+ \theta_k)/2}$), which measures deviation from $|0\rangle$ for the corresponding states.} }
\label{tab_normalisedwieights}
\end{table*}

\begin{figure*}
    \centering
    \includegraphics[width=2\columnwidth]{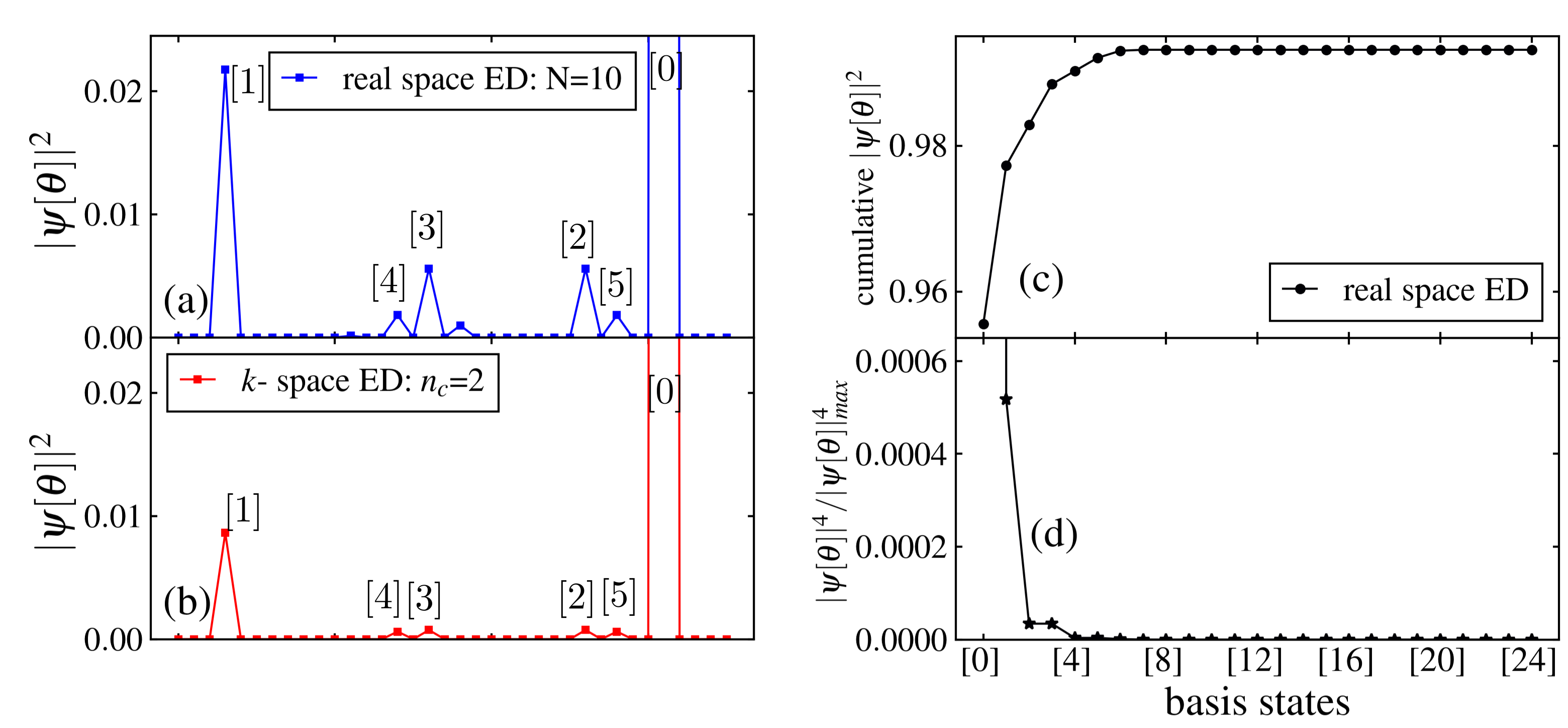}
    \caption{(a) The probabilities of the GS wave-function obtained from ED of the half filled Hubbard model ($N=10, t=1, V_0=0.3$) for spinless Fermions in the basis given by Eq.~\ref{eq_wftheta} (horizontal axis). The relevant momenta in the first BZ are denoted by $k=-\pi+\alpha\pi/5$ with $\alpha=0,1,\cdots,9$. In this labelling the four states with leading contributions are shown with $[0]\equiv [0001111100]$ being the non-interacting GS and the rest corresponding to the excitations above the non-interacting GS. These are $[1]= [0010111010]$, $[2]= [0011011001]$, $[3]=[0100110110]$, $[4]=[0011010110] $ and $[5] = [0100111001]$. The states are arranged in terms of the increasing number of particle-hole excitations which can also be quantified via their root mean square momenta deviation from the FS. Note that the vertical scale has been truncated for improved visibility of the contributing basis states.  (b) The exact diagonalization done using the momentum space cut-off ${\bf q}_c$ ($n_c = 2\left(q_c=\frac{2\pi}{N}n_c\right),~n_0 = 1.5\left(q_0=\frac{2\pi}{N}n_0\right)$, see main text for details). (c) The cumulative weights of the different basis states. (4) The IPR for the different basis states. The IPR has been normalised with respect to the maximum contribution (vertical axis truncated for better visibility) from $[0]$.}
    \label{fig:amp10_app}
\end{figure*}

\begin{figure*}
    \centering
    \includegraphics[width=2\columnwidth]{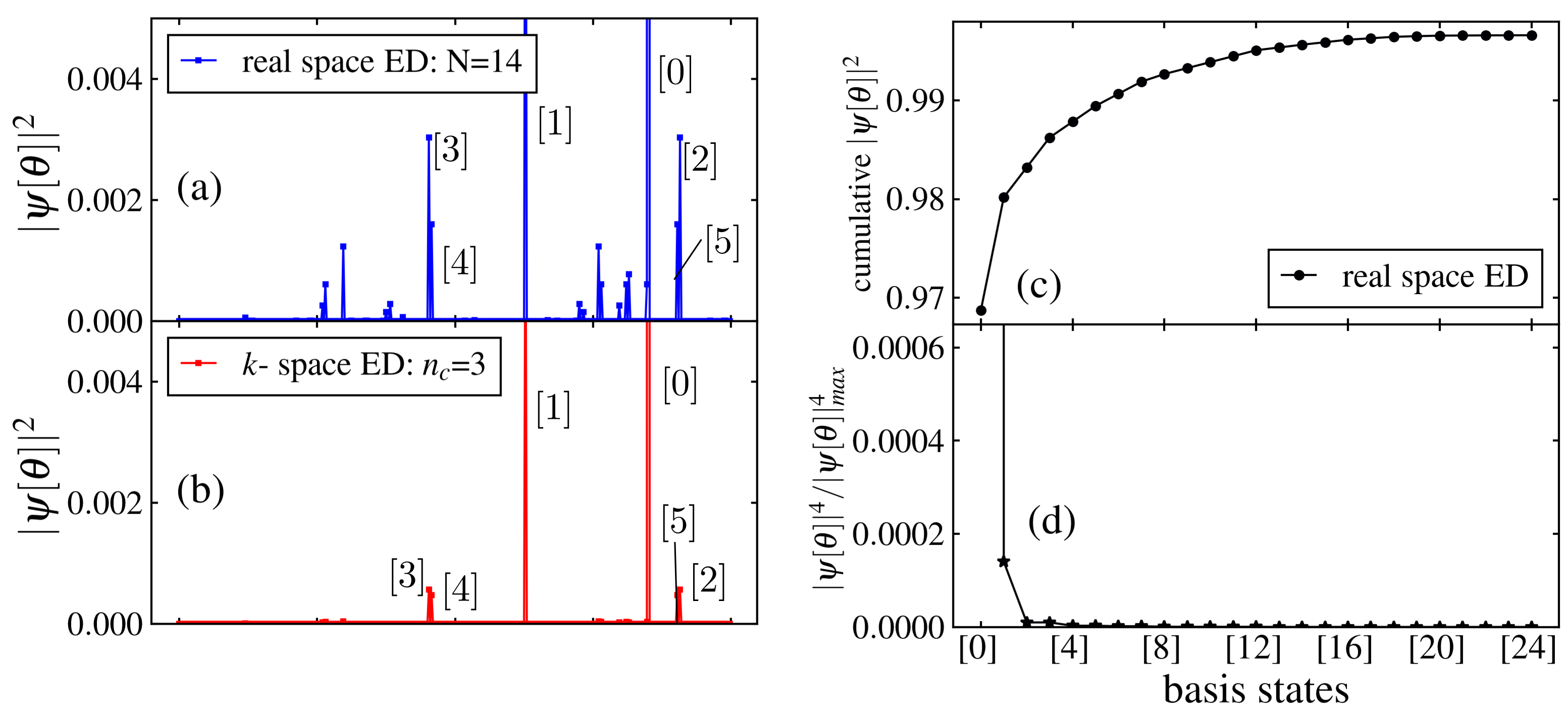}
    \caption{ (a) The probabilities of the GS wave-function obtained from ED of the half filled Hubbard model ($N=14, t=1,  V_0=0.2$) for spinless Fermions in the basis given by Eq.~\ref{eq_wftheta} (horizontal axis). The relevant momenta in the first BZ are denoted by $k=-\pi+\alpha\pi/7$ with $\alpha=0,1,\cdots,13$. In this labelling the four states with leading contributions are shown with $[0]\equiv [00001111111000]$ being the non-interacting GS and the rest corresponding to the excitations above the non-interacting GS. These are $[1]= [00010111110100]$, $[2]= [00011011110010]$, $[3]=[00100111101100]$, $[4]= [00011101110001]$ and $[5] = [01000111011100]$. The states are arranged in terms of the increasing number of particle-hole excitations which can also be quantified via their root mean square momenta deviation from the FS. Note that the vertical scale has been truncated for improved visibility of the contributing basis states.  (b) The exact diagonalization done using the momentum space cut-off ${\bf q}_c$ ($n_c = 3\left(q_c=\frac{2\pi}{N}n_c\right),~n_0 = 1.5\left(q_0=\frac{2\pi}{N}n_0\right)$, see main text for details). (c) The cumulative weights of the different basis states. (4) The IPR for the different basis states. The IPR has been normalised with respect to the maximum contribution (vertical axis truncated for better visibility) from $[0]$.}
    \label{fig:amp14_app}
\end{figure*}

\subsection{Exact Diagonalization in Real space}
{The Hamiltonian given by Eq.~\ref{eq_intham} }can be diagonalized by mapping it to the XXZ model using Jordan-Wigner transformation 
\begin{equation}\label{eqn:eq5}
\begin{split}
S_i^z=\frac{1}{2}-c_i^\dag c_i,~~S_i^+=\frac{1}{2}\tilde{K}_i c_i,~~S_i^-=\frac{1}{2}c_i^\dag \tilde{K}_i,
\end{split}
\end{equation}
where $c_i$ are spinless Fermions and
\begin{equation}\label{eqn:eq4}
    \tilde{K}_i =\exp\left[i\pi\sum_{j=1}^{i-1}c_j^\dag c_j\right].
\end{equation}
{The Hamiltonian (Eq.~\ref{eq_intham}), using this transformation, can be written as} 
\begin{align}\label{eq_XXZ_Ham}
H= -2t\sum_{i}\left( S^x_i~S^x_{i+1}+S^y_i~S^y_{i+1}\right)+V_0\sum_{i} S^z_i~S^z_{i+1}.
\end{align}

The GS of the XXZ Hamiltonian (Eq.\ \ref{eq_XXZ_Ham}) was obtained using ED in real space. The corresponding GS in momentum space is obtained by first computing its Slater determinant and a subsequent Fourier transform. The momentum-space GS so obtained can be understood in the $[\theta]$- basis (Eq.\ \ref{eq_wftheta}) and can be used to compute the connected correlator and the entanglement entropy; these serve as a benchmark and are shown in Fig.~\ref{fig_xxzcompare}.

Since we are interested in understanding the role of interaction in the configuration space spanned by $\vert\Psi[\theta]\rangle$ basis mentioned in Eq.~\ref{eq_wftheta}. We looked at the amplitude of the each basis states in the computed GS as shown in Fig.\  {\ref{fig:amplitude_main}}. {The largest peak, as expected, corresponds to the state with non-interacting FS at a given $\nu$. However, due to interaction between Fermions, there are other states with non-zero amplitudes. } The details of the weights of different $|\Psi[\theta]\rangle$ states with non-zero amplitude in the GS for the system parameters discussed in Fig.~\ref{fig:amplitude_main} is given in Table \ref{tab_normalisedwieights}. Results similar to Fig.~\ref{fig:amplitude_main} for $N=10$ is shown in Fig.~\ref{fig:amp10_app} and for $N=14$ and $V_0=0.2$ in Fig.~\ref{fig:amp14_app}.

\subsection{Exact Diagonalization in momentum space}
\label{appen_thetaed}

{For the momentum space ED, we use a hard cutoff ${\bf q}_c$ and also a soft cutoff ${\bf q_0}$ (we choose $q_0= 2\pi n_0/N$)for cutoff along $k_x$ and $k_y$) using ED in momentum space. This constitute a choice of the interaction potential (Eq.\ \ref{eq_intham})  
\begin{align}V({\bf q})=\left\{\begin{array}{lc}
  V_0 e^{\frac{-q^2}{2q_0^2}}   &  \forall |{\bf q}|\leq |{\bf q}_c|,\\
   0  & {\rm otherwise.}
\end{array}\right.
\end{align}
The Hamiltonian in Eq.~\ref{eq_intham} is then diagonalized by creating a matrix in the $|\Psi[\theta]\rangle$ basis using the matrix elements mentioned in Eq.~\ref{eq_intHam_matrix}. For a given $n_c$ (we choose $q_c= 2 \pi n_c/N$ for cutoff along $k_x$ and $k_y$), the number of states that are kept is given by $N_s=\left(^{2n_c}C_{n_c}\right)^2$ in 2D; this yields a $N_s\times N_s$ matrix form of the Hamiltonian which is diagonalized using ED.}

In 1D at $\nu=1/2$, there are two Fermi-points ${\bf k_{F1}}=-\frac{\pi}{2}\hat{\bf x}$ and ${\bf k_{F2}}=\frac{\pi}{2}\hat{\bf x}$. Then to calculate the different elements in Eq.~\ref{eq_off_diagonal_ham_AC}, we note that the three momenta are restricted to the following domain
  with the possible values of ${\bf k_1, k_2, q}$ being:
    \begin{align}
& {\bf k_{1}} \in \Big({\bf k_{F2}}+\frac{2\pi}{N}\hat{\bf x},~~ {\bf k_{F2}}+{\bf q_c}\Big),\\
&{\bf k_2} \in \Big({\bf k_{F1}-q_c}, ~~{\bf k_{F1}}-\frac{2\pi}{N}\hat{\bf x}\Big),\\
&{\bf q}\in \Big( \frac{2\pi}{N}\hat{\bf x},~~{\bf q_c} \Big), 
 \end{align}

 and 
 
 \begin{align}
& {\bf k_2} \in \Big({\bf k_{F2}}+\frac{2\pi}{N}\hat{\bf x}, ~~{\bf k_{F2}}+{\bf q_c}\Big),\\
&{\bf k_1} \in \Big({\bf k_{F1}}-{\bf q_c}, ~~{\bf k_{F1}}-\frac{2\pi}{N}\hat{\bf x}\Big), \\
&{\bf q}\in \Big(\frac{2\pi}{N}\hat{\bf x},~~{\bf q_c}\Big).
\end{align}

The results for weights of different $|\Psi[\theta]\rangle$ states in the momentum-space diagonalization are shown in Fig.~\ref{fig:amplitude_main}(b), \ref{fig:amp10_app}(b) and \ref{fig:amp14_app}(b) for comparison.

\subsubsection{Details of calculation of entanglement entropy}

\begin{figure}
    \centering
\includegraphics[width=0.9\columnwidth]{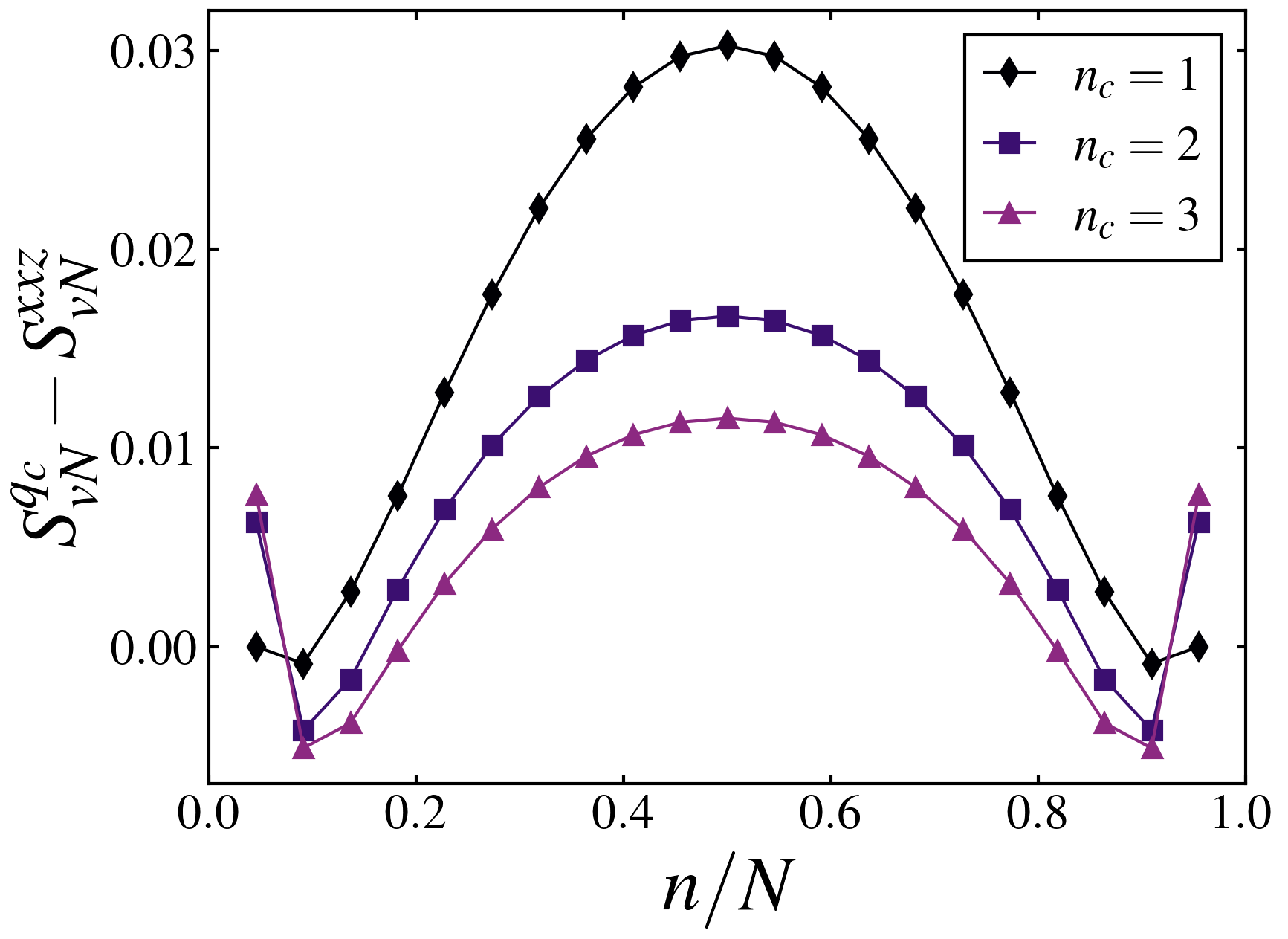}
    \caption{Difference of $S_{vN}$ calculated from {real space ED }and {k-space ED } (for N=22, $V_0$=0.3, $n_0$ = 1.5). As we increase $n_c$, it approaches zero, which means $S_{vN}^{q_c}$ starts to be in more agreement with $S_{vN}^{xxz}$.}
\label{fig:S_systematics}
\end{figure}

For a 1D quantum system at a critical point, the Cardy-Calabrese \cite{Calabrese_2009} formula of the bipartite entanglement entropy for a  system of size  $L$  with subsystem $A$ of length $l$ with periodic boundary conditions imposed on the whole system reads 
\begin{equation}
    S_{vN} = \frac{c}{3}\log{\left(\frac{L}{\pi a} \sin{\pi \frac{l}{L}}\right)} + \text{constant}.
    \label{eq_cc}
\end{equation}
We numerically calculate Von-Neumann Entropy 
\begin{equation}
    S_{vN} = -Tr[\rho_A \ln \rho_A],
\end{equation}
where $\rho_A$  is the reduced density matrix of subsystem $A$ obtained by tracing out the rest of the system ($\bar{A}$). We employed standard singular value decomposition methods (SVD) to calculate the entanglement entropy, $S_{vN}$, and have used Eq.\ \ref{eq_cc} to obtain the central charge $c$ as shown in Fig.\ \ref{fig_xxzcompare1}. 

For comparison, we show, in Fig.\ \ref{fig:S_systematics}, the difference of the entanglement entropy calculated via real space ED and k-space ED for different $n_c$. As $n_c$ is increased, the entanglement calculated by both methods agrees well showing the efficacy of the momentum-space method.

\section{Different Correlations for the RK state}
\label{appen_rk}

\begin{figure}
\includegraphics[width=.9\columnwidth]{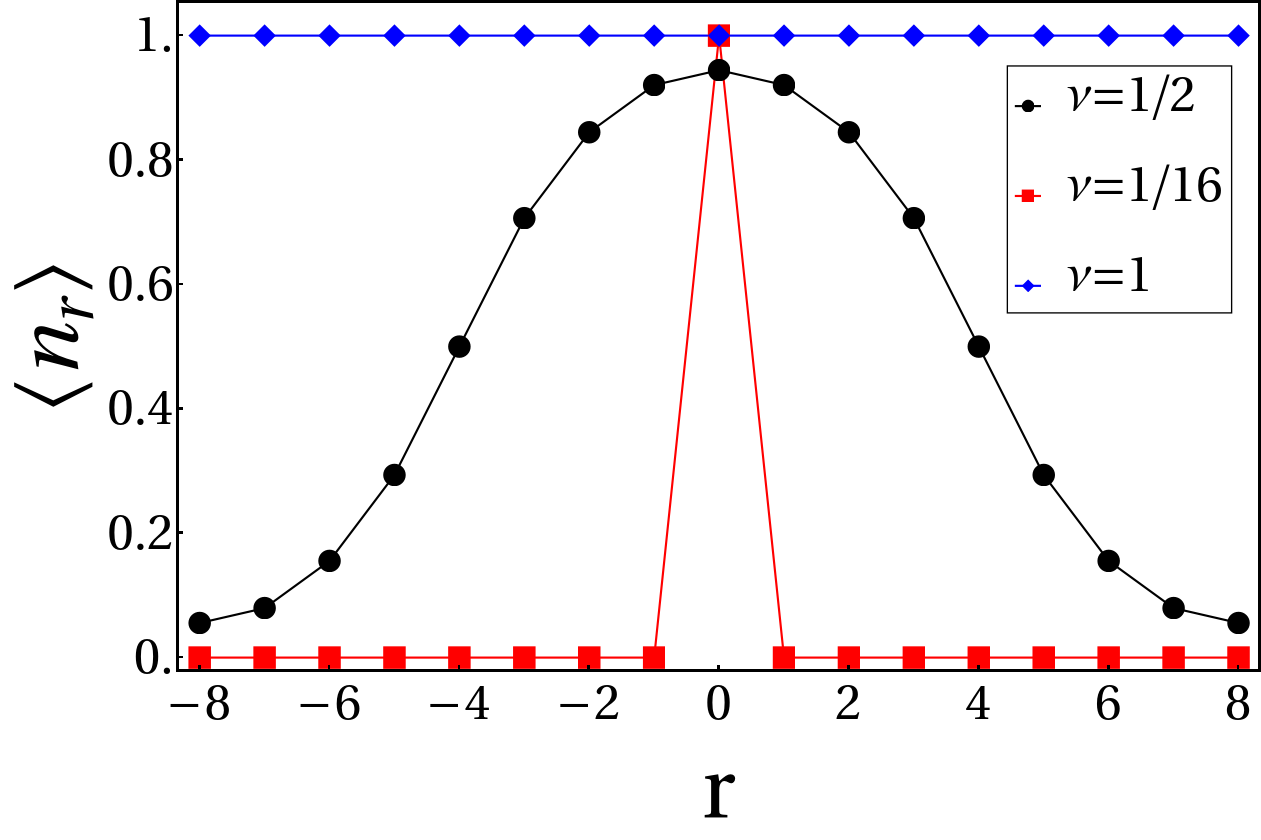}
\caption{{Plot of $\langle n_{\bf r} \rangle$ as a function of ${\bf r}=r\hat{\bf{x}}$ for the RK ground states, where we have taken $16$ sites for different fillings $\nu$ and used periodic boundary conditions (PBC).}}
\label{fig_rk_nr}
\end{figure}

\begin{figure}
\includegraphics[width=.9\columnwidth]{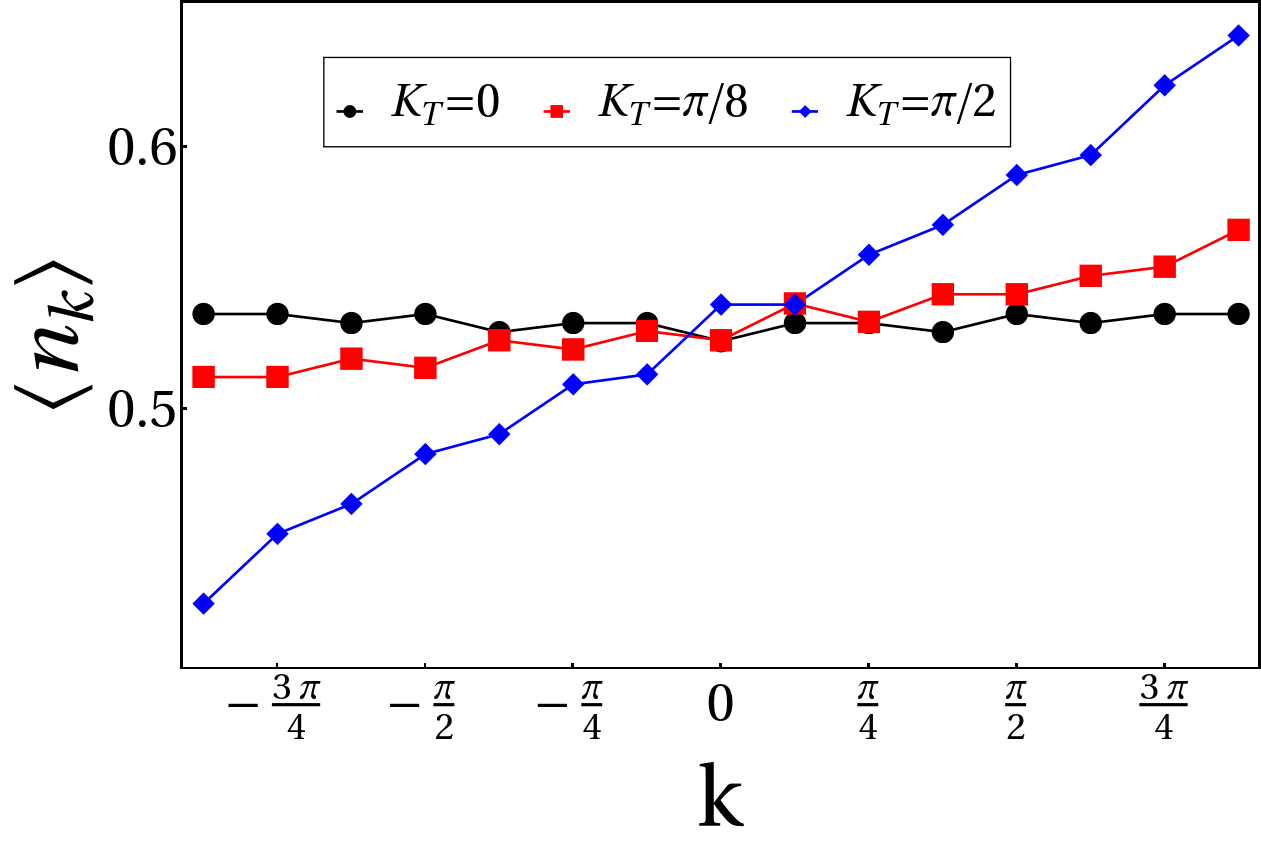}
\caption{{Plot of $\langle n_{\bf k} \rangle $ as a function of $k$ for the constrained RK ground states ($|\psi_{\rm RK}^{{\bf k}_{\rm T}}\rangle$), where we have taken $16$ sites for different  ${\bf k}_{\rm T}$ and used periodic boundary conditions (PBC).}}
\label{fig_constrained_rk_nk_16}
\end{figure}

\begin{figure}
\includegraphics[width=.9\columnwidth]{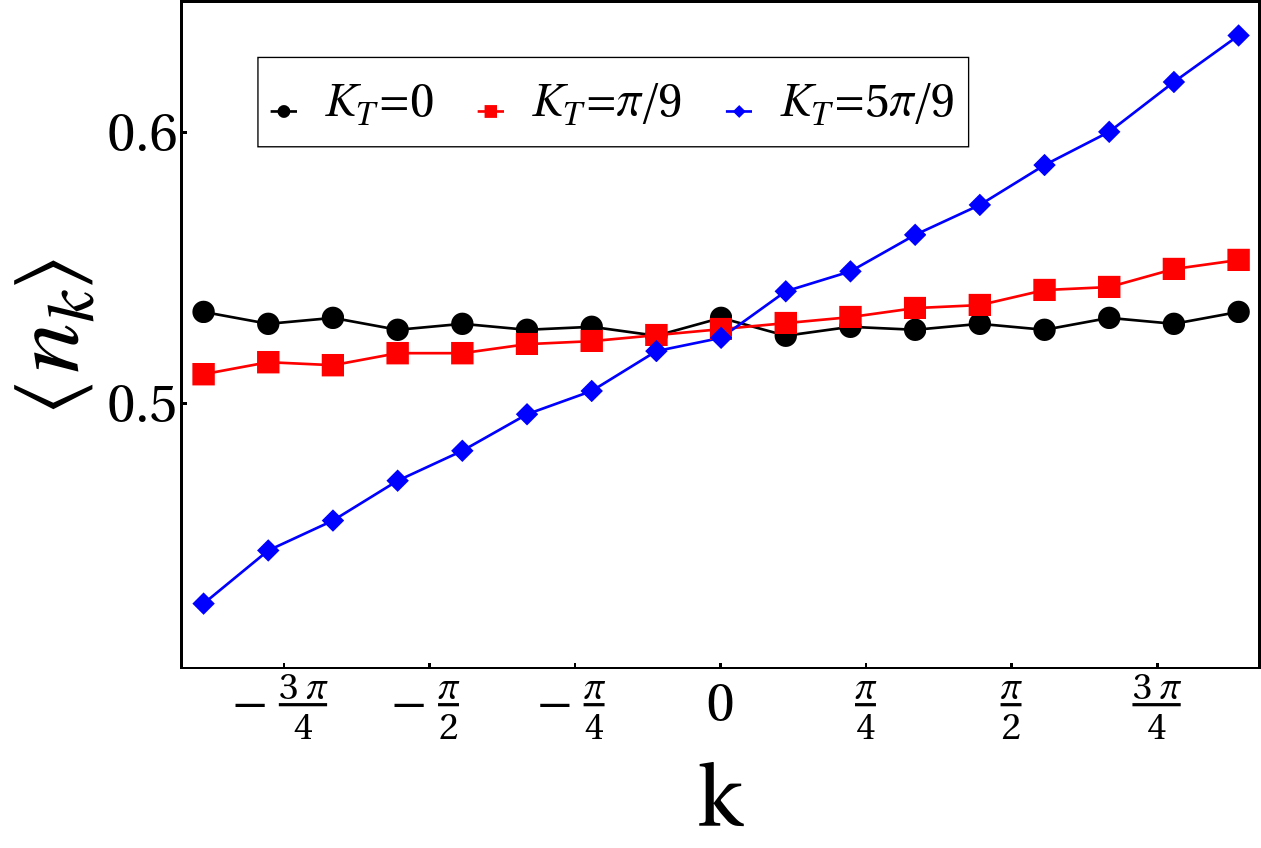}
\caption{{Plot of $\langle n_{\bf k} \rangle $ as a function of $k$ for the constrained RK ground states ($|\psi_{\rm RK}^{{\bf k}_{\rm T}}\rangle$), where we have taken $18$ sites for different  ${\bf k}_{\rm T}$ and used periodic boundary conditions (PBC).}}
\label{fig_constrained_rk_nk_18}
\end{figure}

For Eq.\ \ref{eq_rkstate}, we computed the average real space occupation $\langle\Psi_{\rm RK}|\hat n_{\bf r}|\Psi_{\rm RK}\rangle$ in 1D for finite systems. This is plotted in Fig.\ \ref{fig_rk_nr}. Useful insights are obtained by considering {the single-particle (corresponding to $\nu=1/16$) RK state (Eq.\ \ref{eq_rkstate}) $\sim \sum_{\bf k}|{\bf k}\rangle\sim |{\bf r}=0\rangle$. The expectation value of $n_{\bf r}$ to be sharply peaked at ${\bf r}=0$ for this state (see Fig.\ \ref{fig_rk_nr}) exhibiting breaking of translation. } However, this provides a way to generate Wannier functions centered at other lattice sites by applying the lattice translation operator to the $|{\bf r}=0\rangle$ state. The fully filled ($\nu=1$) band, on the other hand has a flat average with the intermediate fillings interpolating between the two limits via a distribution of $\langle\Psi_{\rm RK}|\hat n_{\bf r}|\Psi_{\rm RK}\rangle$ centered around ${\bf r}=0$. This distribution can be shifted to any lattice site via translation operator as mentioned above. The half-width of the distribution (not shown) increases monotonically with filling. It would be interesting to ask for the spectral properties of $H_{\rm RK}$ which is an interesting future direction.

For the constrained RK state $|\psi_{\rm RK}^{{\bf k}_{\rm T}}\rangle$, we always choose the momentum ${\bf k}=\pi$ to be empty for convenience in the main text. This is because on including this mode, the real space density shows oscillations with wave-vector $\pi$ irrespective of the filling, $\nu$. We think that this is an artifact of this wave-function. Turning back to $|\psi_{\rm RK}^{{\bf k}_{\rm T}}\rangle$, we find \begin{align}\label{eq_real_space_occupancy_constraint_RK}
    \langle \psi_{\rm RK}^{{\bf k}_{\rm T}}\vert n_{\bf r} \vert &\psi_{\rm RK}^{{\bf k}_{\rm T}}\rangle  
    =  \frac{1}{N}\sum_{\bf k_1, k_2} \langle \psi_{\rm RK}^{{\bf k}_{\rm T}} \vert c^{\dagger}_{\bf k_1} c_{\bf k_2} \vert \psi_{\rm RK}^{{\bf k}_{\rm T}}\rangle e^{i (\bf{k_1-k_2})\cdot{\bf r}}.
\end{align}
Since all states in the superposition carries same ${\bf k}_{\rm T}$, this implies $k_1=k_2$ such that the above equation gives
\begin{align}\label{eq_real_space_occupancy_constraint_RK_2}
    \langle \psi_{\rm RK}^{{\bf k}_{\rm T}}\vert n_{\bf r} \vert &\psi_{\rm RK}^{{\bf k}_{\rm T}}\rangle  =  \frac{1}{N}\sum_{\bf k} \langle \psi_{\rm RK}^{{\bf k}_{\rm T}} \vert c^{\dagger}_{\bf k} c_{\bf k} \vert \psi_{\rm RK}^{{\bf k}_{\rm T}}\rangle=\nu.
\end{align}

The average momentum occupancy is shown in Fig.\ \ref{fig:rkmomentumresolved} in the main text. Similar plots for $N=16$ and $18$ are shown in Fig.\ \ref{fig_constrained_rk_nk_16} and \ref{fig_constrained_rk_nk_18}. {As mentioned in the  main text, it is concluded that the oscillation in $\langle n_{\bf k}\rangle$ fades with increasing system size $(N)$ and in the large $N$ limit there is no sharp signature in the momentum space occupation. }

\section{Superposition of Elliptical FS}
\label{appen_ellipses}

\subsection{The Elliptical FS}

The area of BZ is $4 \pi^2$ and for filling $\nu$, the area of FS therefore is $4\pi^2\nu$. If the FS is elliptical, it's equation is given by (denoting states as $\vert \epsilon, \nu,{\bf \hat{k}}\rangle$),
\begin{align}\label{eq_ellipse_eqn}
    \frac{k_{\parallel}^2}{a^2}+\frac{k_{\perp}^2}{a^2(1-\epsilon^2)}=\mu,
    \end{align}
    where $k_\parallel (k_\perp)$ are momentum components resolved along (perpendicular) to the major axis along $\hat{\bf k}$.
For every ellipse (Fig.\ \ref{fig_ellipse}) the relation between the major axis (of length $2a$), minor axis (of length 2$a\sqrt{1-\epsilon^2}$)for given eccentricity $\epsilon$, such that the relation between the area of the ellipse and the filling is given by
\begin{eqnarray}
\mu \, a^2 \sqrt{1-\epsilon^2} &=& 4\pi\nu,
\end{eqnarray}
where $a<\pi$ for closed FS. For given eccentricity at half filling ($\nu=1/2)$, we have taken
\begin{align}
    \mu = 1, ~~~~a=\left(\frac{4\pi^2}{1-\epsilon^2}\right)^{1/4}.
\end{align}

Considering the states mentioned in Eq.~\ref{eq_FL_distribution}, \ref{eq_NFL_distribution_1} and \ref{eq_superposed_ellipses_2}, we would like to emphasize that each individual states in the superposition carries same total momenta (${\bf k}_{\rm T}$ in Eq.~\ref{eq:particleno}). If we compute real space density $\langle n_{\bf r} \rangle$, it will be uniform as shown below, and is equal to $\nu$
\begin{align}\label{eq_real_space_occupancy}
    \langle \psi_{\nu} \vert n_{\bf r} \vert &\psi_{\nu}\rangle = \langle \psi_{\nu} \vert c^{\dagger}_{\bf r} c_{\bf r} \vert  \psi_{\nu}\rangle   \nonumber \\ 
    & =  \frac{1}{N}\sum_{\bf k_1, k_2} \langle \psi_{\nu} \vert c^{\dagger}_{\bf k_1} c_{\bf k_2} \vert  \psi_{\nu}\rangle e^{i (\bf{k_1-k_2})\cdot{\bf r}} .
\end{align}
Since all states carries same total momentum (${\bf k}_{\rm T}$), this implies ${\bf k_1}={\bf k_2}$ in eq.~\ref{eq_real_space_occupancy}. Expanding $\vert \psi_{\nu}\rangle $ using Eq.~\ref{eq_superposed_ellipses}, {we finally obtain} 
\begin{align}\label{eq_real_space_occup_2}
    \langle \psi_{\nu} \vert n_{\bf r} \vert \psi_{\nu}\rangle & \nonumber \\
= & \frac{1}{N}\sum_{\bf k} \int_0^{\pi} d {\hat{\bf k}} \int_0^{\epsilon_{\rm max}} d\epsilon \vert \psi(\epsilon, \hat{\bf k})\vert^2 \langle \epsilon, \nu, \hat{\bf k} \vert  n_{\bf k} \vert \epsilon, \nu, \hat{\bf k} \rangle \nonumber  \\
= & \int_0^{\pi}d{\hat{\bf k}} \int _0^{\epsilon_{\rm max}}d\epsilon \vert \psi(\epsilon, \hat{\bf k})\vert^2 \times \nu = \nu .
\end{align}
\subsection{quasi-particle Residue}\label{sec_quas_residue}
 The quasi-particle residue has the following form~\cite{landau1980course} 
\begin{align}
    Z_{\bf K_F}(\epsilon)=n({\bf k_F-0^+},\epsilon)-n({\bf k_F+0^+},\epsilon),
\end{align}
which {captures the} discontinuity at the FS. So for an ellipse that forms the basis state $|\epsilon,\hat{\bf k}\rangle$, the above formula can be cast into the form given by Eq.\ \ref{eq_quas_residue} using a Lorentzian regulator (which is also used in numerical calculations). For superposed ellipses, each basis state $|\epsilon, \hat{\bf k}\rangle$ then produces its own jump with a strength of $|\psi(\epsilon, \hat{\bf k})|^2$ such that the net {\it average} residue is given by Eq.~\ref{eq_netz}.

\subsubsection{The residue for the FL}

\begin{figure*}
   \begin{subfigure}{
\includegraphics[width=.45\columnwidth]{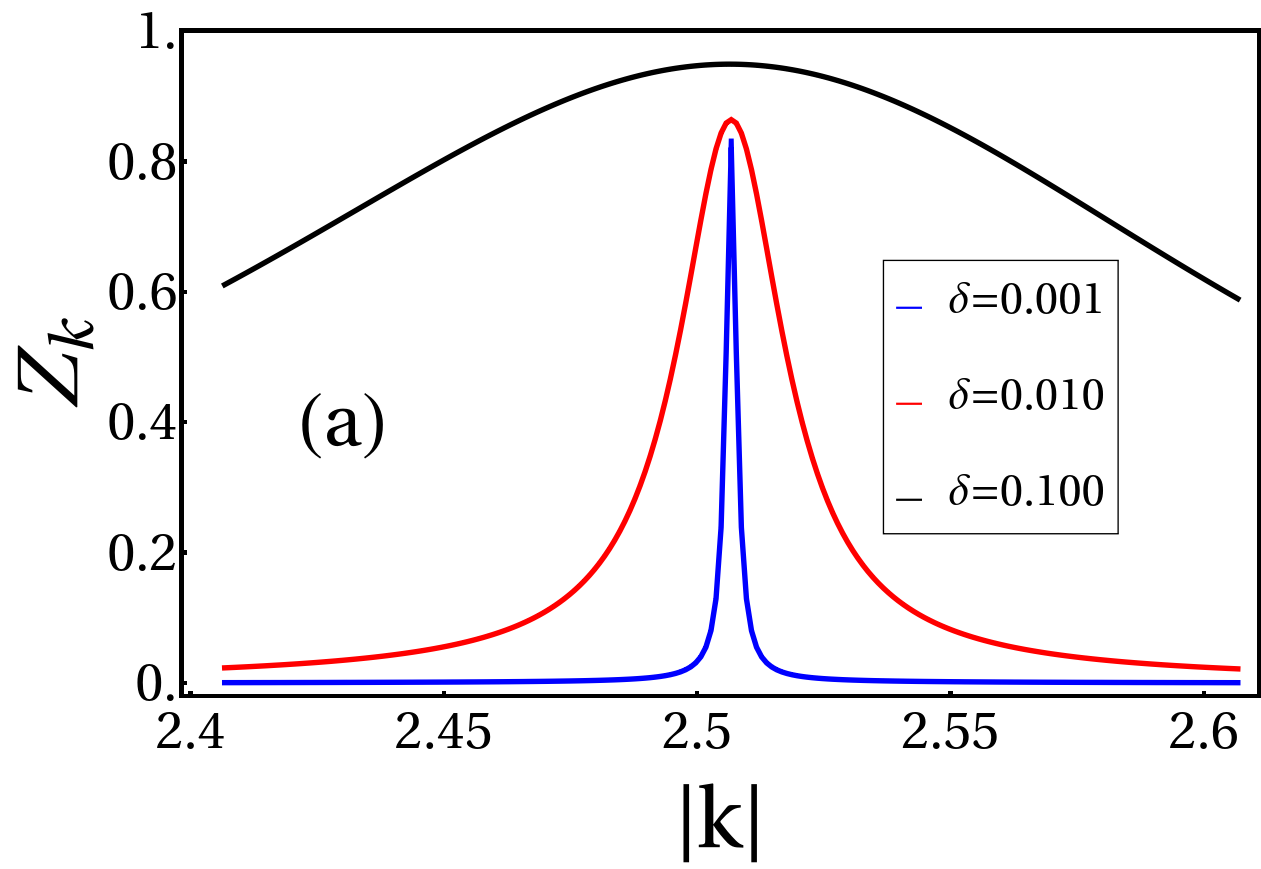}
    \label{fig:fs_Zk_isotropic_FL}}
\end{subfigure}
\begin{subfigure}{
\includegraphics[width=.45\columnwidth]{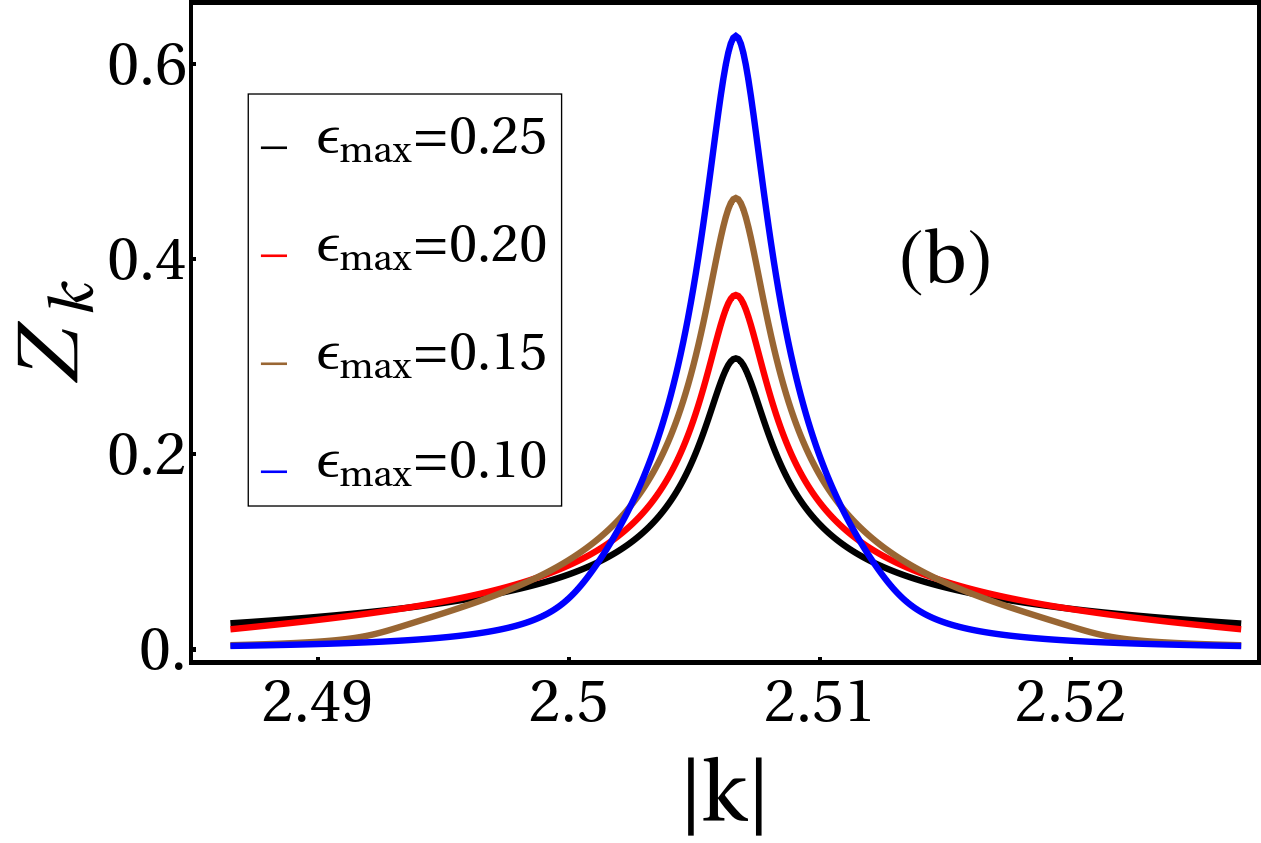}
\label{fig_isotropic_NFL_ZK}}
\end{subfigure}
\begin{subfigure}{
\includegraphics[width=.45\columnwidth]{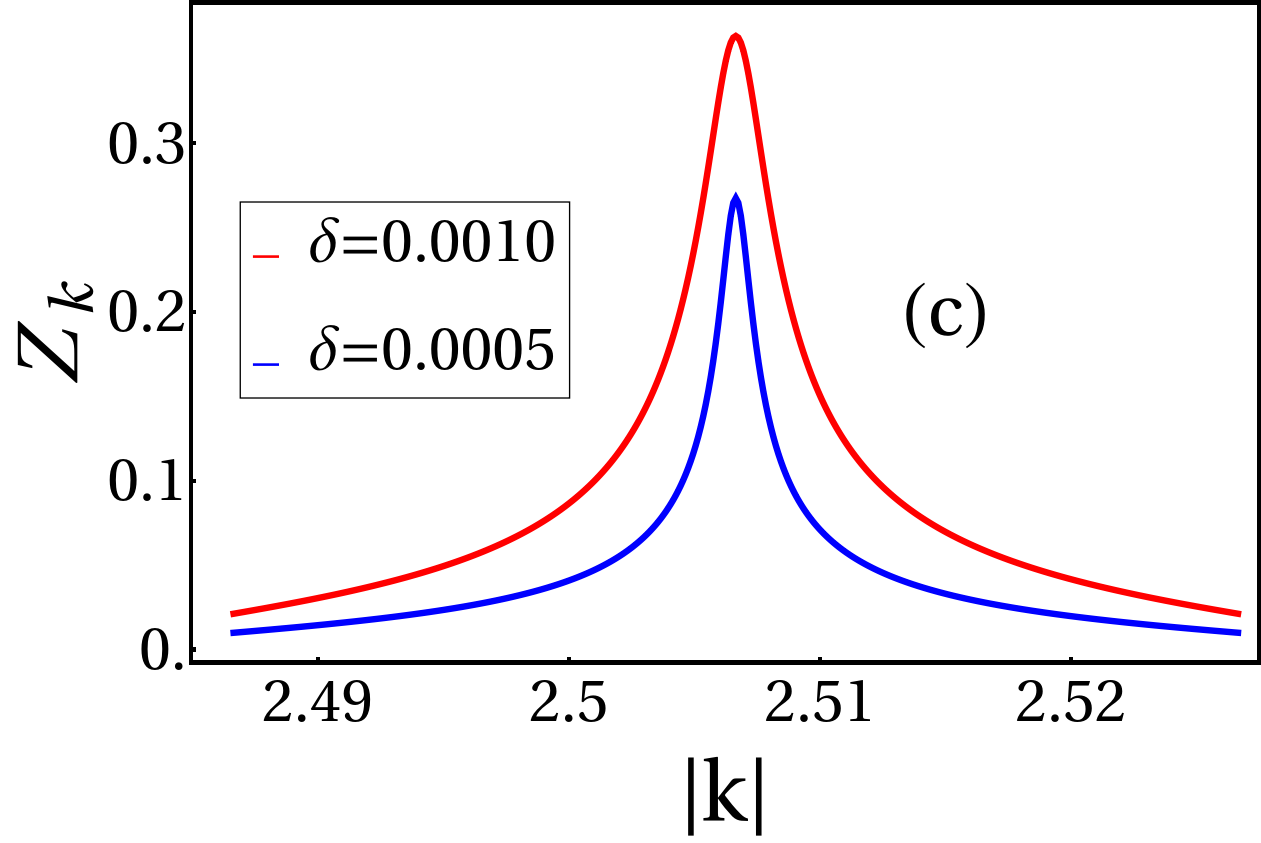}
\label{fig_isotropic_NFL_ZK_2}}
\end{subfigure}
\begin{subfigure}{
\includegraphics[width=.45\columnwidth]{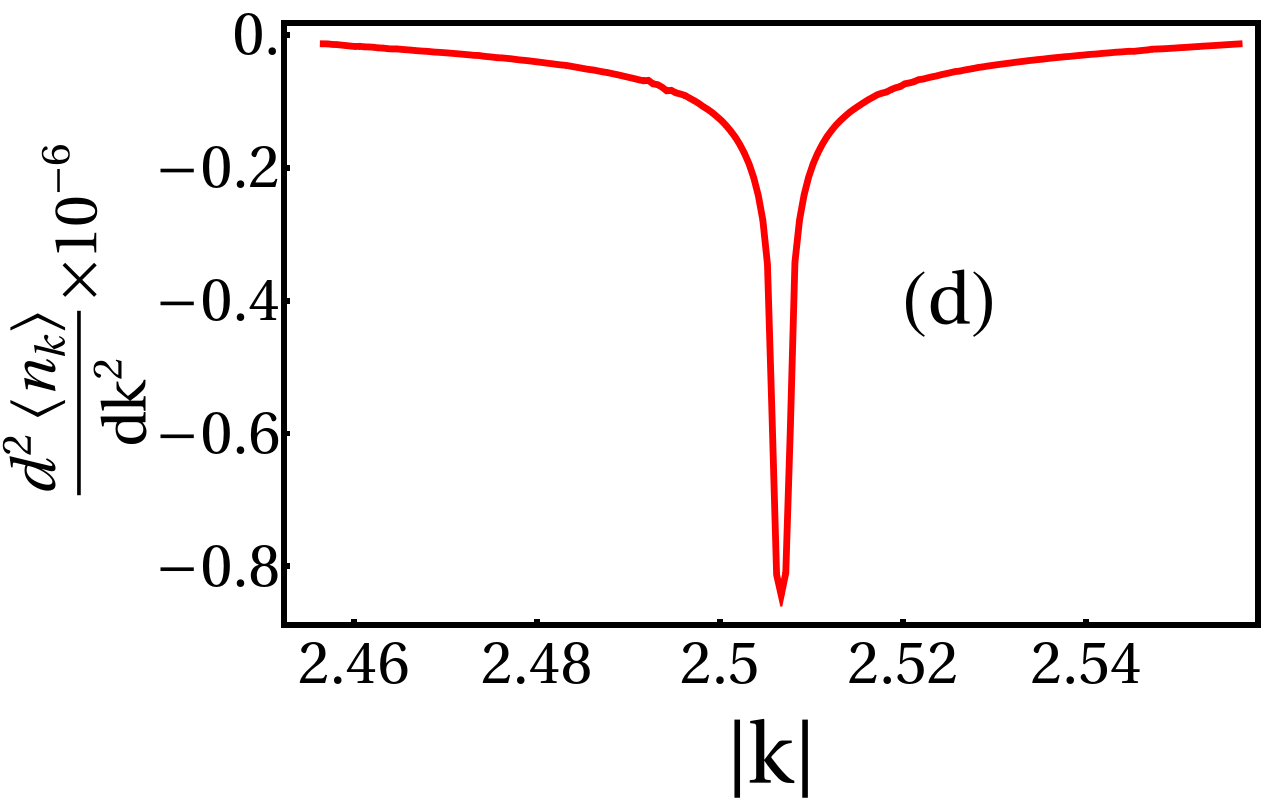}
\label{fig_isotropic_NFL_cusp}}
\end{subfigure}
\caption{(a){Plot of the residue for FL state (Eq.~\ref{eq_FL_distribution}) as a function of $k$, where $\epsilon_0=0.0$, $\epsilon_{\rm max}=0.7$, and $Z_{\epsilon_0,\hat{\bf k}} = 0.8$, 
(b) Plot of $Z_{\bf k}$ as a function of $k$ for the isotropic NFL case, with different $\epsilon_{\rm max}$ and $\delta$=0.001, 
(c) Plot of $Z_{\bf k}$ as a function of $k$ for the isotropic NFL case with the same $\epsilon_{\rm max}=0.2$ and different $\delta$,
(d) Plot of $d^2 \langle n_{\bf k} \rangle/d k^2$ corresponding to Fig.~\ref{fig_isotropic_NFL_nk} as a function of $k$ where $\epsilon_{\rm max}=0.3$}.}
\end{figure*}

\begin{figure*}
\includegraphics[width=2\columnwidth]{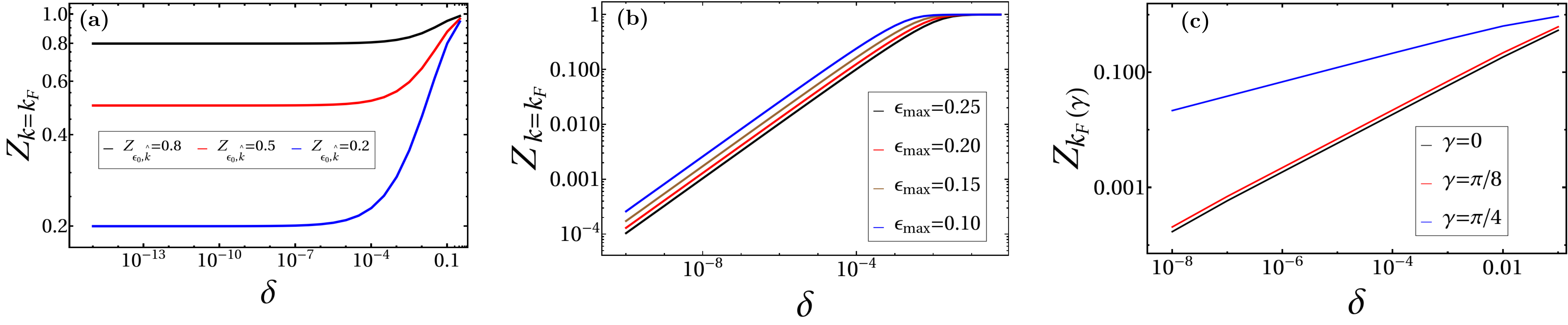}
\caption{{Plot of $Z_{\bf k}$ as a function of $\delta$ for
(a) the FL state (Eq.~\ref{eq_FL_distribution}) at $k = \sqrt{2\pi}$, $\epsilon_0=0.0$ and $\epsilon_{\rm max}=0.7$, (b) for the Isotropic NFL 
(c) for the anisotropic NFL (Eq.~\ref{eq_superposed_ellipses_2})) for different $\gamma$ and $\epsilon_{\rm max}=0.75$.}}\label{fig_Zk_delta}
\end{figure*}

The residue for the FL (Eq.~\ref{eq_FL_distribution}) is obtained using Eq.\ \ref{eq_quas_residue} numerically for different resolution, $\delta$ and taking the limit $\delta\rightarrow 0$. This is shown in Fig.\ \ref{fig:fs_Zk_isotropic_FL} while the variation of the residue at the FS with $\delta$ is shown in Fig.\ \ref{fig_Zk_delta}(a). Note that the residue is completely isotropic in momentum space. 

\subsubsection{Isotropic superposition of elliptic FS : Vanishing Residue}\label{sec_isotropic_NFL}
For the amplitudes in Eq.\ \ref{eq_NFL_distribution_1}, the quantum state (Eq.\ \ref{eq_superposed_ellipses}) is given by
\begin{align}
    \vert \psi_{\nu=1/2} \rangle  = \frac{1}{\sqrt{\pi}\sqrt{\epsilon_{\rm max}}}\int_0^{\pi} d \hat{\bf k}\int_0^{\epsilon_{\rm max}} d\epsilon ~\vert \epsilon, \nu=1/2, \hat{\bf k}\rangle,
\end{align}
such that the expectation value of the occupancy of the momentum modes is given by Fig.\ \ref{fig_isotropic_NFL_nk}. The corresponding residue calculated using Eq.\ \ref{eq_netz} is then shown in Fig.\ \ref{fig_isotropic_NFL_ZK} for a fixed resolution, $\delta=0.001$ for different $\epsilon_{\rm max}$ and Fig.\ \ref{fig_isotropic_NFL_ZK_2} for a fixed $\epsilon_{\rm max}$ and different resolutions.

The residue, shown in Fig~\ref{fig_isotropic_NFL_ZK} and \ref{fig_isotropic_NFL_ZK_2}, has its peak (at finite resolution ($\delta$)) at $ k = \sqrt{2\pi}$ in all directions. This implies that the area of this circle is $2\pi^2$ which is consistent with Luttinger theorem.  As shown in Fig.~\ref{fig_Zk_delta}(b),  $Z_{\bf k_F}\to 0$ as $\delta\to 0$. This is also concluded from the $\langle n_{\bf k}\rangle $ (in Fig \ref{fig_isotropic_NFL_nk}) as the discontinuity at the FS is replaced by the point of inflection. This is shown in Fig.~\ref{fig_isotropic_NFL_cusp}, where the second derivative of $\langle n_{\bf k}\rangle$ is showing a jump at the FS($k=\sqrt{2\pi}$) is plotted a s a function of $k$.

\begin{figure}
     \centering
\includegraphics[width=0.85\linewidth]{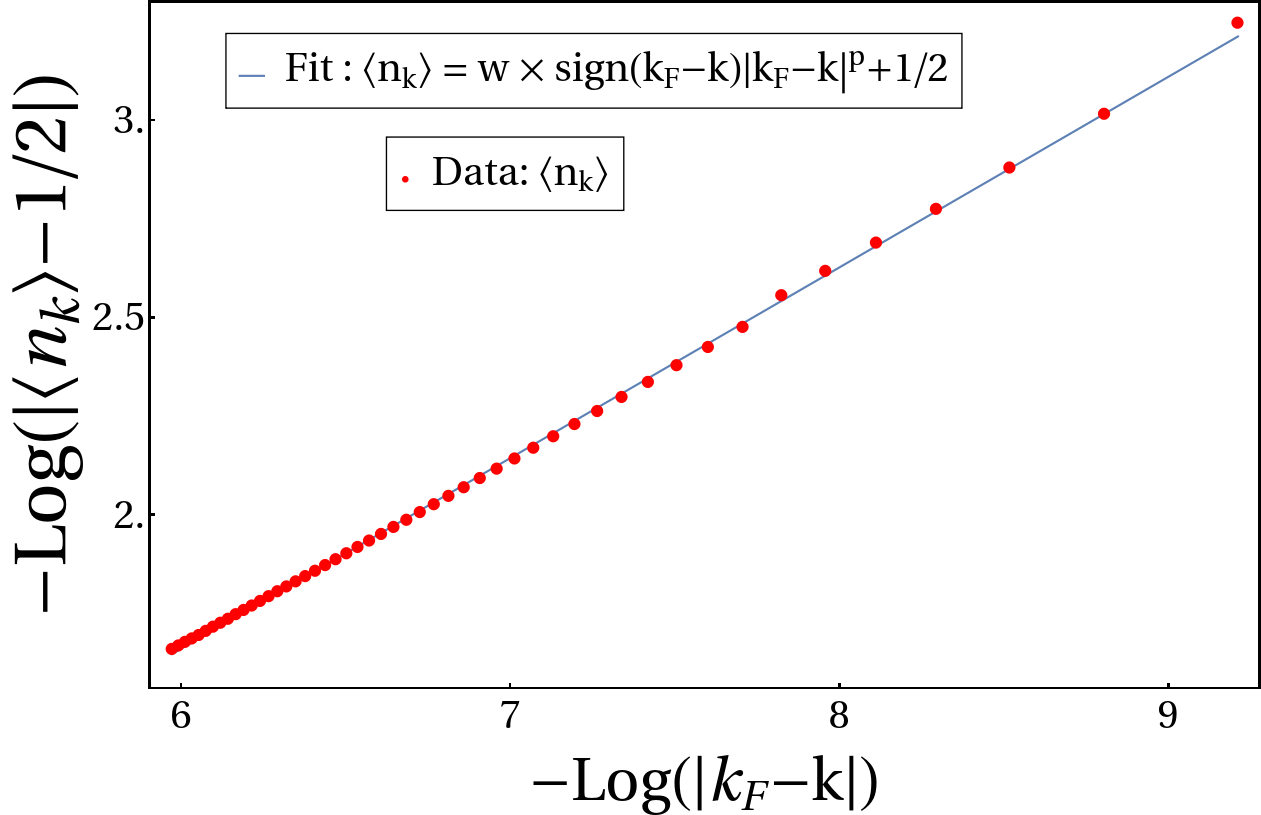}
     \caption{The scaling of $\langle n_{\bf k}\rangle\sim |{\bf k}_{\bf F}-{\bf k}|^p$ with $p=0.4828$ near the FS corresponding to Eq.\ \ref{eq_NFL_distribution_1} with $\epsilon_{\rm max}=0.25$. }
     \label{fig_fitzalpha}
 \end{figure}

Finally in Fig.\ \ref{fig_fitzalpha} we plot the fitting of the average momentum mode occupation to the power law form in agreement with the scaling theory of Ref. \cite{PhysRevB.78.035103} as discussed in the main text.

\subsubsection{Superposition of ellipses on a Square Lattice : Anisotropic residue}\label{fitness_Qp}

\begin{figure*}
\begin{subfigure}{
\includegraphics[width=.65\columnwidth]{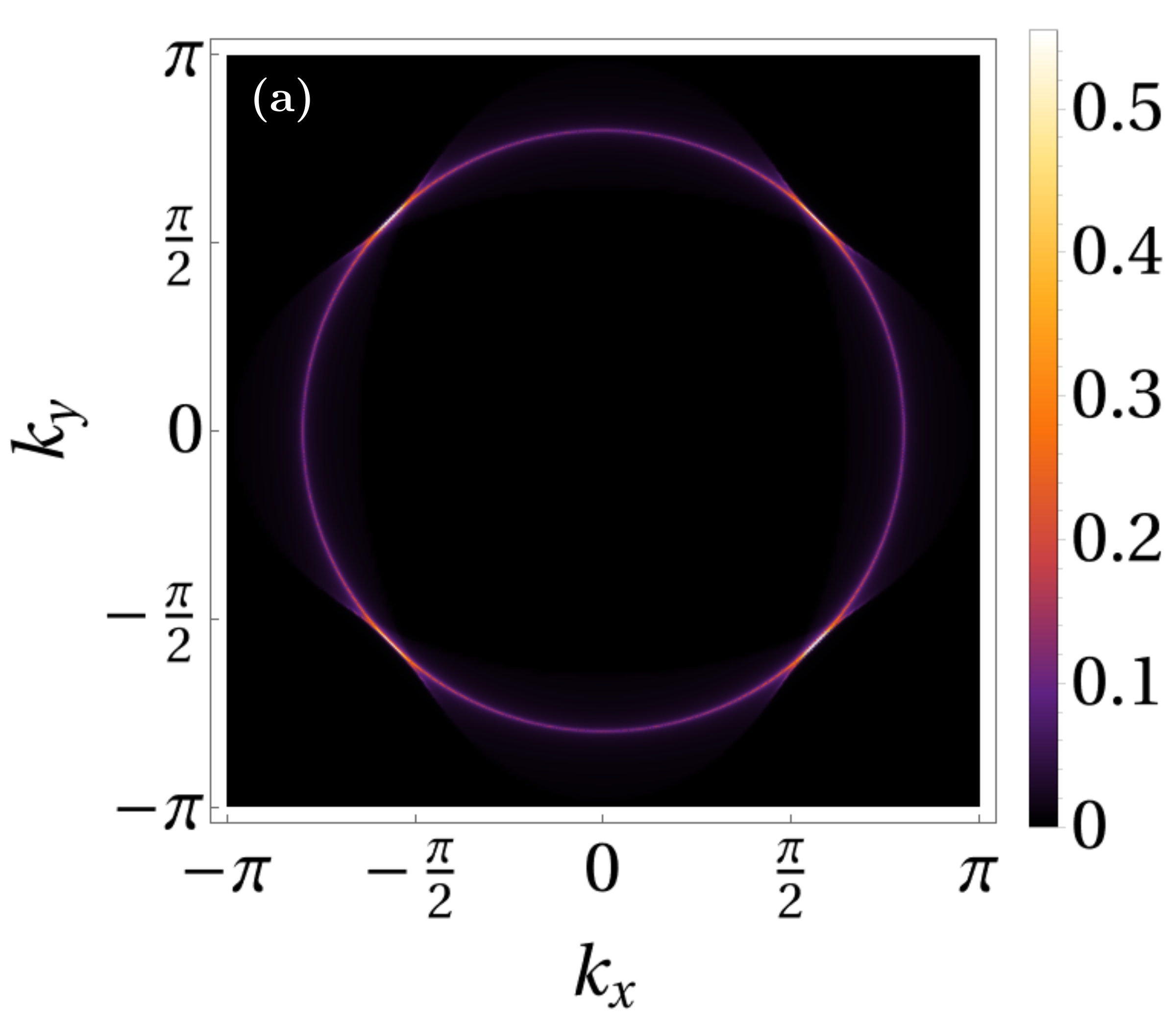}}
\end{subfigure}
\begin{subfigure}{
\includegraphics[width=.65\columnwidth]{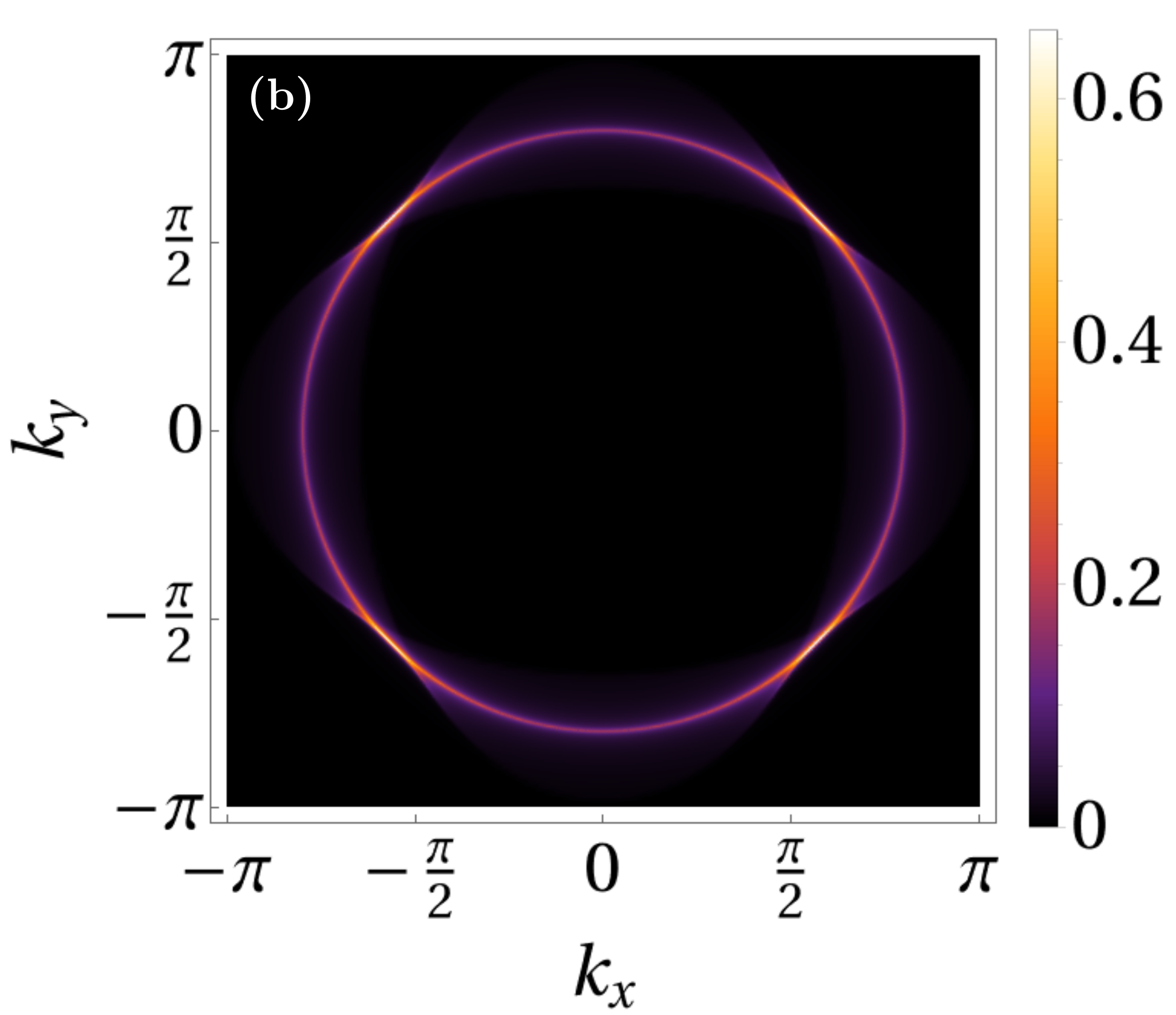} 
}\end{subfigure}
\begin{subfigure}{
\includegraphics[width=.65\columnwidth]{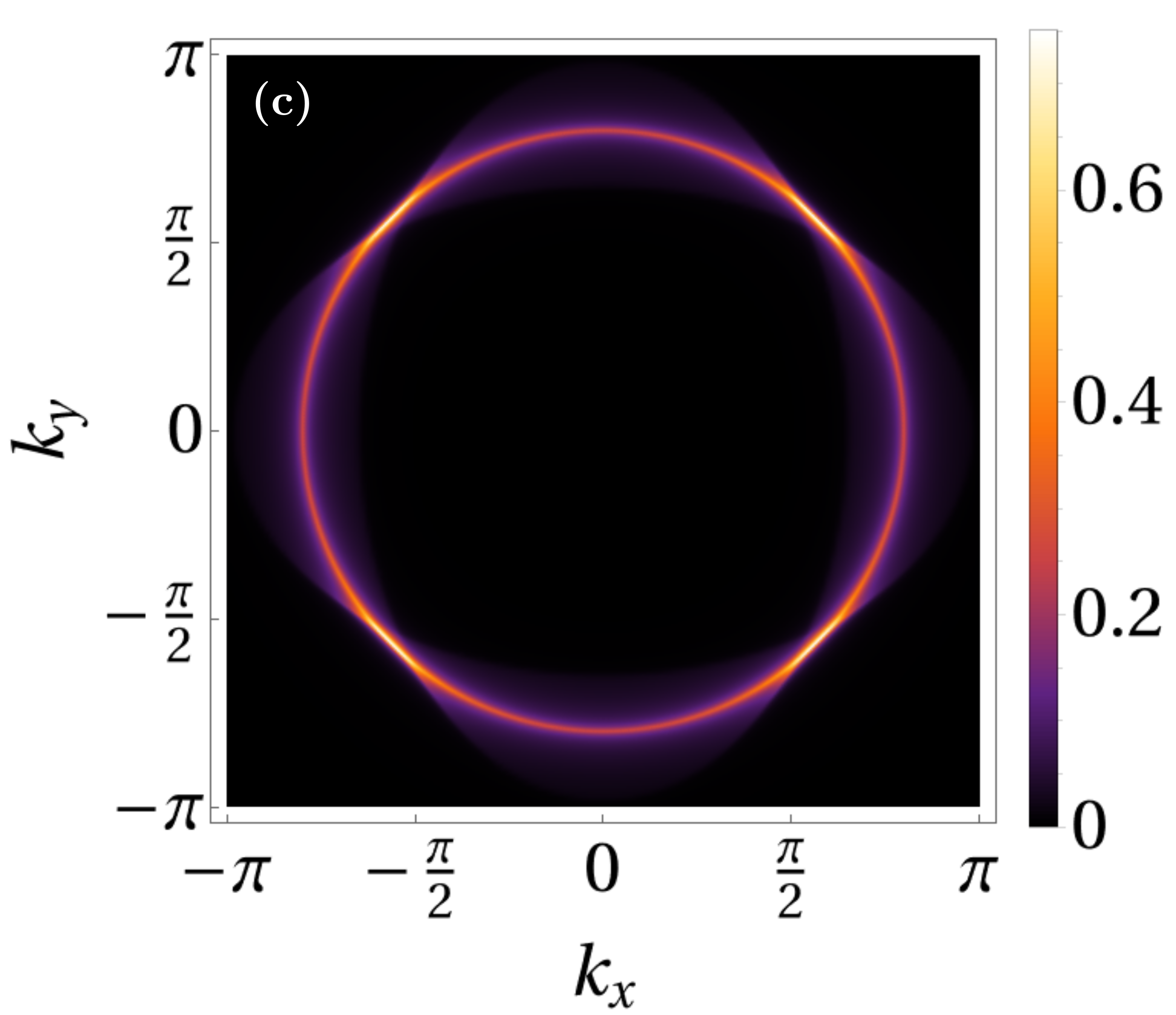} 
}\end{subfigure}
\caption{{Plot of $Z$ as a function of $k_x$ and $k_y$ }for the superposed state mentioned in Eq.~\ref{eq_superposed_ellipses_2} for $\epsilon_{\rm max}=0.75 $  and  (a)~~$\delta=0.005$~~~~(b)~~~$\delta=0.01$~~~(c)~~~$\delta=0.02$. }
\label{fig_quasi-particle_superposed_correlator_diff_delta}
\end{figure*}

\begin{figure}
    \centering
\includegraphics[width=.9\columnwidth]{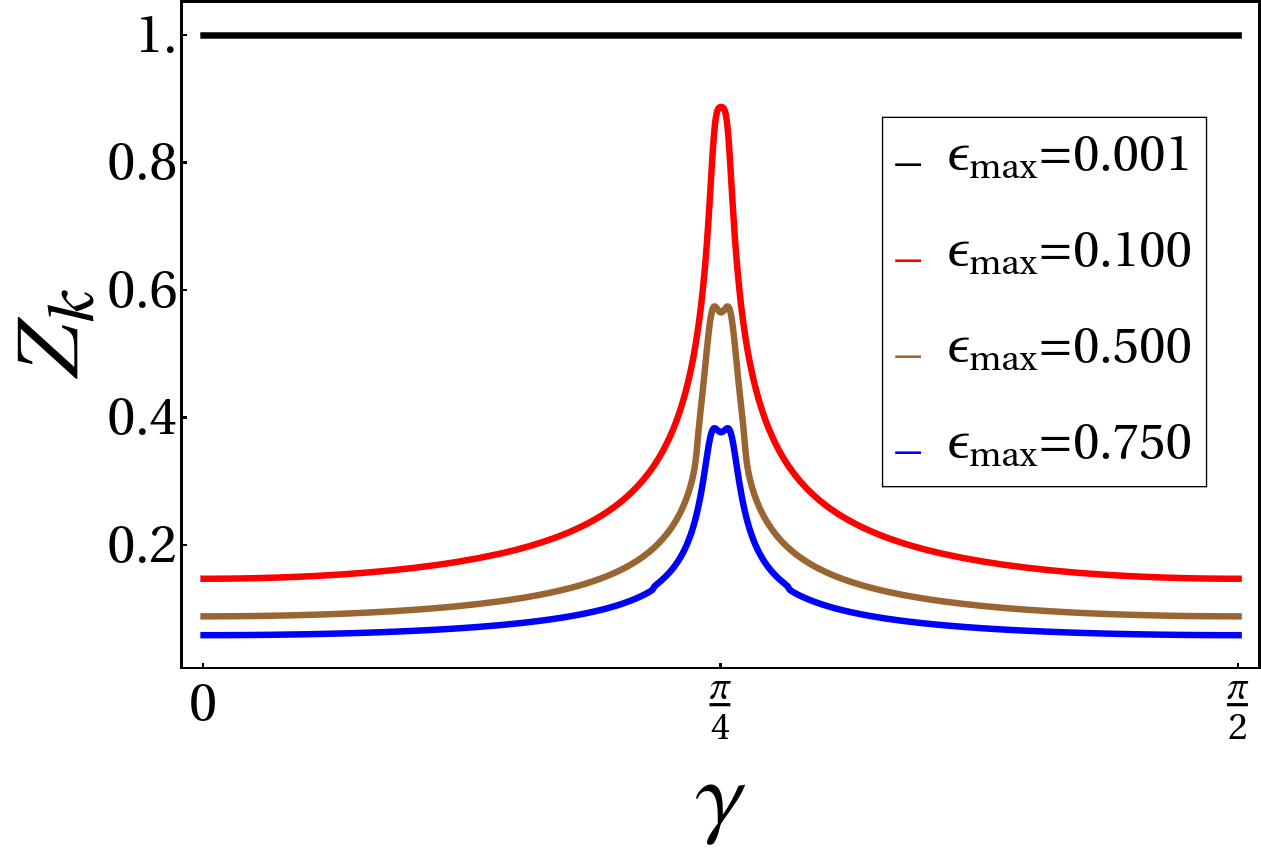}
    \caption{Plot of $Z_{\bf k}$(for $\delta=0.001$) corresponding to state Eq.~\ref{eq_superposed_ellipses_2} {as a function of $\gamma$} where ${\bf k} = (\sqrt{2\pi}\cos(\gamma),\sqrt{2\pi}\sin(\gamma))$ and $\gamma \in \{0,\frac{\pi}{2}\}$.}
\label{Z_delta_.001}
\end{figure}

\begin{figure}
    \centering
\includegraphics[width=.9\columnwidth]{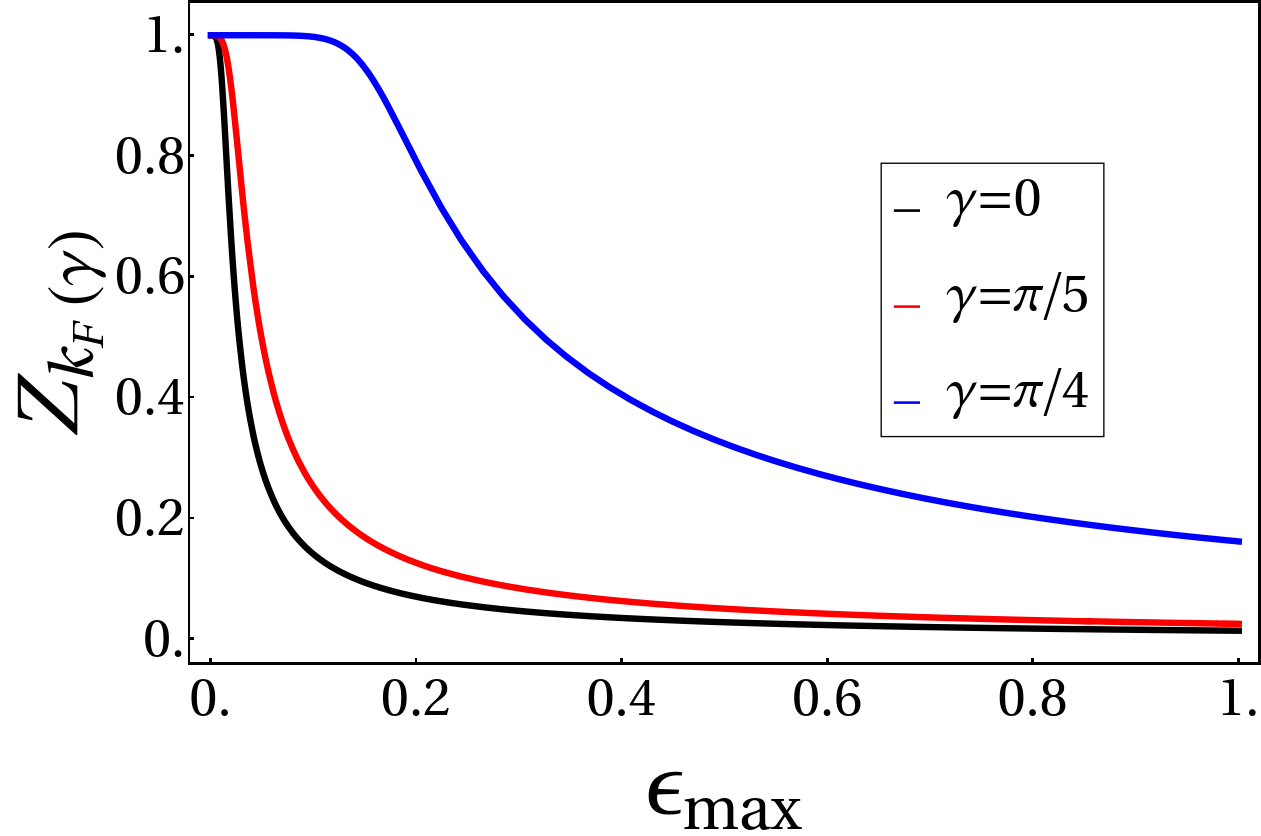}
    \caption{Plot of $ Z_{{\bf k}_{\bf F}(\gamma)}$ for superposed state mentioned in Eq.~\ref{eq_superposed_ellipses_2},  where ${\bf k}_{\bf F}(\gamma)=(\sqrt{2\pi}\cos(\gamma),\sqrt{2\pi}\sin(\gamma)))$ {as a function of $\epsilon_{\rm max}$ for different $\gamma$ }and $\delta=10^{-4}$.}
\label{Z_sqrt_theta}
\end{figure}

\begin{figure}
    \centering
\includegraphics[width=.9\columnwidth]{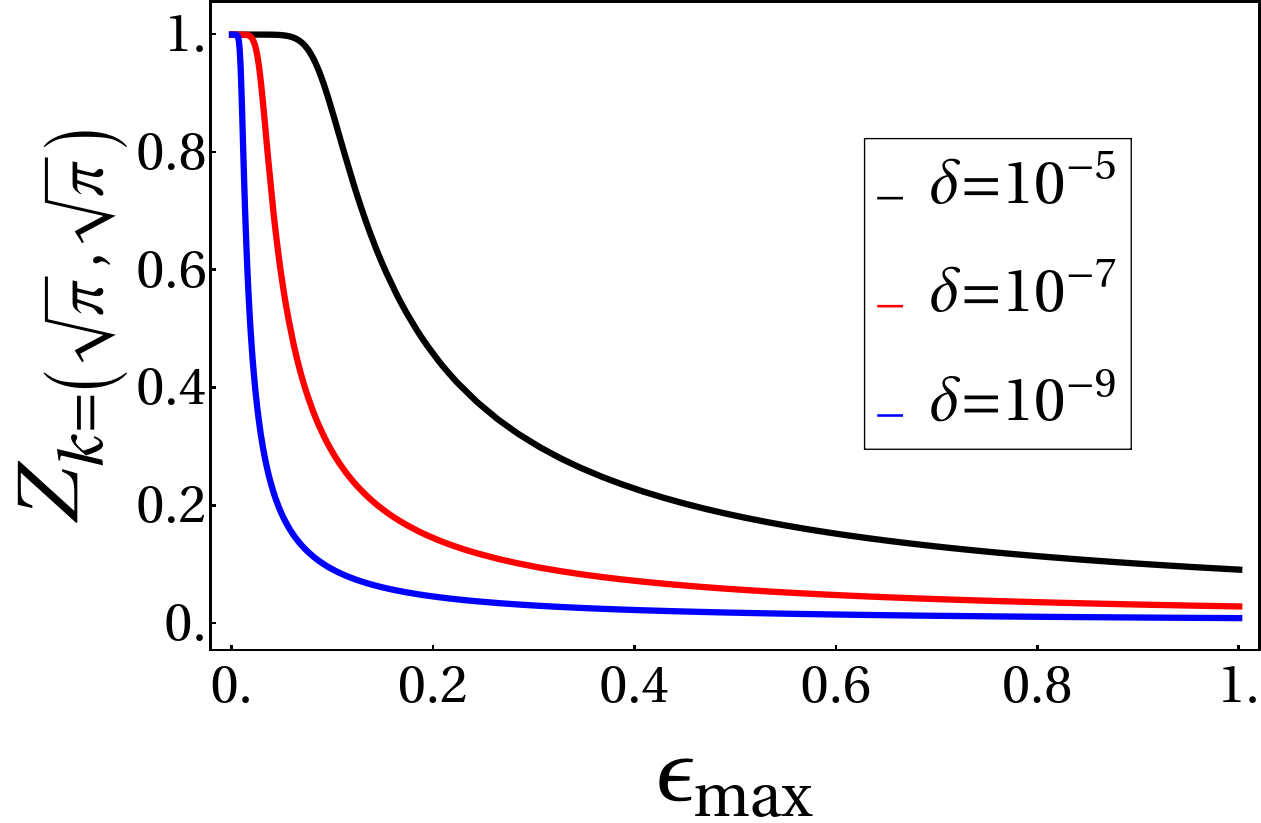}
    \caption{{Plot of $Z_{\bf k}({\bf k}=(\sqrt{\pi},\sqrt{\pi}))$ corresponding to Eq.~\ref{eq_superposed_ellipses_2} as a function of $\epsilon_{\rm max}$ for different $\delta$.}}
\label{Z_sqrt_pi_sqrt_pi}
\end{figure}

\begin{figure}
    \centering
\includegraphics[width=.9\columnwidth]{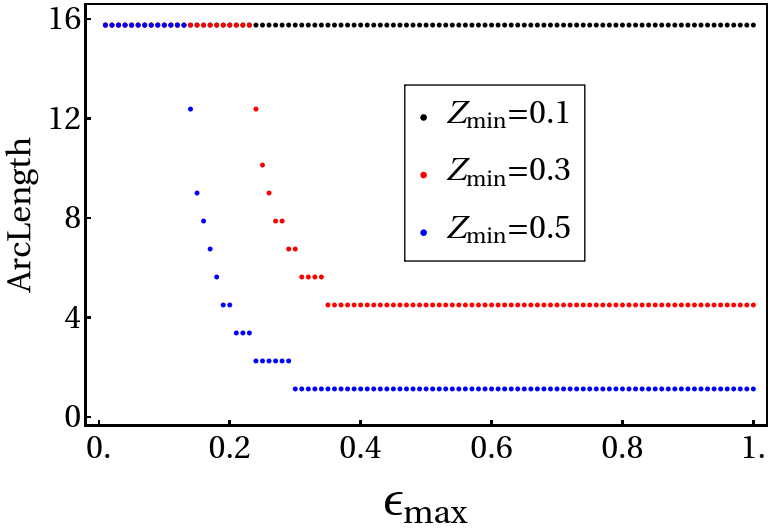}
    \caption{{Plot of the arc length corresponding to  Eq.~\ref{eq_superposed_ellipses_2} as a function of $\epsilon_{\rm max}$ with $\delta=0.001$ and for different $Z_{\rm min}$.}}
\label{arc_length_same_delta}
\end{figure}

For the state in Eq.\ \ref{eq_superposed_ellipses_2},  Fig.~\ref{fig_quasi-particle_superposed_correlator} shows that the superposition of FSs led to anisotropy in the quasi-particle residue (at finite resolution $\delta$). Here we present further details of the same as a function of the two parameters $\epsilon_{\rm max}$ and $\delta$. 

{From the plot of $\langle n_{\bf k} \rangle$ as a function of $k$, shown in Fig.~\ref{nk_theta}, we note that for $\gamma=\frac{\pi}{4}$, $\langle n_{\bf k} \rangle$ vanishes sharply, but smoothly. This is because in $\pi/4$ direction, radius of circle is max distance from origin. For $\epsilon \neq 0$, it lies inside circle where it is smeared; such a smearing does not occur outside the circle.}

{
The finite resolution residue (calculated using Eq.\ \ref{eq_netz}) is plotted in Fig.\ \ref{fig_quasi-particle_superposed_correlator} for different $\epsilon_{\rm max}$ with a given resolution $\delta$ and as a function of $\delta$ in Fig.\ \ref{fig_quasi-particle_superposed_correlator_diff_delta}. Fig.\ \ref{Z_delta_.001} shows a plot of $Z_{\bf {k}}(\delta=0.001)$ as a function of $\gamma$ for different $\epsilon_{\rm max}$ indicating the direction dependence of $Z_{\bf k}$. As we move away from $\gamma=\frac{\pi}{4}$ direction, the value of the residue decreases rapidly. This is also seen by plotting $Z_{\bf k}$ as a function of $\epsilon_{\rm max}$ for given $\delta$ (Fig.\  \ref{Z_sqrt_theta}). At the momentum of highest smeared jump, we also plot the variation of the magnitude of jump as a function of $\epsilon_{\rm max}$ for different resolution in Fig.\ \ref{Z_sqrt_pi_sqrt_pi}. The plot shows that for any appreciable superposition, the residue indeed goes to zero. However, it goes to zero much more slowly compared to other angles as shown in Fig.\ \ref{fig_Zk_delta}(c). }

The {\it length} of each of the segment of finite resolution residue has been computed as follows. We first normalize $Z_{\rm Norm}=Z_{\bf k}/Z_{\rm max}$. Then, while computing the length, we put a lower cut off, $Z_{\rm min}$ on $Z_{\rm Norm}$ and discard all $Z_{\rm Norm}$ below that cutoff. Fig.\ \ref{arc_length_same_delta} shows the width of segment, or the arc length, as a function of $\epsilon_{\rm max}$ for same $\delta$ but with different lower cutoffs $Z_{\rm min}$. We find that the maximum arc length is that of the perimeter of circle ($2\pi\sqrt{2\pi}$). 

\subsection{Density-density Correlation in real space}
\label{appen_density2}

The connected density-density correlator in real space, $\langle\langle n_{{\bf r}_1}n_{{\bf r}_2}\rangle\rangle\equiv W({\bf r_1,r_2})$ is given by
\begin{align}\label{eq_density_density_correlator_sup_ellipse}
 \langle \psi_{\nu} \vert n_{\bf r_1}n_{\bf r_2} \vert \psi_{\nu} \rangle - \langle \psi_{\nu} \vert n_{\bf r_1}\vert \psi_{\nu} \rangle \langle \psi_{\nu} \vert n_{\bf r_2} \vert \psi_{\nu} \rangle ,&
\end{align}
for a many-Fermion state $|\psi_\nu\rangle$ at filling $\nu$. For an elliptical FS given by $|\epsilon, \nu, \hat{\bf k}\rangle$ in the main text, we have
\begin{align}
    W({\bf r_1,r_2}) =  \nu \delta^2({\bf r_1}-{\bf r_2}) -|F(\epsilon,r,\nu,{\bf \hat{k}})|^2,
\end{align}
where,
\begin{align}
&F(\epsilon,r,\nu,{\bf \hat{k}})=2\nu \Bigg(\frac{J_1 \big(r t_{\bf \hat{k}} \sqrt{4 \pi \nu} \big)}{\big(r t_{\bf \hat{k}} \sqrt{4 \pi \nu } \big)}\Bigg),
\end{align}
and
\begin{align}
& t_{\bf \hat{k}} = \frac{\sqrt{1-\epsilon^2\sin^2({u-\gamma})}}{(1-\epsilon^2)^{1/4}},
\end{align}
with
\begin{align}
& \tan{u}= \frac{{\bf \hat{k}}_y}
{{\bf \hat{k}}_x
},~~r=|{\bf r_1-r_2}|, ~\tan{\gamma} = \frac{y_2-y_1}{x_2-x_1}.
\end{align}

Next, turning to superposed FS wave-functions (Eq.\ \ref{eq_superposed_ellipses}) the density-density correlator is given by
\begin{align}
    W =& \nu \delta^2({\bf r_1}-{\bf r_2})\nonumber\\ &~~~~~~+\int_{0}^{\epsilon_{\rm max}} d\epsilon \int_{0}^{\pi} d\hat{\bf k}~\vert \psi(\epsilon,\hat{\bf k})\vert^2 |F(\epsilon,r,\nu,{\bf \hat{k}})|^2.
\end{align}
This completes our discussion of computation of these correlators.

\bibliography{ref}

\end{document}